\documentclass[prd,nofootinbib,showpacs,preprint,superscriptaddress]{revtex4}
\usepackage[T1]{fontenc}
\usepackage{amsmath,amssymb}
\usepackage{epsfig}
\usepackage{graphicx}
\usepackage[usenames,dvipsnames]{color}
\usepackage{slashed}
\usepackage[colorlinks,citecolor=blue]{hyperref}
\usepackage{pdfpages}
\usepackage{color}
\newcommand{\Tr}[1]{\text{Tr}\big[#1\big]}
\begin{document}
\title{Common Origin of Neutrino Mass and Dark Matter from Anomaly Cancellation Requirements of a $U(1)_{B-L}$ Model}
\author{Dibyendu Nanda}
\email{dibyendu.nanda@iitg.ernet.in}
\affiliation{Department of Physics, Indian Institute of Technology Guwahati, Assam 781039, India}
\author{Debasish Borah}
\email{dborah@iitg.ernet.in}
\affiliation{Department of Physics, Indian Institute of Technology Guwahati, Assam 781039, India}

\begin{abstract}
We study a gauged $B-L$ extension of the standard model where the new fermions with fractional $B-L$ charges that play the role of keeping the model anomaly free can also explain the origin of neutrino mass at one loop level as well as dark matter. We discuss two different versions of the model to realise fermion and scalar dark matter, both of which guarantee the dark matter stability by a remnant discrete symmetry to which $U(1)_{B-L}$ gauge symmetry gets spontaneously broken down to. Apart from giving rise to the observed neutrino mass and dark matter abundance, the model also has tantalising signatures at variety of experiments operating at cosmic, intensity and energy frontiers, particularly direct and indirect detection experiments of dark matter, rare decay experiments looking for charged lepton flavour violation as well as collider experiments. The model also predicts vanishing lightest neutrino mass that can be tested at experiments sensitive to the absolute neutrino mass scale.
\end{abstract}
\maketitle

\section{Introduction}
Tiny but non-zero neutrino masses and large leptonic mixing have been confirmed by several experiments for more than a decade till now \cite{Fukuda:2001nk, Ahmad:2002jz, Ahmad:2002ka, Abe:2008aa, Abe:2011sj, Abe:2011fz, An:2012eh, Ahn:2012nd, Adamson:2013ue, Olive:2016xmw}. Specially, the more recent experimental results from the T2K \cite{Abe:2011sj}, Double Chooz \cite{Abe:2011fz}, Daya Bay \cite{An:2012eh}, RENO \cite{Ahn:2012nd} and MINOS \cite{Adamson:2013ue} experiments have not only confirmed the results from earlier experiments but also discovered the non-zero reactor mixing angle $\theta_{13}$. As the latest global fit results \cite{Esteban:2016qun} say, the three leptonic mixing angles and two neutrino mass squared differences are known upto a great accuracy, upto a little uncertainty in the octant of the atmospheric mixing angle $\theta_{23}$. The two presently unknown quantities in the neutrino sector are the mass hierarchy: whether it is normal $(m_3 > m_2 > m_1)$ or inverted $(m_2 > m_1 > m_3)$, and the leptonic Dirac CP phase $\delta$ \footnote{A recent measurement hinted at $\delta \approx -\pi/2$ \cite{Abe:2015awa}.}. Apart from neutrino oscillation experiments, the neutrino sector is constrained by the data from cosmology as well as rare decay experiments also. For example, the latest data from the Planck mission constrains the sum of absolute neutrino masses $\sum_i \lvert m_i \rvert < 0.17$ eV \cite{Ade:2015xua}. On the other hand, search for neutrinoless double beta decay $(0\nu\beta\beta)$ at experiments like KamLAND-Zen \cite{KamLAND-Zen:2016pfg} also constrains the lightest neutrino mass to lie below $0.05-0.1$ eV. In fact, both the Planck and $(0\nu\beta\beta)$ bounds disfavour the quasi-degenerate spectrum of light neutrinos, showing preference for a hierarchical pattern. However, the $(0\nu\beta\beta)$ limit is applicable only for Majorana neutrinos, as such lepton number violating processes are absent if neutrinos are purely Dirac fermion.

Although the issues of neutrino mass hierarchy, Dirac CP phase and the nature of neutrino as a fermion (Dirac or Majorana) are not settled yet, we still have enough evidences to suggest that neutrinos (at least two of them) have tiny but non-zero mass and have large mixing. The Standard Model (SM) of particle physics, in spite of being established as the most successful theory of elementary particles and their interactions (except gravity), can not explain neutrino mass at the renormalisable level. The Higgs field, which lies at the origin of all massive particles in the SM, can not have any Dirac Yukawa coupling with the neutrinos due to the absence of the right handed neutrino. If the right handed neutrinos are included by hand, one needs the Yukawa couplings to be heavily fine tuned to around $10^{-12}$ in order to generate sub-eV neutrino masses from the same Higgs field of the SM. At non-renormalisable level, one can generate a tiny Majorana mass for the neutrinos from the same Higgs field of the SM through the dimension five Weinberg operator \cite{Weinberg:1979sa}. However, the unknown cut-off scale $\Lambda$ in such operators points towards the existence of new physics at some high energy scale. There have been many proposals beyond the Standard Model (BSM) where the effects of such higher dimensional operators can be realised within a renormalisable theory by incorporating the existence of additional fields. They are popularly known as the seesaw mechanism \cite{Minkowski:1977sc, GellMann:1980vs, Mohapatra:1979ia, Schechter:1980gr}. Apart from this conventional or type I seesaw, there exists other variants of seesaw mechanism also namely, type II seesaw \cite{Mohapatra:1980yp, Lazarides:1980nt, Wetterich:1981bx, Schechter:1981cv, Brahmachari:1997cq}, type III seesaw \cite{Foot:1988aq} and so on.

Apart from the problem of neutrino mass and leptonic mixing, another drawback the SM suffers from is its inability to explain the origin of dark matter (DM) in the Universe. In fact, the existence of DM has been known for a much longer time, starting from the galaxy cluster observations by Fritz Zwicky \cite{Zwicky:1933gu} back in 1933, observations of galaxy rotation curves in 1970's \cite{Rubin:1970zza}, the more recent observation of the bullet cluster \cite{Clowe:2006eq} to the latest cosmology data provided by the Planck satellite \cite{Ade:2015xua}. The latest data from the Planck mission suggest that DM gives rise to around $26\%$ of the present Universe's energy density. In terms of density parameter and $h = \text{(Hubble Parameter)}/(100 \;\text{km} \text{s}^{-1} \text{Mpc}^{-1})$, the present dark matter abundance is conventionally reported as \cite{Ade:2015xua}
\begin{equation}
\Omega_{\text{DM}} h^2 = 0.1198 \pm 0.0015.
\label{dm_relic}
\end{equation}
In spite of these irrefutable observational evidences from astrophysics and cosmology confirming the presence of DM in the Universe, the particle nature of DM is still a mystery. No laboratory experiment has so far been able to probe the particle DM directly. Although the particle DM is not yet discovered, the observations suggest that a particle should satisfy certain requirements to be a DM candidate, some of which can be found in \cite{Taoso:2007qk}. These criteria undoubtedly rules out all the particles in the SM from being DM candidates. Although the neutrinos satisfy some of these criteria yet they are too light to contribute $26\%$ of present Universe's energy density as well as to allow large scale structure formation. They only give rise to a tiny fraction of DM in the form of hot dark matter (HDM). This has led to a plethora of BSM proposals suggesting different particle DM candidates. The most widely studied scenario among them is perhaps the weakly interacting massive particle (WIMP) paradigm where the DM particle has mass and interactions typically in the electroweak scale. Due to such sizeable interactions, the particle DM can be in equilibrium with rest of the plasma in the early Universe and can give rise to a relic after thermal freeze-out. The remarkable matching of this relic with the observed DM abundance is popularly known as the \textit{WIMP Miracle}. For a recent review, one may refer to \cite{Arcadi:2017kky}. The sizeable interactions of WIMP DM with the SM particles can not only generate the correct relic abundance through thermal freeze-out, but also leads to its direct detection prospects as such DM particle can be produced at colliders like the Large Hadron Collider (LHC) or it can scatter off nuclei kept in a detector. Several ongoing efforts are dedicated to DM searches at the LHC \cite{Kahlhoefer:2017dnp} as well as direct detection experiments like LUX, PandaX-II and Xenon1T \cite{Akerib:2016vxi, Tan:2016zwf, Cui:2017nnn, Aprile:2017iyp}.

Although the origin of neutrino mass as well as leptonic mixing may be unrelated to the fundamental origin of DM, it is highly motivating to look for a common framework that can explain both the phenomena. This not only keeps the BSM physics minimal, but also allows for its probe in a much wider range of experiments. Motivated by this here we study a very well motivated BSM framework based on the gauged $U(1)_{B-L}$ symmetry, where $B$ and $L$ correspond to baryon and lepton numbers respectively. This minimal and economical model generating non-zero neutrino mass has been studied for a long time \cite{Wetterich:1981bx, Mohapatra:1980qe, Marshak:1979fm, Masiero:1982fi, Mohapatra:1982xz, Buchmuller:1991ce}. The most interesting feature of this model is that the inclusion of three right handed neutrinos, as it is done in type I seesaw mechanism of generating light neutrino masses, is no longer a choice but a necessity due to the requirement of the new $U(1)_{B-L}$ gauge symmetry to be anomaly free. The model has also been studied in the context of dark matter by several groups \cite{Rodejohann:2015lca, Okada:2010wd, Dasgupta:2014hha, Okada:2016tci, Klasen:2016qux}. DM in scale invariant versions of this model was also studied by several authors \cite{Okada:2012sg, Guo:2015lxa}. The interesting feature of such a model from DM point of view is the issue of DM stability that can be ensured in this model if the $U(1)_{B-L}$ gauge symmetry gets spontaneously broken down to a remnant discrete symmetry like $Z_2$, so that the lightest $Z_2$ odd particle can be stable. However, many $U(1)_{B-L}$ models also considered additional discrete symmetries to ensure DM stability, for example \cite{Basak:2013cga, Okada:2016gsh}. Although experimental limits from LEP II constrain such new gauge sector by giving a lower bound on the ratio of new gauge boson mass to the corresponding gauge coupling $M_{Z_{BL}}/g_{BL} \geq 7$ TeV \cite{Carena:2004xs, Cacciapaglia:2006pk}, there can be several interesting consequences of such extended gauge symmetry that can be probed at several ongoing experiments. Motivated by these interesting features of a $U(1)_{B-L}$ model, here we study another version of it, where the extra fermion singlet fields responsible for anomaly cancellations can not only be viable DM candidates, but also take part in generating tiny neutrino masses at one loop level. Such radiative neutrino mass scenarios where DM particles take part in the loop are commonly known as scotogenic models, after the first such proposal by Ma \cite{Ma:2006km}. A recent review of radiative neutrino mass models can be found in \cite{Cai:2017jrq}. The model that we study not only explains simultaneous origin of neutrino mass and DM without introducing additional discrete symmetries but also predicts the lightest neutrino mass to be zero. We discuss the possibility of both scalar doublet and fermion singlet DM in such a model and show how they can give rise to difference consequences, if the scalar sector is kept minimal. We also briefly discuss the new physics contribution to charged lepton flavour violation as well as the prospects of probing such a scenario at indirect detection experiments as well as colliders. 

This paper is organised as follows. In section \ref{sec0}, we briefly discuss the issue of triangle anomalies in a gauged $B-L$ model along with different possible solutions to make the model anomaly free. In section \ref{sec1}, we outline the minimal model to realise fermion dark matter along with radiative neutrino mass. In section \ref{sec2}, we briefly summarise the procedures to calculate the dark matter relic abundance followed by the discussion of fermion DM relic abundance in section \ref{sec3}. We then propose another version of the model in section \ref{sec4} that can have scalar doublet DM. We discuss the constraints from direct and indirect detection of DM in section \ref{sec5} and \ref{sec6} respectively. We then discuss the new physics contribution to charged lepton flavour violation in section \ref{sec7} and add a brief discussion on collider signatures of the model in section \ref{sec8}. We finally summarise our results and conclude in section \ref{sec9}.

\section{Gauged $B-L$ Symmetry}
\label{sec0}
As pointed out above, the $B-L$ gauge extension of the SM is a very natural and minimal possibility as the corresponding charges of all the SM fields under this new symmetry is well known. However, a $U(1)_{B-L}$ gauge symmetry with only the SM fermions is not anomaly free. This is because the triangle anomalies for both $U(1)^3_{B-L}$ and the mixed $U(1)_{B-L}-(\text{gravity})^2$ diagrams are non-zero. These triangle anomalies for the SM fermion content turns out to be
\begin{align}
\mathcal{A}_1 \left[ U(1)^3_{B-L} \right] = \mathcal{A}^{\text{SM}}_1 \left[ U(1)^3_{B-L} \right]=-3  \nonumber \\
\mathcal{A}_2 \left[(\text{gravity})^2 \times U(1)_{B-L} \right] = \mathcal{A}^{\text{SM}}_2 \left[ (\text{gravity})^2 \times U(1)_{B-L} \right]=-3
\end{align}
Remarkably, if three right handed neutrinos are added to the model, they contribute $\mathcal{A}^{\text{New}}_1 \left[ U(1)^3_{B-L} \right] = 3, \mathcal{A}^{\text{New}}_2 \left[ (\text{gravity})^2 \times U(1)_{B-L} \right] = 3$ leading to vanishing total of triangle anomalies. This is the most natural and economical $U(1)_{B-L}$ model where the fermion sector has three right handed neutrinos apart from the usual SM fermions and it has been known for a long time. However, there exists non-minimal ways of constructing anomaly free versions of $U(1)_{B-L}$ model. For example, it has been known for a few years that three right handed neutrinos with exotic $B-L$ charges $5, -4, -4$ can also give rise to vanishing triangle anomalies \cite{Montero:2007cd}. It is clear to see how the anomaly cancels, as follows.
\begin{align}
\mathcal{A}_1 \left[ U(1)^3_{B-L} \right] = \mathcal{A}^{\text{SM}}_1 \left[ U(1)^3_{B-L} \right]+\mathcal{A}^{\text{New}}_1 \left[ U(1)^3_{B-L} \right]=-3 + \left [ -5^3 - (-4)^3 - (-4)^3 \right]=0 \nonumber
\end{align}
\begin{align}
\mathcal{A}_2 \left[(\text{gravity})^2 \times U(1)_{B-L} \right] & = \mathcal{A}^{\text{SM}}_2 \left[ (\text{gravity})^2 \times U(1)_{B-L} \right]+ \mathcal{A}^{\text{New}}_2 \left[ (\text{gravity})^2 \times U(1)_{B-L} \right] \nonumber \\
&=-3 + \left[ -5 - (-4) - (-4) \right]=0
\end{align}
This model was also discussed recently in the context of neutrino mass \cite{Ma:2014qra, Ma:2015mjd} and DM \cite{Sanchez-Vega:2014rka, Sanchez-Vega:2015qva, Singirala:2017see, Nomura:2017vzp} by several groups. Another solution to anomaly conditions with irrational $B-L$ charges of new fermions was proposed by the authors of \cite{Wang:2015saa} where both DM and neutrino mass can have a common origin through radiative linear seesaw.

Very recently, another anomaly free $U(1)_{B-L}$ framework was proposed where the additional right handed fermions possess more exotic $B-L$ charges namely, $-4/3, -1/3, -2/3, -2/3$ \cite{Patra:2016ofq}. The triangle anomalies get cancelled as follows.
\begin{align}
\mathcal{A}_1 \left[ U(1)^3_{B-L} \right] &= \mathcal{A}^{\text{SM}}_1 \left[ U(1)^3_{B-L} \right]+\mathcal{A}^{\text{New}}_1 \left[ U(1)^3_{B-L} \right] \nonumber \\
&=-3 + \left [ -(-4/3)^3 - (-1/3)^3 - (-2/3)^3-(-2/3)^3 \right]=0 \nonumber
\end{align}
\begin{align}
\mathcal{A}_2 \left[(\text{gravity})^2 \times U(1)_{B-L} \right] & = \mathcal{A}^{\text{SM}}_2 \left[ (\text{gravity})^2 \times U(1)_{B-L} \right]+ \mathcal{A}^{\text{New}}_2 \left[ (\text{gravity})^2 \times U(1)_{B-L} \right] \nonumber \\
&=-3 + \left[ -(-4/3) - (-1/3) - (-2/3) - (-2/3) \right]=0
\end{align}
One can have even more exotic right handed fermions with $B-L$ charges $-17/3, 6, -10/3$ so that the triangle anomalies cancel as
\begin{align}
\mathcal{A}_1 \left[ U(1)^3_{B-L} \right] &= \mathcal{A}^{\text{SM}}_1 \left[ U(1)^3_{B-L} \right]+\mathcal{A}^{\text{New}}_1 \left[ U(1)^3_{B-L} \right] \nonumber \\
&=-3 + \left [ -(-17/3)^3 - (6)^3 - (-10/3)^3 \right]=0 \nonumber
\end{align}
\begin{align}
\mathcal{A}_2 \left[(\text{gravity})^2 \times U(1)_{B-L} \right] & = \mathcal{A}^{\text{SM}}_2 \left[ (\text{gravity})^2 \times U(1)_{B-L} \right]+ \mathcal{A}^{\text{New}}_2 \left[ (\text{gravity})^2 \times U(1)_{B-L} \right] \nonumber \\
&=-3 + \left[ -(-17/3) - (6) - (-10/3)  \right]=0
\end{align}
Here we stick to the choice with four right handed neutrinos having $B-L$ charges $-4/3, -1/3, -2/3, -2/3$ and propose a common framework for the origin of neutrino mass and DM. In the original reference where this possibility was proposed \cite{Patra:2016ofq}, the origin of neutrino mass was considered to be through type II seesaw mechanism, which remains decoupled from the DM sector composed of the singlet fermions having exotic $B-L$ charges. The authors in fact constructed two Dirac fermions from the four singlet fermions by appropriately choosing the chirality and studied the corresponding DM phenomenology. Here instead of adding scalar triplet for type II seesaw, we consider the addition of scalar doublets having appropriate $B-L$ charges so that light neutrino mass can arise at one loop level, allowing the possibility of both scalar doublet and fermion singlet DM scenario.

\section{The Minimal Model}
\label{sec1}
In this section, we propose a $U(1)_{B-L}$ model based on the newly suggested anomaly cancellation solution with four right handed neutrinos having $B-L$ charges $-4/3, -1/3, -2/3, -2/3$, but with a minimal scalar content to realise one-loop neutrino mass. The fermion and scalar content of the model are shown in table \ref{tab:data1} and \ref{tab:data2} respectively.
\begin{table}
\begin{center}
\begin{tabular}{|c|c|}
\hline
Particles & $SU(3)_c \times SU(2)_L \times U(1)_Y \times U(1)_{B-L} $   \\
\hline
$q_L=\begin{pmatrix}u_{L}\\
d_{L}\end{pmatrix}$ & $(3, 2, \frac{1}{6}, \frac{1}{3})$  \\
$u_R$ & $(3, 1, \frac{2}{3}, \frac{1}{3})$  \\
$d_R$ & $(3, 1, -\frac{1}{3}, \frac{1}{3})$  \\

$\ell_L=\begin{pmatrix}\nu_{L}\\
e_{L}\end{pmatrix}$ & $(1, 2, -\frac{1}{2}, -1)$  \\
$e_R$ & $(1, 1, -1, -1)$ \\
\hline
$N_1$ & $(1, 1, 0, -\frac{1}{3})$ \\
$N_2$ & $(1, 1, 0, -\frac{2}{3})$ \\
$N_3$ & $(1, 1, 0, -\frac{2}{3})$ \\
$N_4$ & $(1, 1, 0, -\frac{4}{3})$ \\
\hline

\end{tabular}
\end{center}
\caption{Fermion Content of the Model}
\label{tab:data1}
\end{table}
\begin{table}
\begin{center}
\begin{tabular}{|c|c|}
\hline
Particles & $SU(3)_c \times SU(2)_L \times U(1)_Y \times U(1)_{B-L} $   \\
\hline
$H=\begin{pmatrix}H^+\\
H^0\end{pmatrix}$ & $(1,2,\frac{1}{2},0)$  \\
$\eta_1=\begin{pmatrix}\eta^+_1\\
\eta^0_1\end{pmatrix}$ & $(1,2,\frac{1}{2}, -\frac{1}{3})$  \\
$\eta_2=\begin{pmatrix}\eta^+_2\\
\eta^0_2\end{pmatrix}$ & $(1,2,\frac{1}{2}, \frac{1}{3})$  \\
\hline
$\phi_1$ & $(1, 1, 0, 1)$ \\
$\phi_2$ & $(1, 1, 0, 2)$ \\
\hline
\end{tabular}
\end{center}
\caption{Scalar content of the Minimal Model}
\label{tab:data2}
\end{table}

The Yukawa Lagrangian for leptons can be written as 
\begin{align}
\mathcal{L}_Y &= Y_e \overline{\ell_L} H e_R + Y_{i4} (\overline{\ell_L})_i i\tau_2 \eta^*_1 N_4 + Y_{i2} (\overline{\ell_L})_i i\tau_2 \eta^*_2 N_2 + Y_{i3} (\overline{\ell_L})_i i\tau_2 \eta^*_2 N_3  \nonumber \\
& + f_1 \phi_1 N_1 N_2 + f_2 \phi_1 N_1 N_3 + f_3 \phi_2 N_2 N_4 + f_4 \phi_2 N_3 N_4 + \text{h.c.}
\label{yukawa1}
\end{align}

The scalar potential of the model can be written as 
\begin{align}
V &= -\mu^2_H \lvert H \rvert^2 + \lambda_H \lvert H \rvert^4 + \sum_{i=1,2} \left( \mu^2_{\eta_i} \lvert \eta_i \rvert^2 + \lambda_{\eta_i} \lvert \eta_i \rvert^4 \right)+\sum_{i=1,2} \left( -\mu^2_{\phi_i} \lvert \phi_i \rvert^2 + \lambda_{\phi_i} \lvert \phi_i \rvert^4 \right) \nonumber \\
& +\sum_{i=1,2} \lambda_{H\eta_i} (\eta^{\dagger}_i \eta_i) (H^{\dagger} H)+\sum_{i=1,2}\lambda^{\prime}_{H\eta_i} (\eta^{\dagger}_i H) (H^{\dagger} \eta_i)  + \sum_{i=1,2}\lambda_{H\phi_i} (\phi^{\dagger}_i \phi_i) (H^{\dagger} H) \nonumber \\
& + \lambda_{H \eta} \left ( (\eta^T_1 H^{\dagger})(\eta^T_2 H^{\dagger}) +\text{h.c.} \right) + \sum_{i,j=1,2} \lambda_{\eta_i \phi_j} (\eta^{\dagger}_i \eta_i)(\phi^{\dagger}_j \phi_j) +\lambda_{\phi} (\phi^{\dagger}_1 \phi_1)(\phi^{\dagger}_2 \phi_2) \nonumber \\
& + \mu_{\phi} \left ( \phi_1 \phi_1 \phi^{\dagger}_2 + \text{h.c.} \right)
\end{align}

We choose the mass squared terms of $\eta_{1,2}$ to be positive so that the neutral components of only $H, \phi_1, \phi_2$ acquire non-zero vacuum expectation value (vev). We denote these vev's as 
$$\langle H \rangle = \frac{v}{\sqrt{2}}\begin{pmatrix}0\\
1\end{pmatrix}, \;\; \langle \phi_1 \rangle = \frac{u_1}{\sqrt{2}}, \;\;  \langle \phi_2 \rangle = \frac{u_2}{\sqrt{2}} $$
The minimisation conditions of the above scalar potential corresponds to 
$$ \mu^2_H = \lambda_H v^2 + \lambda_{H \phi_1} \frac{u^2_1}{2} +  \lambda_{H \phi_2} \frac{u^2_2}{2} $$
$$ \mu^2_{\phi_1} = \lambda_{\phi_1} u^2_1 +\lambda_{\phi} \frac{u^2_2}{2} + \lambda_{H \phi_1} \frac{v^2}{2} + \sqrt{2} \mu_{\phi} u_2 $$
$$ \mu^2_{\phi_2} = \lambda_{\phi_2} u^2_2 +\lambda_{\phi} \frac{u^2_1}{2} + \lambda_{H \phi_2} \frac{v^2}{2} + \frac{1}{\sqrt{2}} \mu_{\phi} \frac{u^2_1}{u_2} $$
Writing down the kinetic terms of the relevant scalar fields as 
\begin{align}
\mathcal{L}_{\text{kinetic}} & \supset \lvert (\partial_{\mu} + i g \vec{T} \cdot \vec{W} + i g^{\prime} Y B_{\mu}) H \rvert^2 + \lvert (\partial_{\mu} +i g_{BL} Z^{\prime}_{\mu}) \phi_1 \rvert^2 \nonumber \\
& +  \lvert (\partial_{\mu} +i 2 g_{BL} Z^{\prime}_{\mu}) \phi_2 \rvert^2
\end{align}
where $T_i = \sigma_i/2$, we can find out the masses of gauge bosons as
$$ M_W = \frac{1}{2}g v, \;\; M_Z = \frac{1}{2} g v \sqrt{1+\left(\frac{g^{\prime}}{g} \right)^2}, \;\; M_{Z_{BL}} =  g_{BL} \sqrt{u^2_1+4u^2_2} $$
The neutral scalar mass matrix constructed from the singlet Higgs fields is given by
\begin{equation}
M^2_{\phi_r}=\begin{pmatrix}
 2\lambda_{\phi_1} u^2_1 & u_1(\lambda_{\phi}u_2+\sqrt{2} \mu_{\phi})  \\
 u_1(\lambda_{\phi}u_2+\sqrt{2} \mu_{\phi}) & 2\lambda_{\phi_2} u^2_2-\frac{\mu_{\phi} u^2_1}{\sqrt{2} u_2}
 \end{pmatrix} 
  \end{equation}
 The neutral pseudoscalar mass matrix constructed from the singlet Higgs fields is given by
\begin{equation}
M^2_{\phi_i}=\begin{pmatrix}
-2\sqrt{2} u_2 \mu_{\phi} & \sqrt{2} u_1 \mu_{\phi} \\
\sqrt{2} u_1 \mu_{\phi} & -\frac{u^2_1 \mu_{\phi}}{\sqrt{2} u_2}
\end{pmatrix}
 \end{equation}
which clearly gives rise to a vanishing eigenvalue, corresponding to the Goldstone boson that gets converted into the longitudinal mode of the $U(1)_{B-L}$ gauge boson. After the electroweak symmetry breaking, these mass matrices become $3\times 3$ due to mixing with the components of the Higgs doublet $H$. The neutral scalar mass matrix becomes 
\begin{equation}
M^2_{(H \phi)_r}=\begin{pmatrix}
2\lambda_H v^2 & \lambda_{H \phi_1} u_1 v & \lambda_{H \phi_2} u_2 v \\
\lambda_{H \phi_1} u_1 v & 2\lambda_{\phi_1} u^2_1 & u_1(\lambda_{\phi}u_2+\sqrt{2} \mu_{\phi})  \\
 \lambda_{H \phi_2} u_2 v & u_1(\lambda_{\phi}u_2+\sqrt{2} \mu_{\phi}) & 2\lambda_{\phi_2} u^2_2-\frac{\mu_{\phi} u^2_1}{\sqrt{2} u_2}
 \end{pmatrix}. 
  \end{equation}
Assuming the third neutral Higgs to be very heavy and decoupled, we can find the mixing between the light and next to lightest neutral Higgs (in the small mixing limit) as 
\begin{equation}
\tan{2 \theta_1} \approx 2 \sin{\theta_1} \approx 2 \xi = \frac{2 \lambda_{H \phi_1} u_1 v}{2\lambda_{\phi_1} u^2_1-2\lambda_H v^2}.
\label{Hphimixing}
\end{equation}
The mixing parameter $\xi$ plays a non-trivial role in DM phenomenology as we discuss later. Such a mixing can be tightly constrained by LEP as well as LHC Higgs exclusion searches as shown recently by \cite{Dupuis:2016fda}. These constraints are more strong for low mass scalar and the upper bound on the mixing angle can be as low as $ \sin{\theta} < 0.1$ \cite{Dupuis:2016fda}. We consider a conservative upper limit on the mixing parameter $\xi \leq 0.1$ for our analysis. This can be easily satisfied by suitable tuning of the parameters involved in the expression for mixing given in \eqref{Hphimixing}.

The neutral scalar mass matrix constructed from the doublets $\eta_{1,2}$ is given by
\begin{equation}
M^2_{\eta_r}=\begin{pmatrix}
M^2_{r11} & M^2_{r12} \\
M^2_{r21} & M^2_{r22}
\end{pmatrix}
 \end{equation}
 where 
 $$ M^2_{r11}= \mu^2_{\eta_1} + (\lambda_{H \eta_1}+ \lambda^{\prime}_{H \eta_1}) \frac{v^2}{2} + \frac{1}{2}(\lambda_{\eta_1 \phi_1}u^2_1+\lambda_{\eta_1 \phi_2} u^2_2)$$
 $$ M^2_{r12} = M^2_{r21} = \lambda_{H \eta} \frac{v^2}{2} $$
 $$ M^2_{r22}= \mu^2_{\eta_2} + (\lambda_{H \eta_2}+ \lambda^{\prime}_{H \eta_2}) \frac{v^2}{2} + \frac{1}{2}(\lambda_{\eta_2 \phi_1}u^2_1+\lambda_{\eta_2 \phi_2} u^2_2)$$
  The pseudoscalar mass matrix constructed from the doublets $\eta_{1,2}$ is given by
\begin{equation}
M^2_{\eta_i}=\begin{pmatrix}
M^2_{i11} & M^2_{i12} \\
M^2_{i21} & M^2_{i22}
\end{pmatrix}
 \end{equation}
 where 
 $$ M^2_{i11}= \mu^2_{\eta_1} + (\lambda_{H \eta_1}+ \lambda^{\prime}_{H \eta_1}) \frac{v^2}{2} + \frac{1}{2}(\lambda_{\eta_1 \phi_1}u^2_1+\lambda_{\eta_1 \phi_2} u^2_2)$$
 $$ M^2_{i12} = M^2_{i21} = -\lambda_{H \eta} \frac{v^2}{2} $$
 $$ M^2_{i22}= \mu^2_{\eta_2} + (\lambda_{H \eta_2}+ \lambda^{\prime}_{H \eta_2}) \frac{v^2}{2} + \frac{1}{2}(\lambda_{\eta_2 \phi_1}u^2_1+\lambda_{\eta_2 \phi_2} u^2_2)$$
One can find the mass eigenstates of scalars and pseudoscalars using orthogonal rotations. For example, the scalar mass eigenstates are
$$ \eta^{\prime}_{1r} = \eta_{1r} \cos{\theta} -\eta_{2r} \sin{\theta}, \; \eta^{\prime}_{2r} = \eta_{1r} \sin{\theta} +\eta_{2r} \cos{\theta} $$
where 
$$\theta = \frac{1}{2}\tan^{-1} \left ( \frac{2M^2_{r12}}{M^2_{r22}-M^2_{r11}} \right)$$
Similarly one can find the mass eigenstates of the pseudoscalars with a rotation angle $\theta^{\prime}=-\theta$. It is straightforward to see that the scalars are degenerate with their pseudoscalar counterparts which can have serious consequences for dark matter physics as we discuss later.
\begin{figure}[!h]
\centering
\epsfig{file=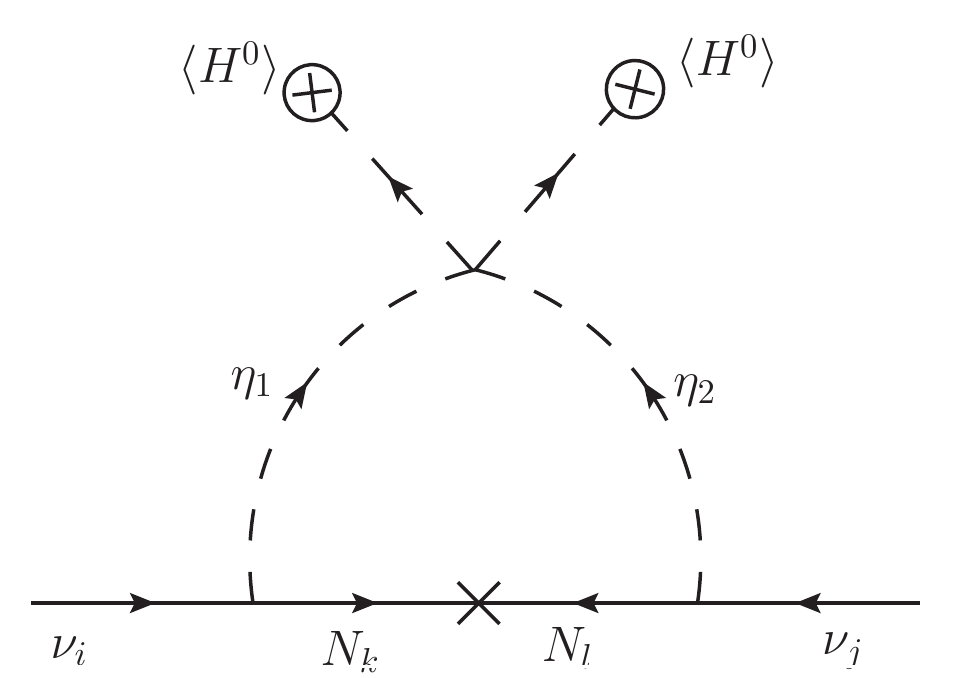,width=0.5\textwidth,clip=}
\caption{One loop neutrino mass for the particle content shown in table \ref{tab:data1}, \ref{tab:data2}}
\label{fig1}
\end{figure}
The one loop neutrino mass arising from the diagram in figure \ref{fig1} can be estimated as \cite{Ma:2006km}
\begin{equation}
(m_{\nu})_{ij} = \frac{Y_{ik}Y_{jk} M_{k}}{16 \pi^2} \left ( \frac{m^2_R}{m^2_R-M^2_k} \text{ln} \frac{m^2_R}{M^2_k}-\frac{m^2_I}{m^2_I-M^2_k} \text{ln} \frac{m^2_I}{M^2_k} \right)
\label{numass1}
\end{equation}
Here $m^2_{R,I}=m^2_{H,A}$ are the masses of scalar and pseudo-scalar part of $\eta^0_{1,2}$ and $M_k$ is the mass of singlet fermion $N_k$ in the internal line. The index $i, j = 1,2,3$ runs over the three fermion generations as well as three copies of $N_i$. For $m^2_{H}+m^2_{A} \approx M^2_k$, the above expression can be simply written as
\begin{equation}
(m_{\nu})_{ij} \approx  \frac{m^2_A-m^2_H}{32 \pi^2}\frac{Y_{ik}Y_{jk} }{M_k}
\label{numass2}
\end{equation}
From the discussion of scalar and pseudoscalar masses above, one can find the mass difference between $\eta^{\prime}_{1r}$ and $\eta^{\prime}_{2i}$ to be 
$$m^2_A-m^2_H = \sqrt{(M^2_{r22}-M^2_{r11})^2+4M^4_{r12}}$$
The light neutrino mass matrix, in the simplified approximation above \eqref{numass2} has a type I seesaw structure upto a loop suppression factor. The structure of this mass matrix can be obtained as 
$$ m_{\nu} = c_1 Y M^{-1} _RY^T, \;\; c_1 = \frac{m^2_A-m^2_H}{32 \pi^2}$$
where $Y, M_R$ can be found from the Yukawa Lagrangian \eqref{yukawa1} as
\begin{equation}
Y = \begin{pmatrix}
0 & Y_{12} & Y_{13} & Y_{14} \\
0 & Y_{22} & Y_{23} & Y_{24} \\
0 & Y_{32} & Y_{33} & Y_{34} 
\end{pmatrix}, \;\; M_R = \begin{pmatrix}
0 & f_1u_1 & f_2u_1 & 0\\
f_1u_1 & 0 & 0 & f_3u_2 \\
f_2 u_1 & 0 & 0 & f_4 u_2 \\
0 & f_3 u_2 & f_4 u_2 & 0 
\end{pmatrix}
\label{fermionmatrix1}
\end{equation}
The light neutrino mass matrix constructed from these mass matrices has one vanishing eigenvalue predicting the lightest neutrino mass to be zero. The non-vanishing masses can be kept within experimentally observed limits ($\sim 0.1$~eV), either by tuning the Yukawa couplings or the scalar-pseudoscalar mass difference while keeping the right handed
neutrino mass around the TeV scale. From the right handed neutrino mass matrix written above, it is also clear that there is a two-fold degeneracy in the masses with two pairs of right handed neutrinos having degenerate masses.

\section{Dark Matter}
\label{sec2}
The relic abundance of a dark matter particle $\rm DM$, which was in thermal equilibrium at some earlier epoch can be calculated by solving the Boltzmann equation
\begin{equation}
\frac{dn_{\rm DM}}{dt}+3Hn_{\rm DM} = -\langle \sigma v \rangle (n^2_{\rm DM} -(n^{\rm eq}_{\rm DM})^2)
\end{equation}
where $n_{\rm DM}$ is the number density of the dark matter particle $\rm DM$ and $n^{\rm eq}_{\rm DM}$ is the number density when $\rm DM$ was in thermal equilibrium. $H$ is the Hubble expansion rate of the Universe and $ \langle \sigma v \rangle $ is the thermally averaged annihilation cross section of the dark matter particle $\rm DM$. In terms of partial wave expansion $ \langle \sigma v \rangle = a +b v^2$. Numerical solution of the Boltzmann equation above gives \cite{Kolb:1990vq,Scherrer:1985zt}
\begin{equation}
\Omega_{\rm DM} h^2 \approx \frac{1.04 \times 10^9 x_F}{M_{\text{Pl}} \sqrt{g_*} (a+3b/x_F)}
\end{equation}
where $x_F = M_{\rm DM}/T_F$, $T_F$ is the freeze-out temperature, $M_{\rm DM}$ is the mass of dark matter, $g_*$ is the number of relativistic degrees of freedom at the time of freeze-out and and $M_{\text{Pl}} \approx 2.4\times 10^{18}$ GeV is the Planck mass. Dark matter particles with electroweak scale mass and couplings freeze out at temperatures approximately in the range $x_F \approx 20-30$. More generally, $x_F$ can be calculated from the relation 
\begin{equation}
x_F = \ln \frac{0.038gM_{\text{Pl}}M_{\rm DM}<\sigma v>}{g_*^{1/2}x_F^{1/2}}
\label{xf}
\end{equation}
which can be derived from the equality condition of DM interaction rate $\Gamma = n_{\rm DM} \langle \sigma v \rangle$ with the rate of expansion of the Universe $H \approx g^{1/2}_*\frac{T^2}{M_{Pl}}$. There also exists a simpler analytical formula for the approximate DM relic abundance \cite{Jungman:1995df}
\begin{equation}
\Omega_{\rm DM} h^2 \approx \frac{3 \times 10^{-27} cm^3 s^{-1}}{\langle \sigma v \rangle}
\label{eq:relic}
\end{equation}
The thermal averaged annihilation cross section $\langle \sigma v \rangle$ is given by \cite{Gondolo:1990dk}
\begin{equation}
\langle \sigma v \rangle = \frac{1}{8m^4T K^2_2(m/T)} \int^{\infty}_{4m^2}\sigma (s-4m^2)\surd{s}K_1(\surd{s}/T) ds
\end{equation}
where $K_i$'s are modified Bessel functions of order $i$, $m$ is the mass of Dark Matter particle and $T$ is the temperature.

If there exists some additional particles having mass difference close to that of DM, then they can be thermally accessible during the epoch of DM freeze out. This can give rise to additional channels through which DM can coannihilate with such additional particles and produce SM particles in the final states. This type of coannihilation effects on dark matter relic abundance were studied by several authors in \cite{Griest:1990kh, Edsjo:1997bg, Bell:2013wua}. Here we summarise the analysis of \cite{Griest:1990kh} for the calculation of the effective annihilation cross section in such a case. The effective cross section can given as 
\begin{align}
\sigma_{eff} &= \sum_{i,j}^{N}\langle \sigma_{ij} v\rangle r_ir_j \nonumber \\
&= \sum_{i,j}^{N}\langle \sigma_{ij}v\rangle \frac{g_ig_j}{g^2_{eff}}(1+\Delta_i)^{3/2}(1+\Delta_j)^{3/2}e^{\big(-x_F(\Delta_i + \Delta_j)\big)} \nonumber \\
\end{align}
where $x_F = \frac{m_{DM}}{T_F}$ and $\Delta_i = \frac{m_i-M_{\text{DM}}}{M_{\text{DM}}}$  and 
\begin{align}
g_{eff} &= \sum_{i=1}^{N}g_i(1+\Delta_i)^{3/2}e^{-x_F\Delta_i}
\end{align}
The masses of the heavier components of the inert Higgs doublet are denoted by $m_{i}$. The thermally averaged cross section can be written as
\begin{align}
\langle \sigma_{ij} v \rangle &= \frac{x_F}{8m^2_im^2_jM_{\text{DM}}K_2((m_i/M_{\text{DM}})x_F)K_2((m_j/M_{\text{DM}})x_F)} \times \nonumber \\
& \int^{\infty}_{(m_i+m_j)^2}ds \sigma_{ij}(s-2(m_i^2+m_j^2)) \sqrt{s}K_1(\sqrt{s}x_F/M_{\text{DM}}) \nonumber \\
\label{eq:thcs}
\end{align}

\section{Fermion Dark Matter in the Minimal Model}
\label{sec3}
The lightest of the four Majorana fermions in the minimal model is the DM in this scenario if it has mass smaller than the scalar doublets $\eta_{1,2}$. Since the singlet fermion mass matrix is not diagonal, as can be seen from \eqref{fermionmatrix1}, we first diagonalise this mass matrix for some benchmark structure. For example, we choose $ f_{1}=f, f_{2}=2f, f_{3}=2f, f_{4}=f, u_1=u_2$ so that the mass eigenstates of the right handed neutrinos become 
$$ \chi_{1}=\frac{1}{2}(N_{1} - N_{2} - N_{3}+N_{4})$$
$$ \chi_{2}=\frac{1}{2}(N_{1} + N_{2} + N_{3}+N_{4})$$
$$ \chi_{3}=\frac{1}{2}(N_{1} - N_{2} + N_{3}+N_{4})$$
$$ \chi_{4}=\frac{1}{2}(N_{1} + N_{2} - N_{3}+N_{4})$$
having masses $-3fu_1, 3fu_1, -fu_1, fu_1$ respectively. Using the known interactions of $N_{1,2,3,4}$ we then find out the possible interactions of the lightest among $\chi$'s in order to calculate its relic abundance. Here we show the results for $\chi_4$ fermion dark matter though the results for $\chi_3$ having same mass is exactly similar.

\begin{figure}[!h]
\centering
\begin{tabular}{cc}
\epsfig{file=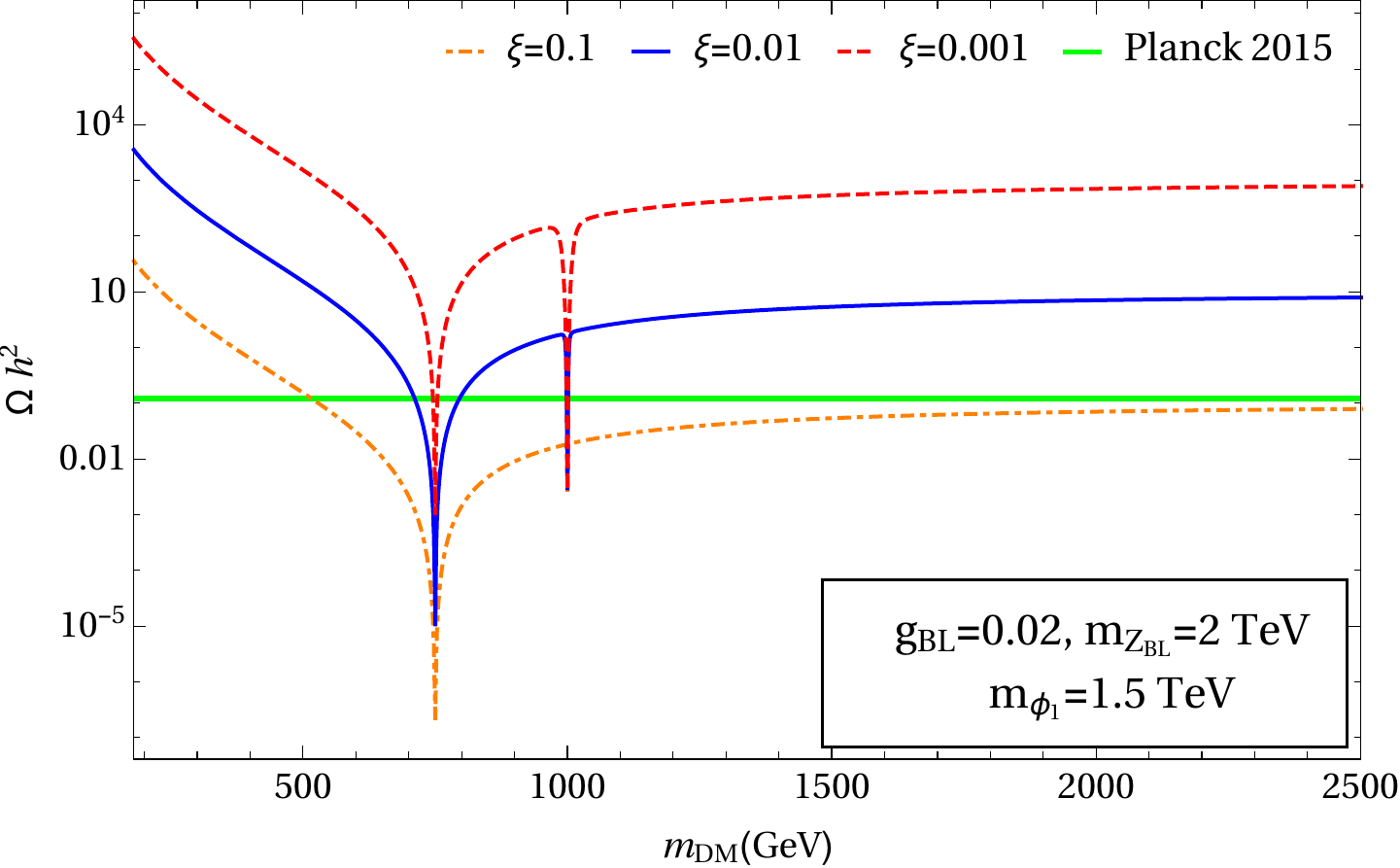,width=0.50\textwidth,clip=}
\epsfig{file=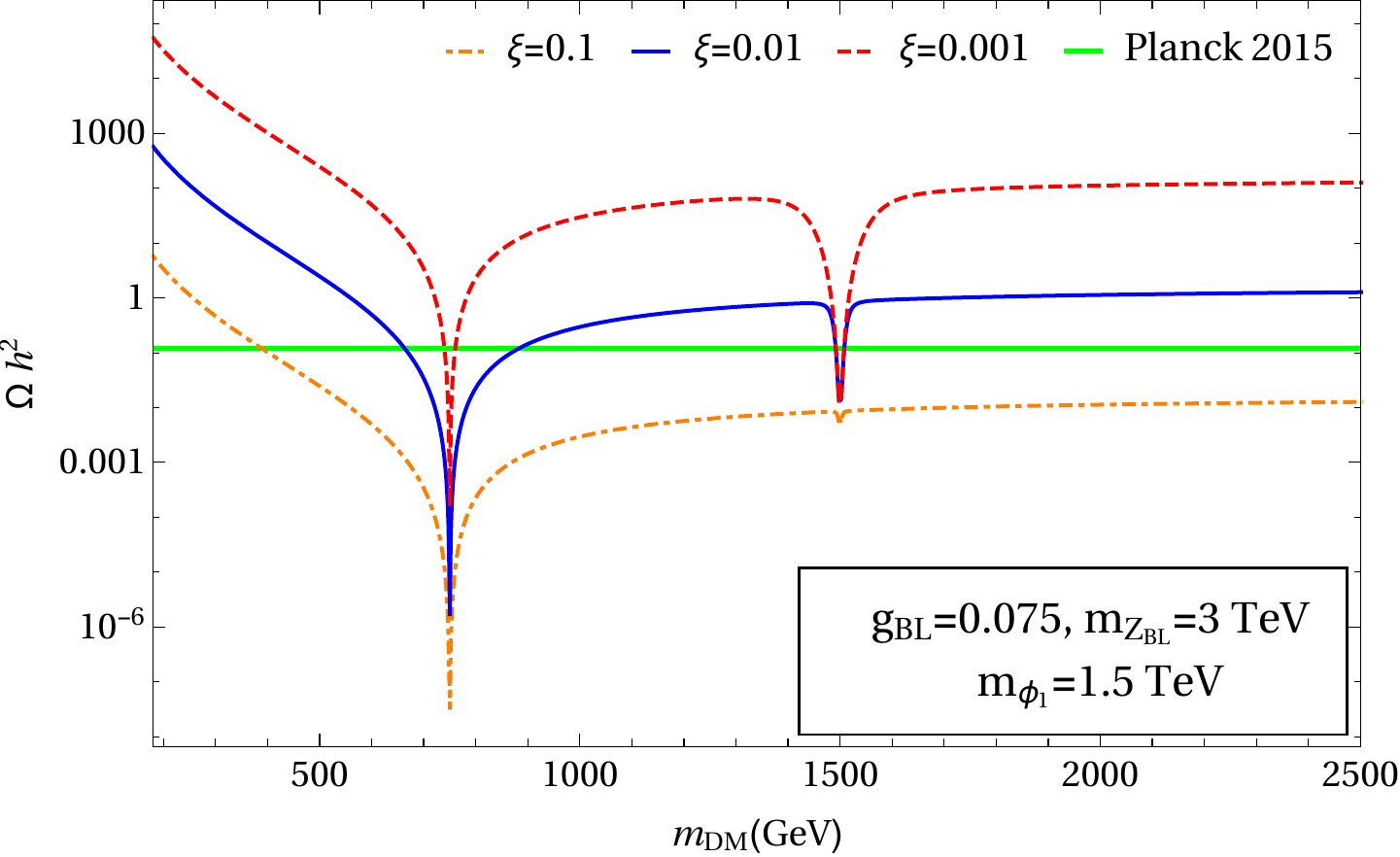,width=0.50\textwidth,clip=}
\end{tabular}
\caption{Fermion DM relic abundance as a function of DM mass for different benchmark values of free parameters.}
\label{fig2}
\end{figure}

Fermion DM can annihilate into SM particles either through $Z_{BL}$ gauge boson or through singlet scalar (denoted by $\phi_1$, the lightest of the neutral singlet scalars) by virtue of its mixing with the SM Higgs, parametrised by $\xi$. The relevant interaction vertices are given in appendix \ref{appen1}. We choose benchmark values for $M_{Z_{BL}}, g_{BL}, m_{\phi_1}, \xi$ and show the relic abundance of fermion DM as a function of its mass in figure \ref{fig2}. The resonances due to $2M_{DM} = M_{Z_{BL}}, 2M_{DM}=m_{\phi_1}$ are clearly visible in the plots, leading to minimum of the relic abundance as expected. The shallow nature of the $2M_{DM} = M_{Z_{BL}}$ resonance is due to the larger decay width of $Z_{BL}$ compared to the singlet scalar $\phi_1$.

\section{Scalar Dark Matter in a non-Minimal Model}
\label{sec4}
We can not have scalar dark matter in the minimal model discussed above, due to constraints from direct detection experiments. This is because the neutral scalar and pseudoscalars of both the doublets $\eta_{1,2}$ remain degenerate. This will give rise to a large scattering of scalar DM off nuclei through Z boson mediation, which is ruled out by the strong constraints coming from direct detection experiments, which we discuss in details in the next section. Here we consider a slight modification of the previous model so that a mass splitting can be introduced between the neutral scalar and pseudoscalar components of the scalar doublet. To break the degeneracy between scalar and pseudoscalar components of the doublet scalar, required to avoid $Z$ boson mediated direct detection scattering, we simply incorporate the presence of a scalar triplet $\Delta$ with hypercharge 1 and $U(1)_{B-L}$ charge $\pm 2/3$. This will allow a term $\mu \eta_1 \eta_1 \Delta^{\dagger}$ or $\mu \eta_2 \eta_2 \Delta^{\dagger}$ breaking the degeneracy between scalar-pseudoscalar components of $\eta_1, \eta_2$ respectively. However this also allows other terms like $\eta_1 H \Delta \phi_1$ making the scalar DM unstable. Therefore we change the singlet scalar charges of the minimal model apart from incorporating the triplet. Also, it turns out that in this case, the correct neutrino mass can be generated at one loop with just one inert scalar doublet $\eta$. The modified scalar content of the model is shown in table \ref{tab:data3}.
\begin{table}
\begin{center}
\begin{tabular}{|c|c|}
\hline
Particles & $SU(3)_c \times SU(2)_L \times U(1)_Y \times U(1)_{B-L} $   \\
\hline
$H=\begin{pmatrix}H^+\\
H^0\end{pmatrix}$ & $(1,2,\frac{1}{2},0)$  \\
$\eta=\begin{pmatrix}\eta^+\\
\eta^0\end{pmatrix}$ & $(1,2,\frac{1}{2}, \frac{1}{3})$  \\
$\Delta = \begin{pmatrix}
 \delta^+/\surd 2 & \delta^{++} \\
 \delta^0 & -\delta^+/\surd 2
 \end{pmatrix}$ & $(1, 3, 1, \frac{2}{3})$  \\
\hline
$\phi_1$ & $(1, 1, 0, \frac{4}{3})$ \\
$\phi_2$ & $(1, 1, 0, 2)$ \\
\hline
\end{tabular}
\end{center}
\caption{Scalar content of the non-Minimal Model}
\label{tab:data3}
\end{table}

The Yukawa Lagrangian for leptons in this model can be written as 
\begin{align}
\mathcal{L}_Y &= Y_e \overline{\ell_L} H e_R + Y_{i2} (\overline{\ell_L})_i i\tau_2 \eta^* N_2 + Y_{i3} (\overline{\ell_L})_i i\tau_2 \eta^* N_3 + f_1 \phi_1 N_2 N_2 \nonumber \\
& + f_2 \phi_1 N_3 N_3 + f_3 \phi_1 N_2 N_3 + f_4 \phi_2 N_2 N_4+f_5 \phi_2 N_3 N_4 + \text{h.c.}
\label{yukawa2}
\end{align}
It is clear from this Yukawa Lagrangian that the heavy neutrino $N_1$ remains massless at renormalisable level. One can however, introduce non-renormalisable operators contributing to the right handed neutrino masses as 
$$ f_6 \frac{1}{\Lambda} \phi^{\dagger}_1 \phi_2 N_1 N_1 + f_7 \frac{1}{\Lambda} \phi^2_1 \phi_2 N_4 N_4 $$
where $\Lambda$ is the unknown cut-off scale above the scale of $U(1)_{B-L}$ symmetry. The scalar potential of the model can be written as 
\begin{align}
V &= -\mu^2_H \lvert H \rvert^2 + \lambda_H \lvert H \rvert^4 + \left( \mu^2_{\eta} \lvert \eta \rvert^2 + \lambda_{\eta} \lvert \eta \rvert^4 \right)+\sum_{i=1,2} \left( -\mu^2_{\phi_i} \lvert \phi_i \rvert^2 + \lambda_{\phi_i} \lvert \phi_i \rvert^4 \right) \nonumber \\
& + \lambda_{H\eta} (\eta^{\dagger} \eta) (H^{\dagger} H)+\lambda^{\prime}_{H\eta} (\eta^{\dagger} H) (H^{\dagger} \eta)  + \sum_{i=1,2}\lambda_{H\phi_i} (\phi^{\dagger}_i \phi_i) (H^{\dagger} H) \nonumber \\
& + \mu_{\Delta \eta}\left ( (\eta^T \Delta^{\dagger} \eta) +\text{h.c.} \right) + \sum_{j=1,2} \lambda_{\eta \phi_j} (\eta^{\dagger} \eta)(\phi^{\dagger}_j \phi_j) +\lambda_{\phi} (\phi^{\dagger}_1 \phi_1)(\phi^{\dagger}_2 \phi_2) \nonumber \\
& + \mu^2_{\Delta} \Tr{\Delta^{\dagger} \Delta} +\lambda_{\Delta1} \big[\Tr{\Delta^{\dagger} \Delta}\big]^2+\lambda_{\Delta2} \Tr{\big[\Delta^{\dagger} \Delta\big]^2}+ \Tr{\Delta^{\dagger} \Delta} \big[ \lambda_{\Delta H} (H^{\dagger}H) \nonumber \\
& + \lambda_{\Delta \eta} (\eta^{\dagger}\eta) +\sum_{i=1,2} \lambda_{\Delta \phi_i} (\phi^{\dagger}_i \phi_i) \big]
\label{scalarpot2}
\end{align}

Also, the smallness of the vev of the neutral component of $\Delta$ does not arise naturally in the form of an induced vev after electroweak symmetry breaking. This is due to the absence of trilinear coupling of the form $H H \Delta^{\dagger}$ in the model. However, one needs to keep the vev of left triplet scalar small as the constraints from electroweak $\rho$ parameter restricts it to $v_{\delta} \leq 2$ GeV \cite{Olive:2016xmw}. In the Standard Model, the $\rho$ parameter is unity at tree level, given by 
$$ \rho = \frac{M^2_{W}}{M^2_{Z} \cos^2 \theta_W} $$
where $\theta_W$ is the Weinberg angle. But in the presence of left scalar triplet vev, there arises additional contribution to the electroweak gauge boson masses which results in a departure of the $\rho$ parameter from unity at tree level.
$$ \rho = \frac{1+\frac{2v^2_{\delta}}{v^2}}{1+\frac{4v^2_{\delta}}{v^2}} $$
Experimental constraints on the $\rho$ parameter $\rho = 1.00040 \pm 0.00024$ \cite{Olive:2016xmw} forces one to have $v_{\delta} \leq 2$ GeV. Since, this can not be generated as an induced vev (which can be naturally small), one has to fine tune the quartic couplings and bare mass term of $\Delta$ scalar in order to generate such a small vev. However, if we introduce higher dimensional operators, then it is possible to generate an induced vev due to the existence of $\frac{1}{\Lambda} (H^T \Delta^{\dagger} H) (\phi^{\dagger}_2 \phi_1)$ term. Using the notations for vev as before, we can write down the induced vev as
$$ \langle \delta^0 \rangle =\frac{v_{\delta}}{\sqrt{2}}=\frac{v^2 u_1 u_2}{4 M^2_{\Delta} \Lambda} $$
This gives additional contribution to $W$ boson mass as 
$$M^2_W = \frac{g^2v^2}{4}+\frac{g^2 v^2_{\delta}}{2} $$
The neutral gauge boson mass matrix in the basis $(W_{3\mu}, B_{\mu}, B^{\prime}_{\mu})$ is given by
\begin{equation}
M^2_0 = \begin{pmatrix}
\frac{1}{4}g^2v^2+g^2 v^2_{\delta} & -\frac{1}{4}g g^{\prime} v^2-g g^{\prime} v^2_{\delta} & \frac{2}{3} g g_{BL} v^2_{\delta} \\
-\frac{1}{4}g g^{\prime} v^2-g g^{\prime} v^2_{\delta} &  \frac{1}{4}  g'^2 v^2 + g'^2 v^2_{\delta} & -\frac{2}{3} g^{\prime} g_{BL} v^2_{\delta} \\
\frac{2}{3} g g_{BL} v^2_{\delta} & -\frac{2}{3} g^{\prime} g_{BL} v^2_{\delta} & \frac{4}{9} g^2_{BL} v^2_{\delta} + g^2_{BL} \left( \frac{16}{9} u^2_1+ 4u^2_2 \right)
\end{pmatrix}
\end{equation}
In the limit $v_{\delta} \ll v \ll u_{1,2}$, the non-zero eigenvalues are 
$$M^2_Z \approx \frac{1}{4}(g^2+g'^2)(v^2+4v^2_{\delta})+\frac{1}{4}(g^2+g'^2) \frac{v^2v^2_{\delta}}{4u^2_1+9u^2_2} $$
$$M^2_{Z_{BL}} \approx \frac{4}{9} g^2_{BL} (4u^2_1+9u^2_2+v^2_{\delta}) $$
The vanishing eigenvalue corresponds to the massless photon. The other two mixing angles are given by
$$\cos^2{\theta_W} =\frac{g^2v^2+4g^2v^2_{\delta}}{4M^2_Z} $$
$$ \tan{\theta_m} = \frac{8g_{BL} v^2_{\delta}}{3 g' (v^2+4v^2_{\delta})}$$
where $\theta_W$ corresponds to the usual Weinberg angle whereas $\theta_m$ corresponds to the mixing between $Z$ and $Z_{BL}$ gauge bosons. Clearly, this mixing is zero in the limit of $v_{\delta} \rightarrow 0$. The required value of $v_{\delta}$ is obtained from the mass splitting between scalar-pseudoscalar components of $\eta$. From the scalar potential given in \eqref{scalarpot2}, the mass splitting between $\eta^{0r}, \eta^{0i}$ where $\eta^0 =(\eta^{0r}+i\eta^{0i})/\sqrt{2}$ is given by
$$M^2_{\eta^{0i}}-M^2_{\eta^{0r}} = 2\sqrt{2} \mu_{\Delta \eta} v_{\delta}$$
Thus, even if the mass splitting is as large as 50 GeV, considered in this work, we can generate a small $v_{\delta} < 2$ GeV, by tuning the trilinear mass parameter $\mu_{\eta \Delta}$ accordingly. Even for such maximum possible value of $v_{\delta}$ and $g_{BL} \sim g'$, the mixing between $Z-Z_{BL}$ comes out to be 
$$ \tan{\theta_m} \approx 1.7 \times 10^{-4}$$
We therefore, ignore the effects due to such gauge mixing in our calculations.

\begin{figure}[!h]
\centering
\epsfig{file=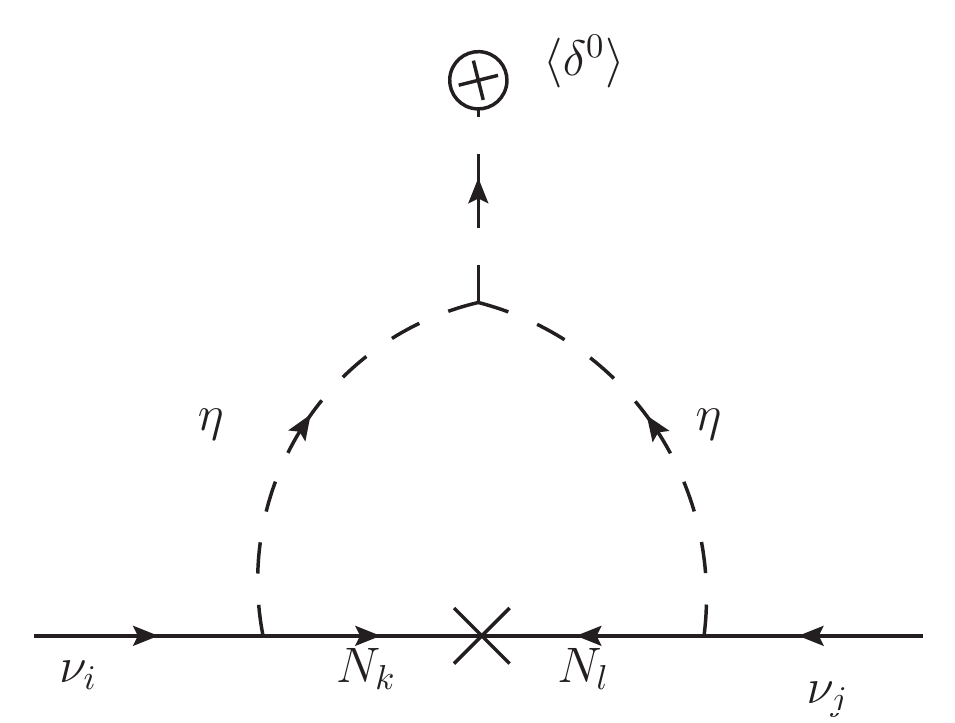,width=0.5\textwidth,clip=}
\caption{One loop neutrino mass for the particle content shown in table \ref{tab:data1}, \ref{tab:data3}}
\label{fignu2}
\end{figure}
Light neutrino masses can be generated at one loop level through the diagram shown in figure \ref{fignu2}. Similar to earlier analysis, here also the approximate structure of the light neutrino mass matrix can be obtained as 
$$ m_{\nu} = c_1 Y M^{-1} _RY^T, \;\; c_1 = \frac{M^2_{\eta^{0i}}-M^2_{\eta^{0r}}}{32 \pi^2}=\frac{2\sqrt{2} \mu_{\Delta \eta} v_{\delta}}{32\pi^2}$$
where $Y, M_R$ are can be found from the Yukawa Lagrangian \eqref{yukawa2} in $(\nu_e, \nu_{\mu}, \nu_{\tau}), (N_2, N_3, N_4)$ basis as
\begin{equation}
Y = \begin{pmatrix}
Y_{12} & Y_{13} & 0 \\
Y_{22} & Y_{23} & 0\\
Y_{32} & Y_{33} & 0 
\end{pmatrix}, \;\; M_R = \begin{pmatrix}
f_1u_1 & f_3u_1 & f_4 u_2\\
f_3u_1 & f_2 u_1 & f_4u_2 \\
f_4u_2 & f_4 u_2 &  f_7 \frac{u^2_1}{\Lambda} 
\end{pmatrix}
\end{equation}
It is clear that the light neutrino mass matrix constructed from these give rise to a vanishing eigenvalue that is, a scenario with vanishing lightest neutrino mass. This is a similar prediction like the minimal model discussed before. On the other hand, one of the right handed neutrinos $N_1$ can remain light in this scenario, as its mass is generated only by dimension five operators. Such a light right handed neutrino can have interesting cosmological consequences.

The components of dark scalar doublet $\eta$ acquire masses as
\begin{align}
M_{\eta^\pm}^2 = \mu^2_{\eta} + \frac{1}{2}\lambda_{H\eta} v^2+\frac{1}{2}\lambda_{\eta \phi_1} u^2_1+\frac{1}{2}\lambda_{\eta \phi_2} u^2_2 , \nonumber 
\end{align}
\begin{align}
M_{\eta^{0r}}^2 &=  \mu^2_{\eta} +\frac{1}{2} (\lambda_{H\eta}+\lambda'_{H\eta})v^2+\sqrt{2} \mu_{\Delta \eta} v_{\delta}+\frac{1}{2}\lambda_{\eta \phi_1} u^2_1+\frac{1}{2}\lambda_{\eta \phi_2} u^2_2 \nonumber \\
&=M_{\eta^\pm}^2+\frac{1}{2} \lambda'_{H\eta}v^2+\sqrt{2} \mu_{\Delta \eta} v_{\delta}, \nonumber
\end{align}
\begin{align}
M_{\eta^{0i}}^2 &= \mu^2_{\eta} +\frac{1}{2} (\lambda_{H\eta}+\lambda'_{H\eta})v^2-\sqrt{2} \mu_{\Delta \eta} v_{\delta}+\frac{1}{2}\lambda_{\eta \phi_1} u^2_1+\frac{1}{2}\lambda_{\eta \phi_2} u^2_2 \nonumber \\
& =M_{\eta^\pm}^2+\frac{1}{2} \lambda'_{H\eta}v^2-\sqrt{2} \mu_{\Delta \eta} v_{\delta}.
\label{massETA}
\end{align}
Without loss of generality, we consider $\eta^{0r}$ as the DM candidate which implies $\mu_{\Delta \eta} <0, \frac{1}{2}\lambda'_{H\eta}v^2+\sqrt{2} \mu_{\Delta \eta} v_{\delta} <0$ so that the CP even neutral scalar $\eta^{0r}$ is the lightest among the components of $\eta$ and hence a stable DM candidate.

The new scalar fields discussed above can be constrained from the LEP I precision measurement of the $Z$ boson decay width. In order to forbid the decay channel $Z \rightarrow \eta^{0r} \eta^{0i}$, one arrives at the constraint $M_{\eta^{0r}} + M_{\eta^{0r}} > M_Z$. In addition to this, the LEP II constraints roughly rule out the triangular
region \cite{Lundstrom:2008ai}
\[
	M_{\eta^{0r}} < 80\ {\rm\ GeV},\quad M_{\eta^{0i}} < 100{\rm\ GeV},\quad
	M_{\eta^{0i}} - M_{\eta^{0r}} > 8{\rm\ GeV}
\]
The LEP collider experiment data restrict the charged scalar mass to $m_{\eta^\pm} > 70-90$ GeV \cite{Pierce:2007ut}. The Run 1 ATLAS dilepton limit on such charged component of additional scalar doublets have also been discussed in \cite{Belanger:2015kga} taking into consideration of specific masses of charged Higgs. Another important restriction on $M_{\eta^\pm}$ comes from the electroweak precision data (EWPD). Since the contribution of the additional scalar doublet $\eta$ to electroweak S parameter is always small \cite{Barbieri:2006dq}, we only consider the contribution to the electroweak T parameter here. The relevant contribution is given by \cite{Barbieri:2006dq}
\begin{equation}
\Delta T = \frac{1}{16 \pi^2 \alpha v^2} [F(M_{\eta^\pm}, M_{\eta^{0i}})+F(M_{\eta^\pm}, M_{\eta^{0r}}) -F(M_{\eta^{0i}}, M_{\eta^{0r}})]
\end{equation}
where 
\begin{equation}
F(m_1, m_2) = \frac{m^2_1+m^2_2}{2}-\frac{m^2_1m^2_2}{m^2_1-m^2_2} \text{ln} \frac{m^2_1}{m^2_2}
\end{equation}
The EWPD constraint on $\Delta T$ is given as \cite{LopezHonorez:2010tb}
\begin{equation}
-0.1 < \Delta T + T_h < 0.2
\end{equation}
where $T_h \approx -\frac{3}{8 \pi \cos^2{\theta_W}} \text{ln} \frac{m_h}{m_Z}$ is the SM Higgs contribution to the T parameter \cite{Peskin:1991sw}.

Another important bound on such additional stable scalars can come from the LHC measurements of the SM Higgs invisible decay width. However, this constraint is applicable only for dark matter mass $M_{DM} < m_h/2$. The invisible decay width is given by
\begin{equation}
\Gamma (h \rightarrow \text{Invisible})= {\lambda^2_L v^2\over 64 \pi m_h} 
\sqrt{1-4\,M^2_{DM}/m^2_h}
\end{equation}
Here, $\lambda_L = (\lambda_{H\eta}+\lambda'_{H\eta})$. The latest constraint on invisible Higgs decay from the ATLAS experiment at the LHC is \cite{Aad:2015pla}
$$\text{BR} (h \rightarrow \text{Invisible}) = \frac{\Gamma (h \rightarrow \text{Invisible})}{\Gamma (h \rightarrow \text{Invisible}) + \Gamma (h \rightarrow \text{SM})} < 22 \%$$. We incorporate this bound on light scalar DM mass in the next section.

\begin{figure}[!h]
\centering
\begin{tabular}{cc}
\epsfig{file=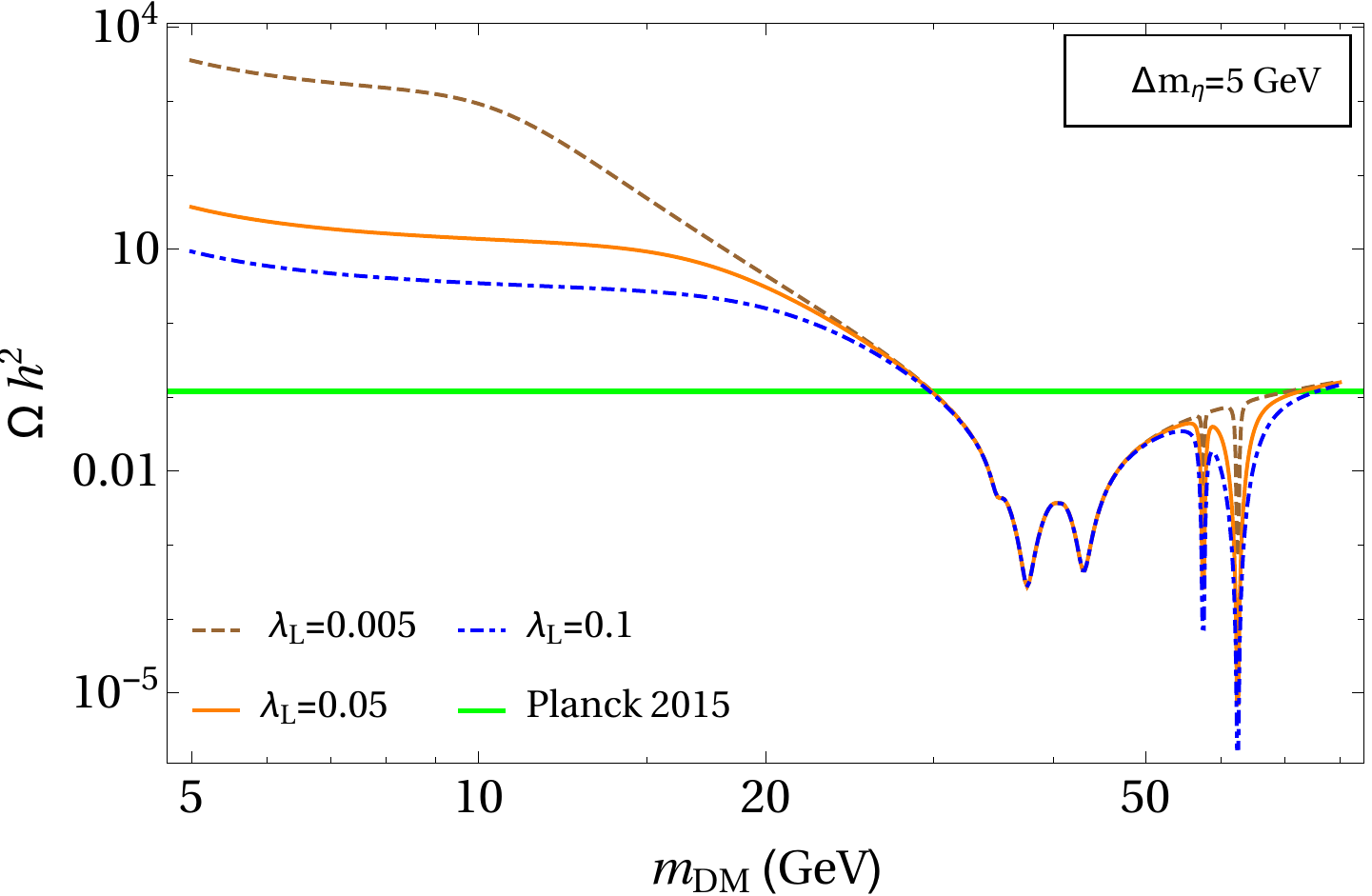,width=0.50\textwidth,clip=}
\epsfig{file=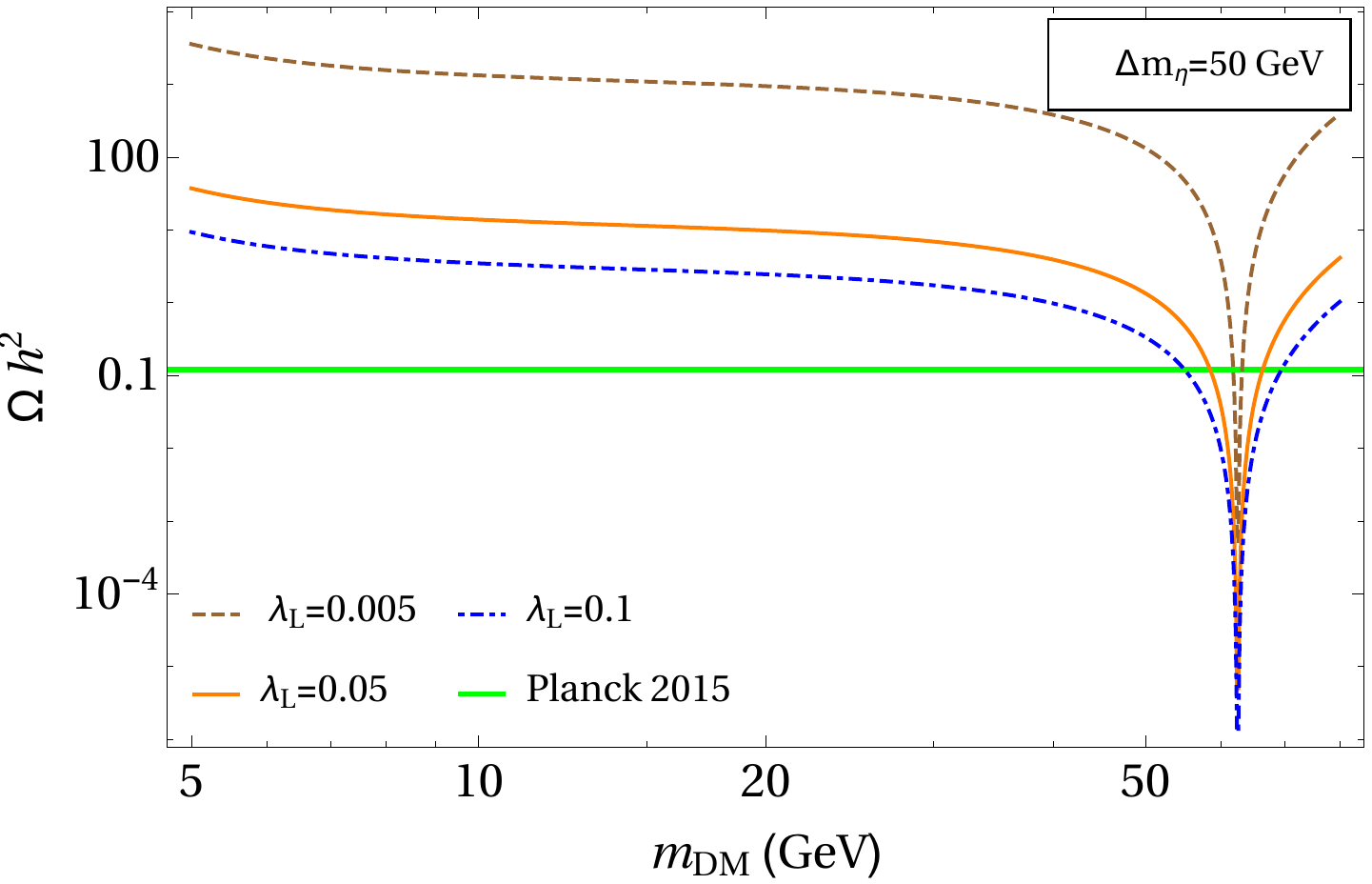,width=0.50\textwidth,clip=}
\end{tabular}
\caption{Relic abundance of scalar doublet dark matter $\eta^{0r}$ as a function of its mass for different benchmark values of DM-Higgs coupling, in the low mass regime. The mass splitting between $\eta^{0i}, \eta^{\pm}$ and $\eta^{0r}$ is fixed at 5 GeV, 50 GeV in the left and right panel plots respectively.}
\label{fig5a}
\end{figure}
\begin{figure}[!h]
\centering
\epsfig{file=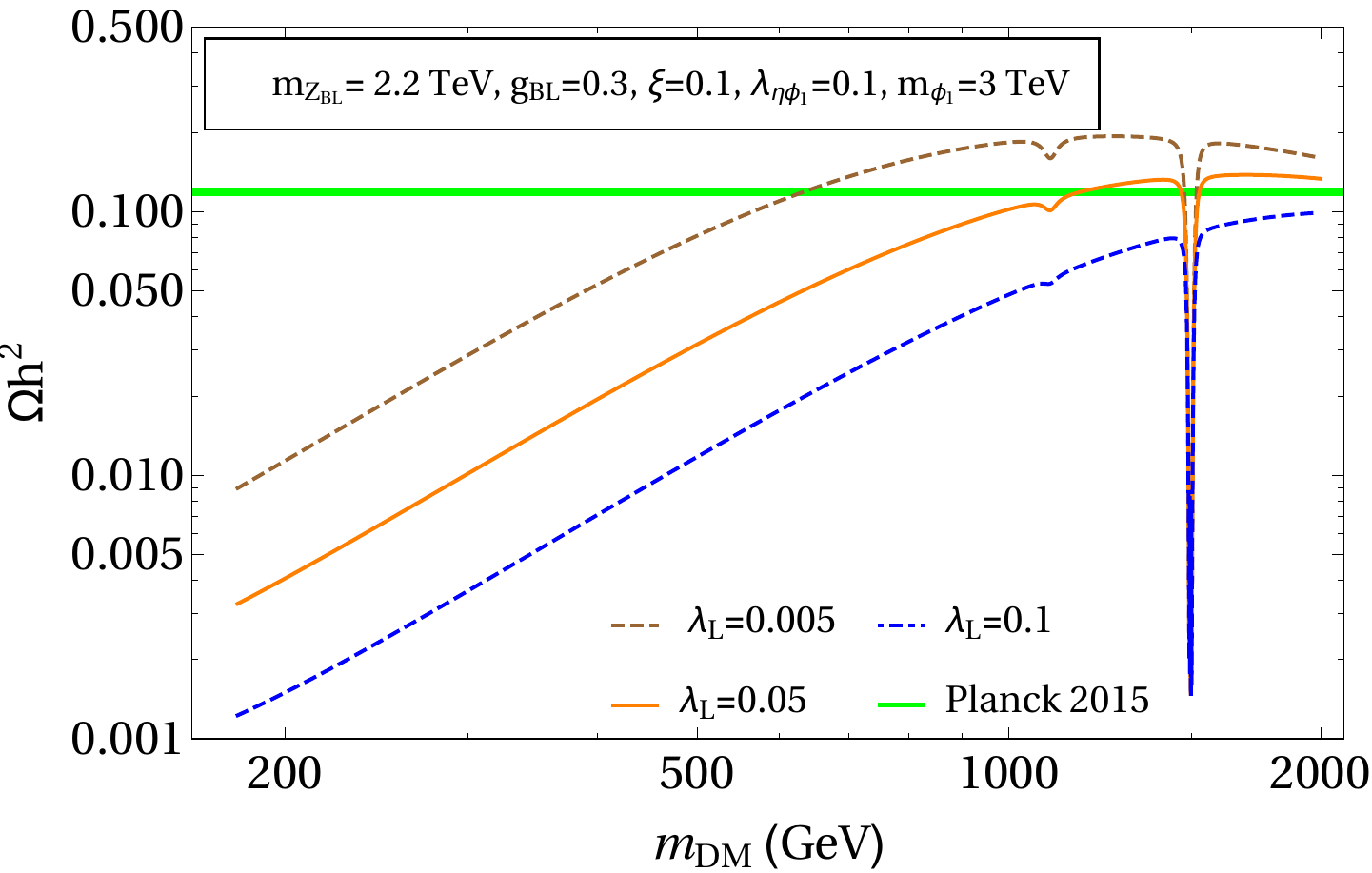,width=0.95\textwidth,clip=}
\caption{Relic abundance of scalar doublet dark matter $\eta^{0r}$ as a function of its mass for different benchmark values of DM-Higgs coupling, in the high mass regime. The mass splitting between $\eta^{0i}, \eta^{\pm}$ and $\eta^{0r}$ is fixed at 5 GeV.}
\label{fig6a}
\end{figure}

The relic abundance of scalar DM for some benchmark values of mass splitting, DM-Higgs couplings are shown in figure \ref{fig5a}, \ref{fig6a} for low and high mass regimes respectively. The effects of coannihilations between different components of the scalar doublet $\eta$ are very dominant for small mass splitting $\Delta m_{\eta} = 5$ GeV, as can be seen from the left panel of the figure \ref{fig5a}. The multiple resonances shown in the left panel of figure \ref{fig5a} corresponds to $W^{\pm}, Z$ and SM Higgs (h) mediated (co)annihilations. As we increase the mass splitting to 50 GeV, such coannihilation effects disappear and only the DM annihilation through SM Higgs remains, as can be seen from the single resonance shown in the right panel plot of figure \ref{fig5a}. In fact, for low mass DM such small mass splitting is disfavoured from LEP II data as mentioned above. However, we calculate the relic abundance in this regime to see the interesting differences originating from enhanced coannihilations. In the high mass regime, apart from the usual Higgs portal and electroweak gauge boson portal interactions, we also include the $Z_{BL}$ mediated annihilations for benchmark values of $M_{Z_{BL}}, g_{BL}$. As seen from the figure \ref{fig6a}, the impact of this $Z_{BL}$ portal on scalar DM relic abundance is very minimal. The relic abundance in the high mass regime is mainly dominated by the usual DM-Higgs coupling $\lambda_L$ and the mass splitting $\Delta m_{\eta}$. We also take the singlet scalar $\phi_1$ mediated annihilations of scalar DM and its contribution to the relic abundance. As can be seen from figure \ref{fig6a}, this gives rise to a new resonance due to s channel mediation of $\phi_1$ resulting in a new allowed region of scalar doublet DM in the high mass regime. Due to smaller decay width of $\phi_1$ compared to $Z_{BL}$, this resonance is efficient enough to bring the usually overproduced DM in high mass regime to the allowed range.

\section{Direct Detection of Dark Matter}
\label{sec5}
Apart from the relic abundance constraints from Planck experiment, there exists strict bounds on the dark matter nucleon cross section from direct detection experiments like LUX \cite{Akerib:2016vxi}, PandaX-II \cite{Tan:2016zwf, Cui:2017nnn} and Xenon1T \cite{Aprile:2017iyp}. For fermion dark matter, there are two ways through which it can scatter off nuclei: one is mediated by $Z_{BL}$ and the other mediated by scalars. The scalar mediated interactions occur due to mixing of singlet scalars of the model with the SM Higgs boson. Both these interactions give rise to spin independent DM-nuclei scattering only due to the absence of axial or pseudoscalar type couplings of the mediators with the quarks. Adopting the general formalism given in \cite{Berlin:2014tja}, the $Z_{BL}$ mediated DM-nucleon cross section is found to be
\begin{equation}
\sigma_{SI} = \frac{g^4_{BL} v^2 m^2_{\chi} m^4_n}{2 \pi M^4_{Z_{BL}} (m_n+m_{\chi})^4}
\label{DDBL}
\end{equation}
where $m_n$ is the mass of nucleon, $m_{\chi}$ is the DM mass and $v \approx 0.1\%c$ is the typical speed of DM particle. The scalar mediated DM-nucleon scattering cross section is 
\begin{equation}
\sigma_{SI}= \frac{\lambda^2_{\chi \phi} \mu_{\chi n}^2 \xi^2}{\pi m_{\phi_1}^4} f^2_{n,p}\end{equation}
where $\mu_{\chi n}$ is the reduced mass of the DM-nucleon system, $m_{\phi_1}$ is the mass of the mediator and $\xi$ is the mixing parameter between $\phi_1$ and the SM Higgs boson. Also, $\lambda_{\chi \phi}$ is the coupling between DM and scalar $\phi_1$ and the parameters $f_{n,p}$ correspond to scalar nucleon couplings, taken as input from QCD calculations \cite{Junnarkar:2013ac}.
\begin{figure}[!h]
\centering
\begin{tabular}{cc}
\epsfig{file=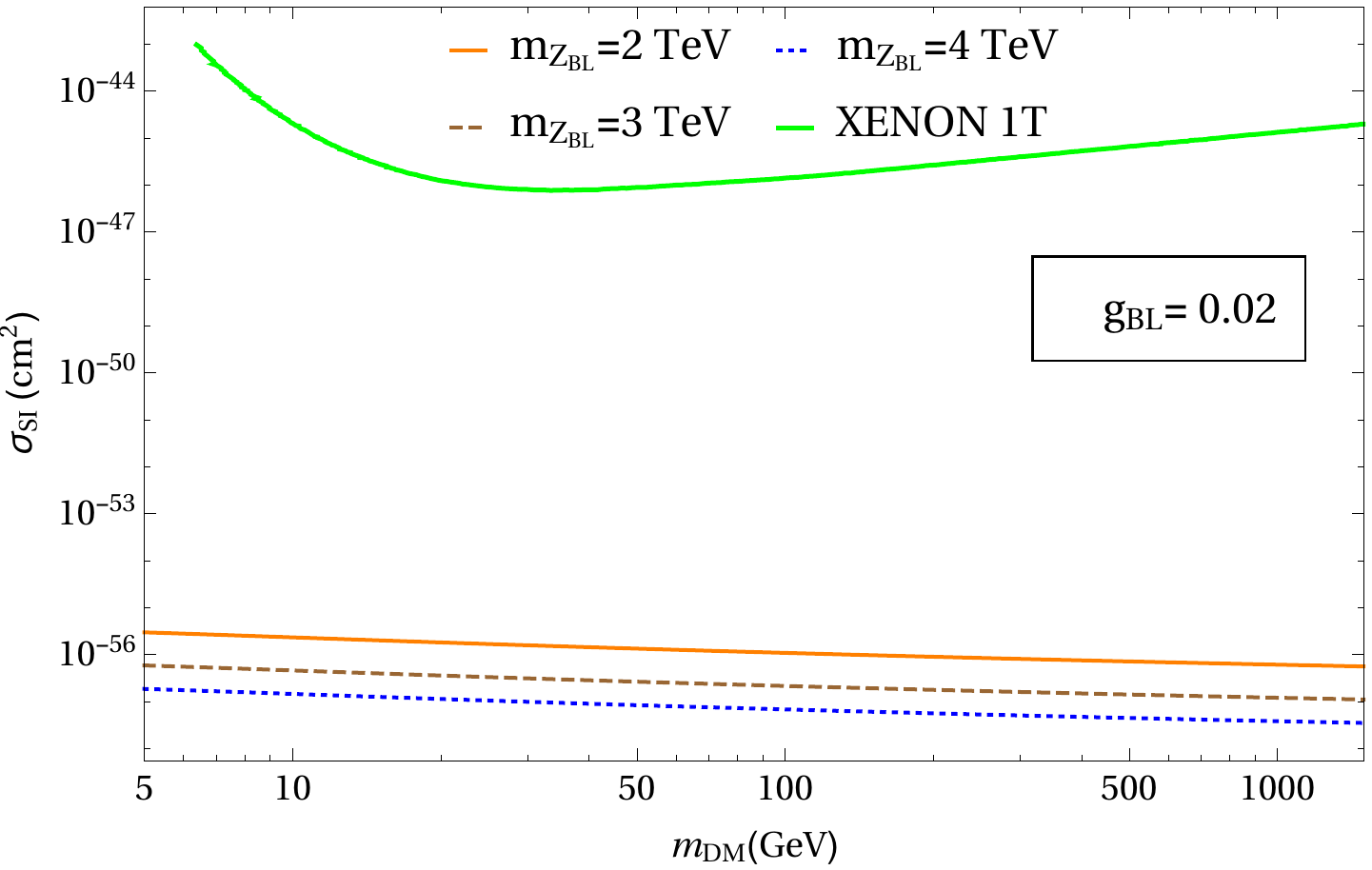,width=0.51\textwidth,clip=}
\epsfig{file=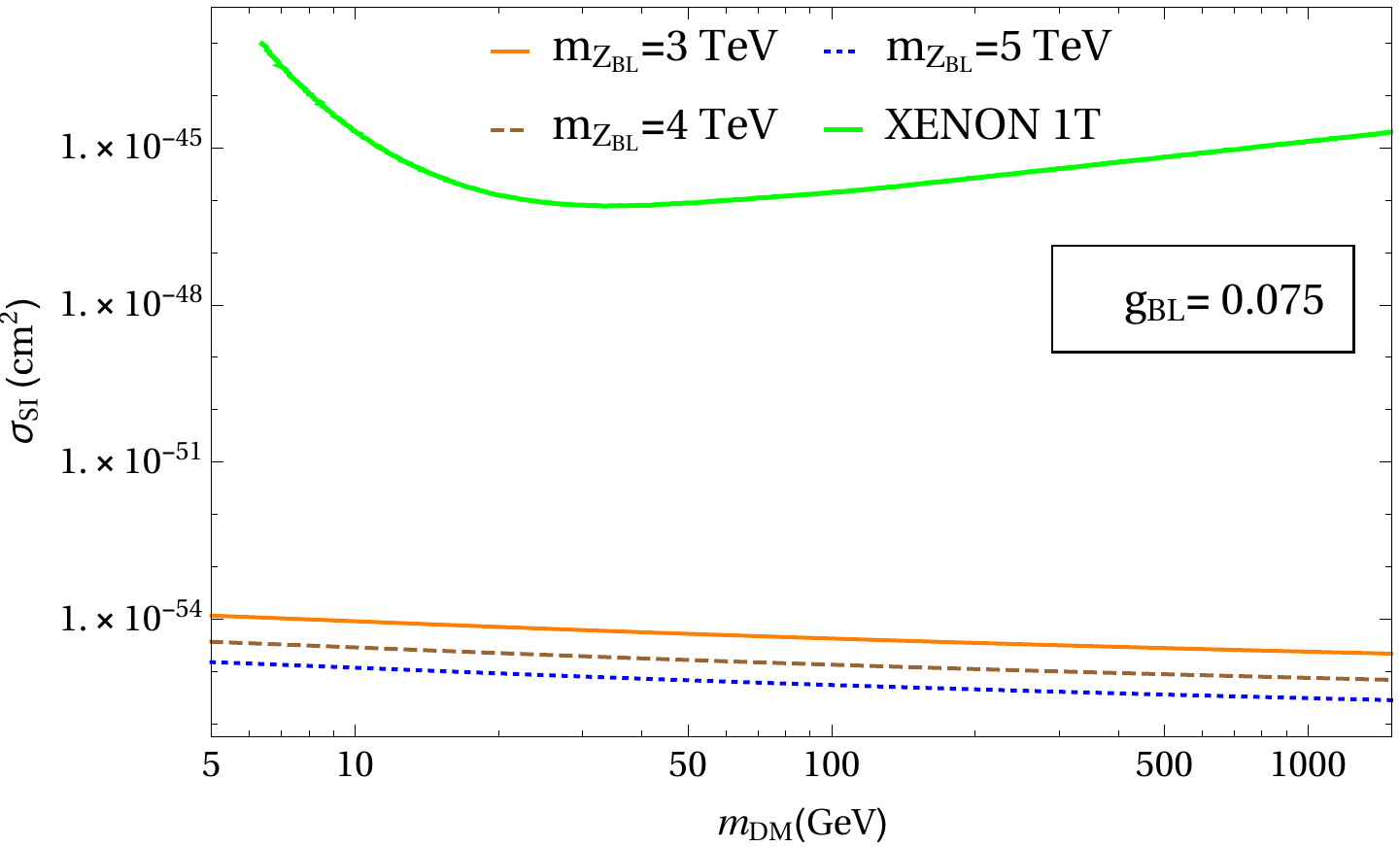,width=0.49\textwidth,clip=}
\end{tabular}
\caption{Spin independent direct detection cross section for fermion DM mediated by $Z_{BL}$ for different benchmark values of $g_{BL}, M_{Z_{BL}}$.}
\label{fig4}
\end{figure}

\begin{figure}[!h]
\centering
\epsfig{file=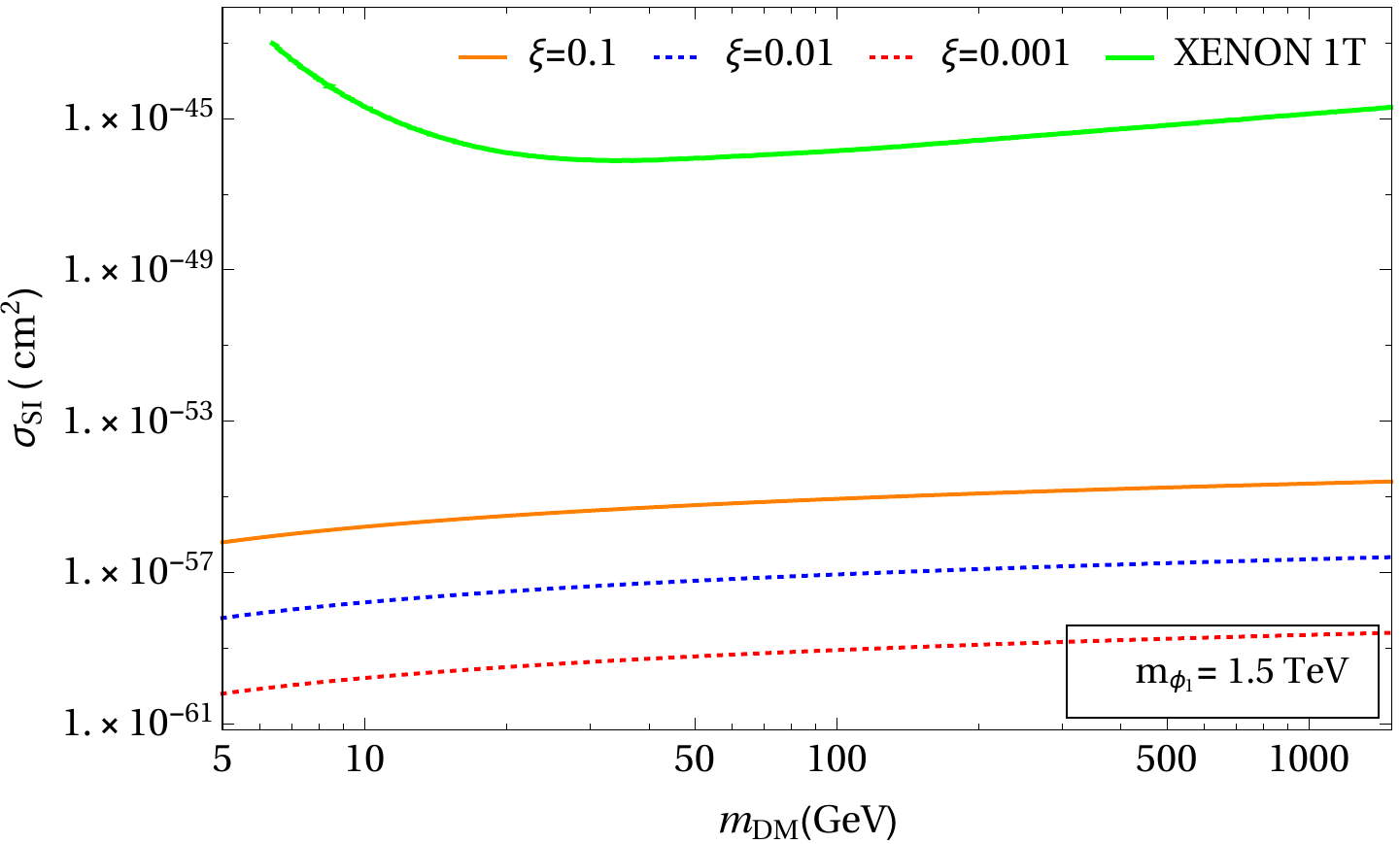,width=0.75\textwidth,clip=}
\caption{Spin independent direct detection cross section for fermion DM mediated by singlet scalar $\phi_1$ for different benchmark values of $\phi_1$-Higgs mixing $\xi$.}
\label{fig6}
\end{figure}

\begin{figure}[!h]
\centering
\epsfig{file=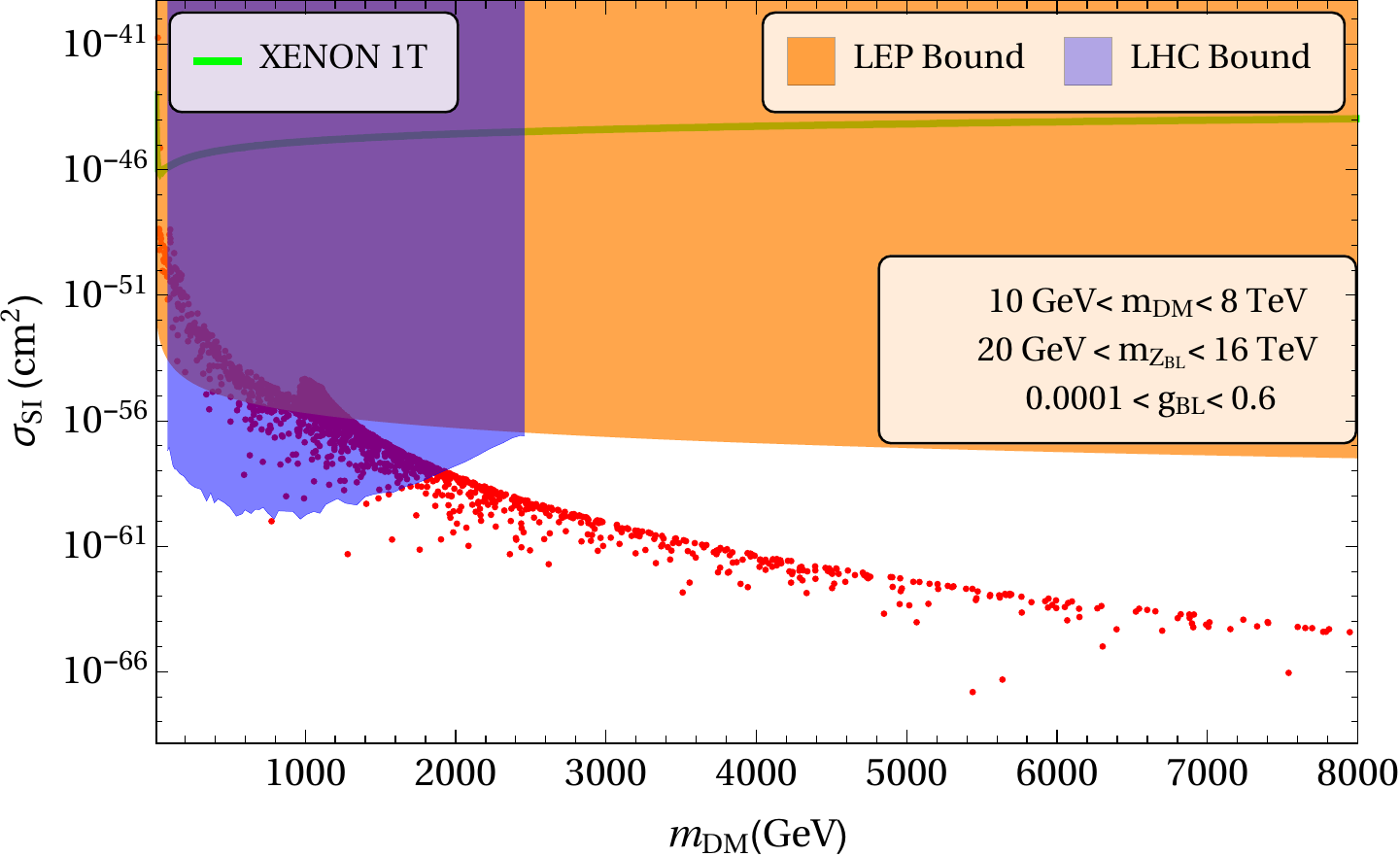,width=0.75\textwidth,clip=}
\caption{Spin independent direct detection cross section for fermion DM mediated by $Z_{BL}$ for randomly varied $M_{DM}, M_{Z_{BL}}, g_{BL}$ in the range denoted by the labels.}
\label{fig5}
\end{figure}

\begin{figure}[!h]
\centering
\epsfig{file=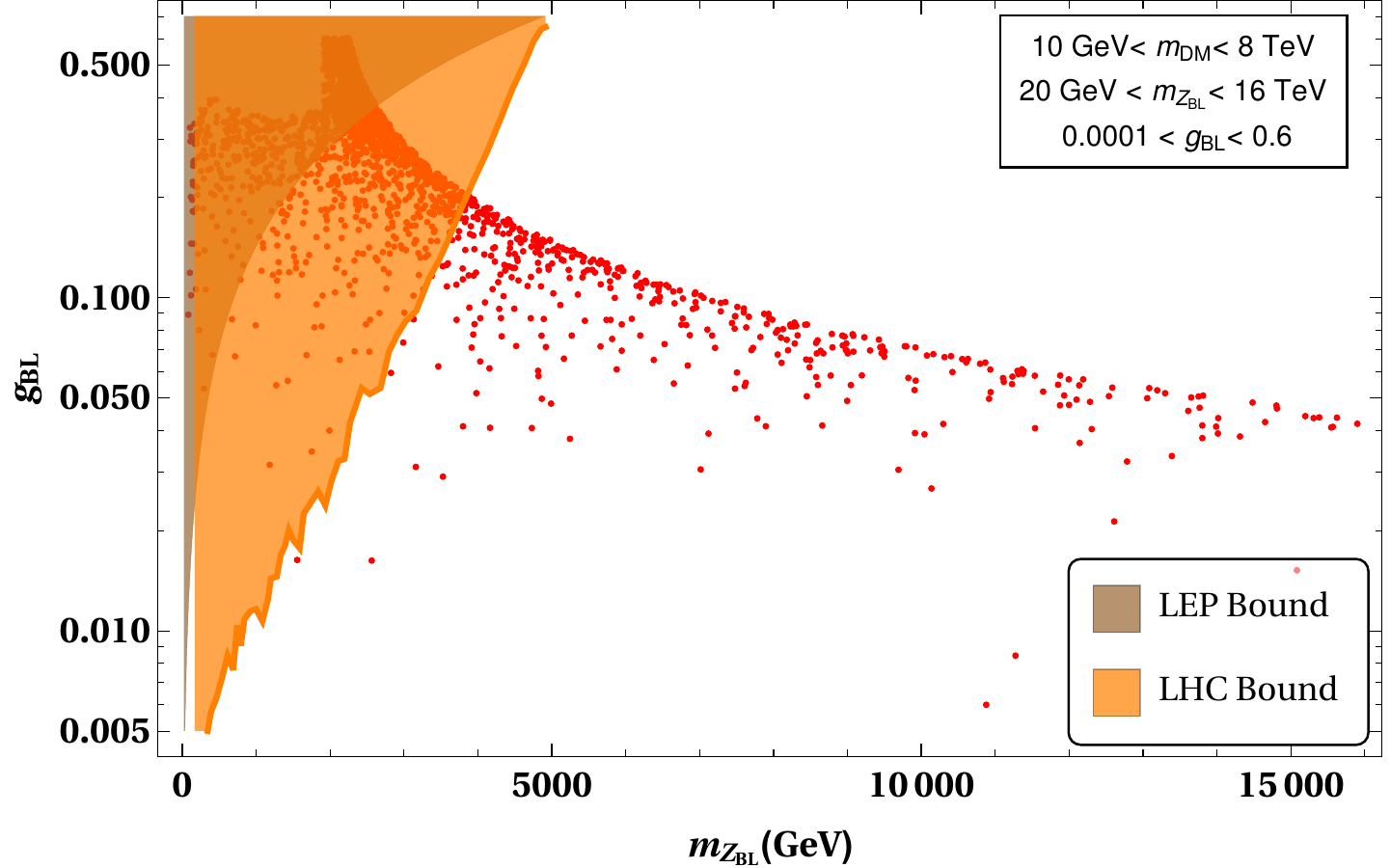,width=0.75\textwidth,clip=}
\caption{Parameter space allowed from relic abundance and direct detection criteria of fermion DM annihilating purely through $Z_{BL}$ portal for randomly varied $M_{DM}, M_{Z_{BL}}, g_{BL}$ in the range denoted by the labels. The excluded parts of parameter space corresponding to LEP, LHC are shown as shaded regions.}
\label{fig5b}
\end{figure}

\begin{figure}[!h]
\centering
\epsfig{file=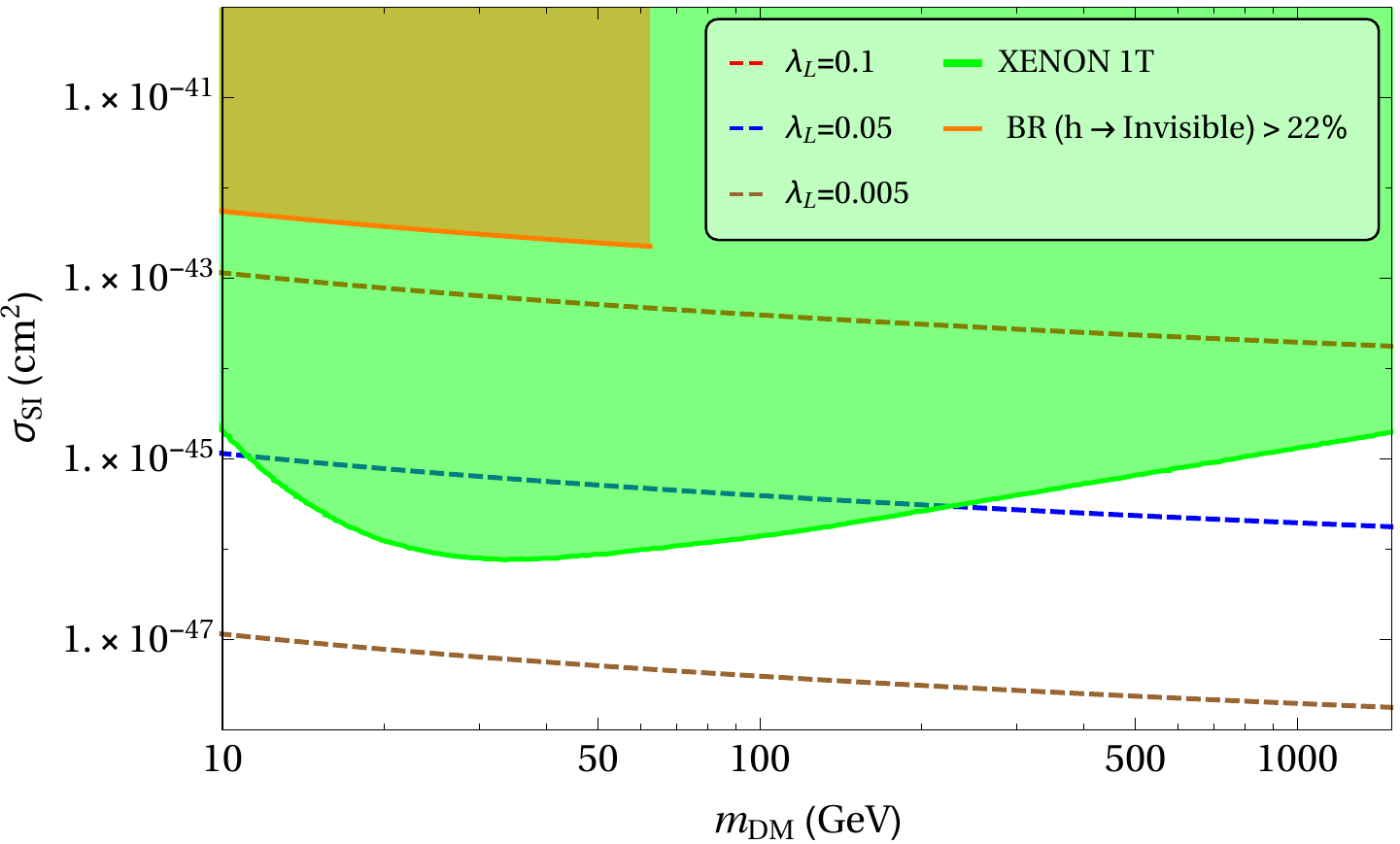,width=0.75\textwidth,clip=}
\caption{Spin independent direct detection cross section for scalar DM for different benchmark values of DM-Higgs couplings. The region ruled out from Xenon1T and LHC results are shown as shaded regions.}
\label{fig7}
\end{figure}
We first show the direct detection cross section mediated by the $Z_{BL}$ boson for different benchmark values of $Z_{BL}$ masses and couplings $g_{BL}$. As can be seen from figure \ref{fig4}, the model can survive the Xenon1T bound even if the $Z_{BL}$ mass is close to a few TeV and $g_{BL} \sim 0.075, 0.02$. These bounds are in fact weaker than the LEP II bound $M_{Z_{BL}}/g_{BL} \geq 7$ TeV mentioned earlier. This can be realised by taking even smaller values of $Z_{BL}$ mass and larger values of gauge coupling $g_{BL}$ as the spin independent direct detection cross section remains very much suppressed compared to the Xenon1T bound. However, the latest LHC bound can rule out such combinations of $M_{Z_{BL}}, g_{BL}$ as we discuss below. We also show the scalar mediated direct detection cross section for fermion DM, using different benchmark values of $\phi_1-$Higgs mixing $\xi$ in figure \ref{fig6} and found them to be lying within the Xenon1T upper limit.

To find the allowed parameter space of fermion DM model, we perform a random scan of three parameters $M_{DM}, M_{Z_{BL}}, g_{BL}$ and plot the direct detection cross section for only those parameters that satisfy the relic abundance criteria. The plot is shown in figure \ref{fig5}. The scalar mediated interactions are not taken into account in this scan, so that the strongest constraint on the new gauge sector can be obtained. It shows that DM masses in the entire mass range scanned $M_{DM} \in (10, 8000)$ GeV remain allowed by the Xenon1T constraint for the chosen range of gauge sector parameters $g_{BL} \in (10^{-4}, 0.6), M_{Z_{BL}} \in (20, 16000) $ GeV. We also show the LEP II bound for comparison in figure \ref{fig5}. The exclusion line corresponding to this is derived by using $M_{Z_{BL}}/g_{BL} = 7$ TeV in the direct detection formula given by \eqref{DDBL}. It is interesting that the LEP II bound $M_{Z_{BL}}/g_{BL} \geq 7$ TeV remains stronger than the Xenon1T bound and has the potential of ruling few points in the low mass regime of DM. We also include the latest bound from the LHC on the mass of $Z_{BL}$ gauge boson and corresponding coupling $g_{BL}$ given in \cite{Aaboud:2017buh} for $Z_{BL}$ masses all the way upto 5 TeV. Since fermion DM relic is satisfied for $m_{\text{DM}} \approx M_{Z_{BL}}/2$, there arises a sharp cut-off at $m_{\text{DM}} = 2.5$ TeV, since the LHC bound is available only till $M_{Z_{BL}} = 5$ TeV. Thus, the latest LHC bound is stronger than both Xenon1T and LEP bounds in the low mass regime of DM. We also show the corresponding parameter space in $g_{BL}-M_{Z_{BL}}$ plane in figure \ref{fig5b}. For simplicity, in the scanned plot of figure \ref{fig5b}, we apply the most conservative Xenon1T bound, corresponding to DM mass of 35 GeV \cite{Aprile:2017iyp} and minimum of the exclusion line shown before.

For scalar dark matter considered in this work, the relevant spin independent scattering cross section mediated by the SM Higgs boson is given as \cite{Barbieri:2006dq}
\begin{equation}
 \sigma_{\text{SI}} = \frac{\lambda^2_L f^2}{4\pi}\frac{\mu^2 m^2_n}{m^4_h m^2_{DM}}
\label{sigma_dd}
\end{equation}
where $\mu = m_n m_{DM}/(m_n+m_{DM})$ is the DM-nucleon reduced mass and $\lambda_L$ is the quartic coupling involved in DM-Higgs interaction. A recent estimate of the Higgs-nucleon coupling $f$ gives $f = 0.32$ \cite{Giedt:2009mr} although the full range of allowed values is $f=0.26-0.63$ \cite{Mambrini:2011ik}. We show the Higgs mediated direct detection cross section for scalar doublet dark matter in figure \ref{fig7} for different benchmark values of $\lambda_L$. We find that for large values of quartic coupling $\lambda_L \sim 0.1$, scalar DM mass upto a TeV can be ruled out by the Xenon1T bound. We also show the region ruled out by the LHC limit on the Higgs invisible decay width and found this bound to be much weaker than the latest Xenon1T limits.

\section{Indirect Detection of Dark Matter}
\label{sec6}
Apart from direct detection experiments, DM can also be probed at different indirect detection experiments that are looking for SM particles produced either through DM annihilations or via DM decay in the local Universe. Among these final states, photon and neutrinos, being neutral and stable can reach the indirect detection experiments without getting affected much by intermediate regions of space on the way from source to to the detector. If the DM is of WIMP type, like the one we are discussing in the present work, these photons lie in the gamma ray regime that can be measured at space based telescopes like the Fermi Large Area Telescope (LAT) or ground based telescopes like MAGIC. Here we constrain the DM parameters from the indirect detection bounds arising from the global analysis of the Fermi-LAT and MAGIC observations of dSphs \cite{Ahnen:2016qkx}.

The observed differential gamma ray flux produced due to DM annihilations is given by
\begin{equation}
\frac{d\Phi}{dE} (\triangle \Omega) = \frac{1}{4\pi} \langle \sigma v \rangle \frac{J (\triangle \Omega)}{2M^2_{\text{DM}}} \frac{dN}{dE}
\end{equation}
where $\triangle \Omega$ is the solid angle corresponding to the observed region of the sky, $\langle \sigma v \rangle$ is the thermally averaged DM annihilation cross section, $dN/dE$ is the average gamma ray spectrum per annihilation process and the astrophysical $J$ factor is given by
\begin{equation}
J(\triangle \Omega) = \int_{\triangle \Omega} d\Omega' \int_{LOS} dl \rho^2(l, \Omega').
\end{equation}
In the above expression, $\rho$ is the DM density and LOS corresponds to line of sight. Thus, measuring the gamma ray flux and using the standard astrophysical inputs, one can constrain the DM annihilation into different final states like $\mu^+ \mu^-, \tau^+ \tau^-, W^+ W^-, b\bar{b}$. Since DM can not couple to photons directly, gamma rays can be produced from such charged final states. Using the bounds on DM annihilation to these final states \cite{Ahnen:2016qkx}, we show the status of our model for different benchmark values of parameters.

Since the constraint on DM annihilation given in \cite{Ahnen:2016qkx} was for $100\%$ annihilations in to a particular final state, we use the appropriate weight factor $w <1$ while comparing the model prediction with the constraints. For example, to compare the DM annihilation to $\mu^+ \mu^-$ final state with the Fermi-LAT constraint, we multiply $\sigma v (\text{DM}\; \text{DM} \rightarrow \mu^+ \mu^-)$ by the weight factor 
$$ w= \frac{\sigma v (\text{DM} \; \text{DM} \rightarrow \mu^+ \mu^-)}{\sigma v (\text{DM} \;\text{DM} \rightarrow \text{All})}. $$
This is equivalent to dividing the Fermi-LAT bound by $w<1$ and comparing with $\sigma v (\text{DM} \; \text{DM} \rightarrow \mu^+ \mu^-)$. Since there are multiple annihilation channels to different final states, the Fermi-LAT constraints on individual final states are weak for most of the cases. Only in the case of scalar DM annihilations into $W^+W^-$ final states, the constraints can be very severe as scalar DM into $W^+W^-$ is the most dominant annihilation channel for certain mass range of DM. 

\begin{figure}[!h]
\centering
\begin{tabular}{cc}
\epsfig{file=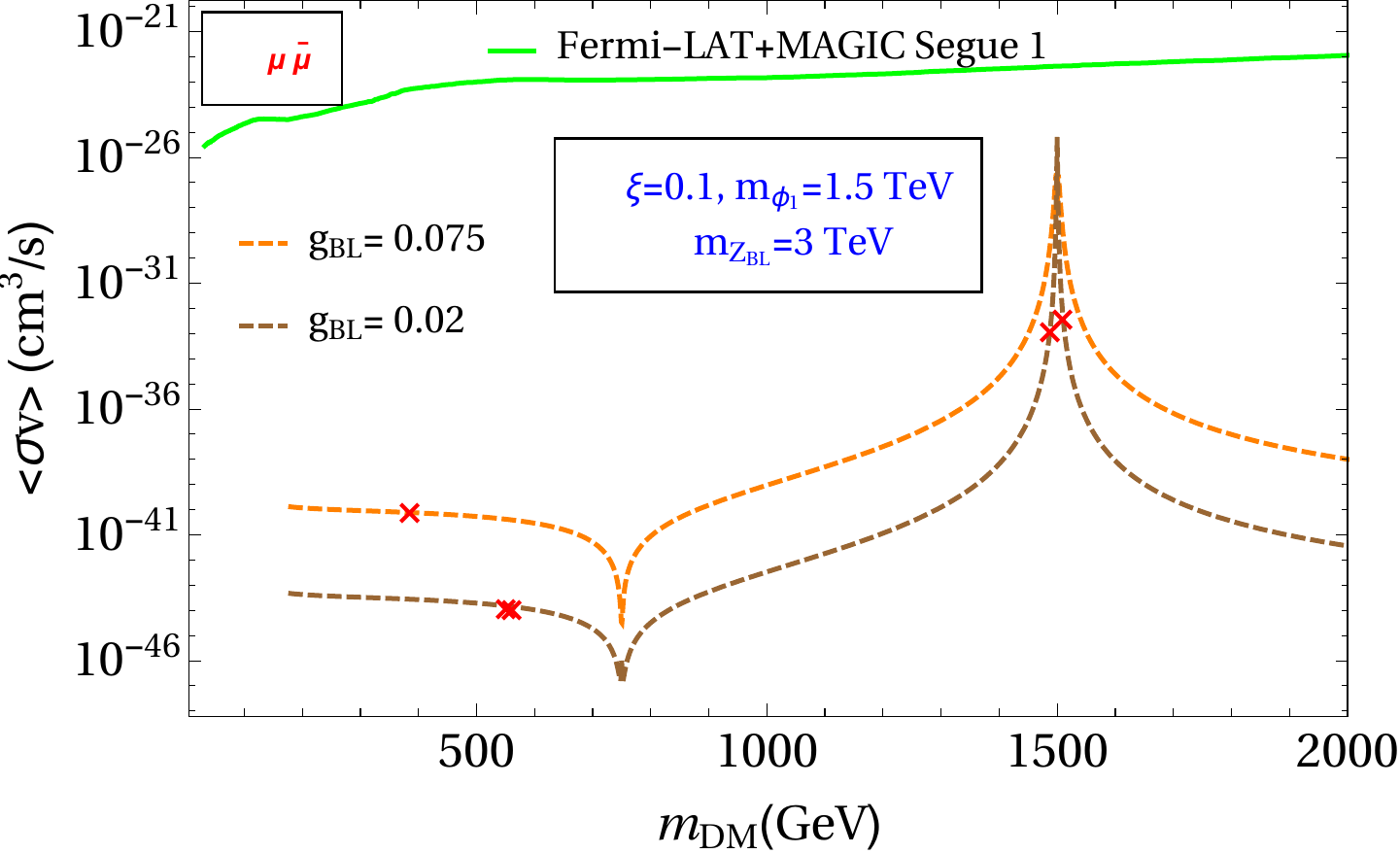,width=0.50\textwidth,clip=}
\epsfig{file=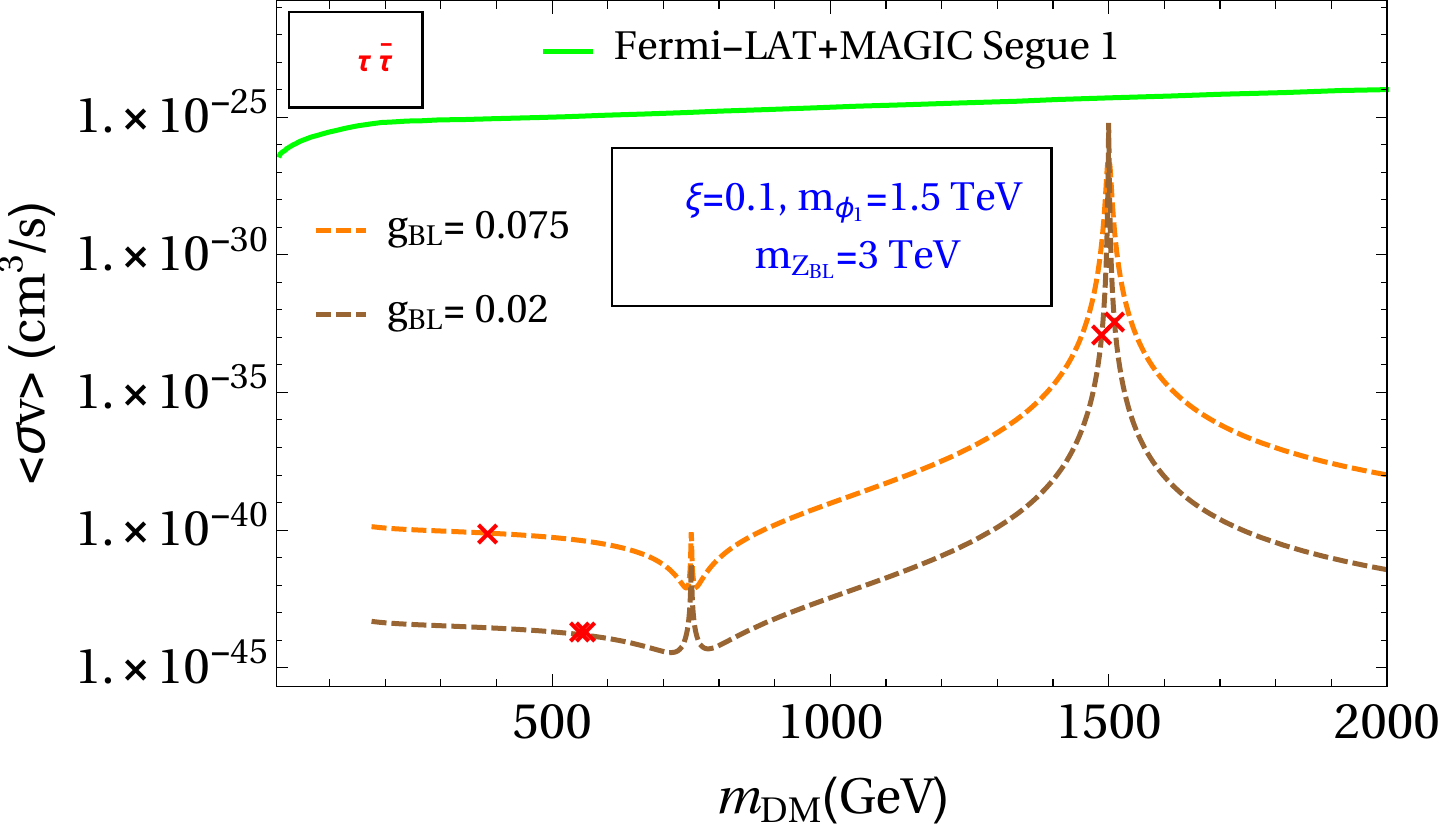,width=0.50\textwidth,clip=}
\end{tabular}
\caption{Fermion DM annihilations into $\mu^+ \mu^-, \tau^+ \tau^-$ compared against the indirect detection bounds. The points denoted by $\times$ correspond to the ones satisfying relic abundance criteria.}
\label{fig8}
\end{figure}

\begin{figure}[!h]
\centering
\begin{tabular}{cc}
\epsfig{file=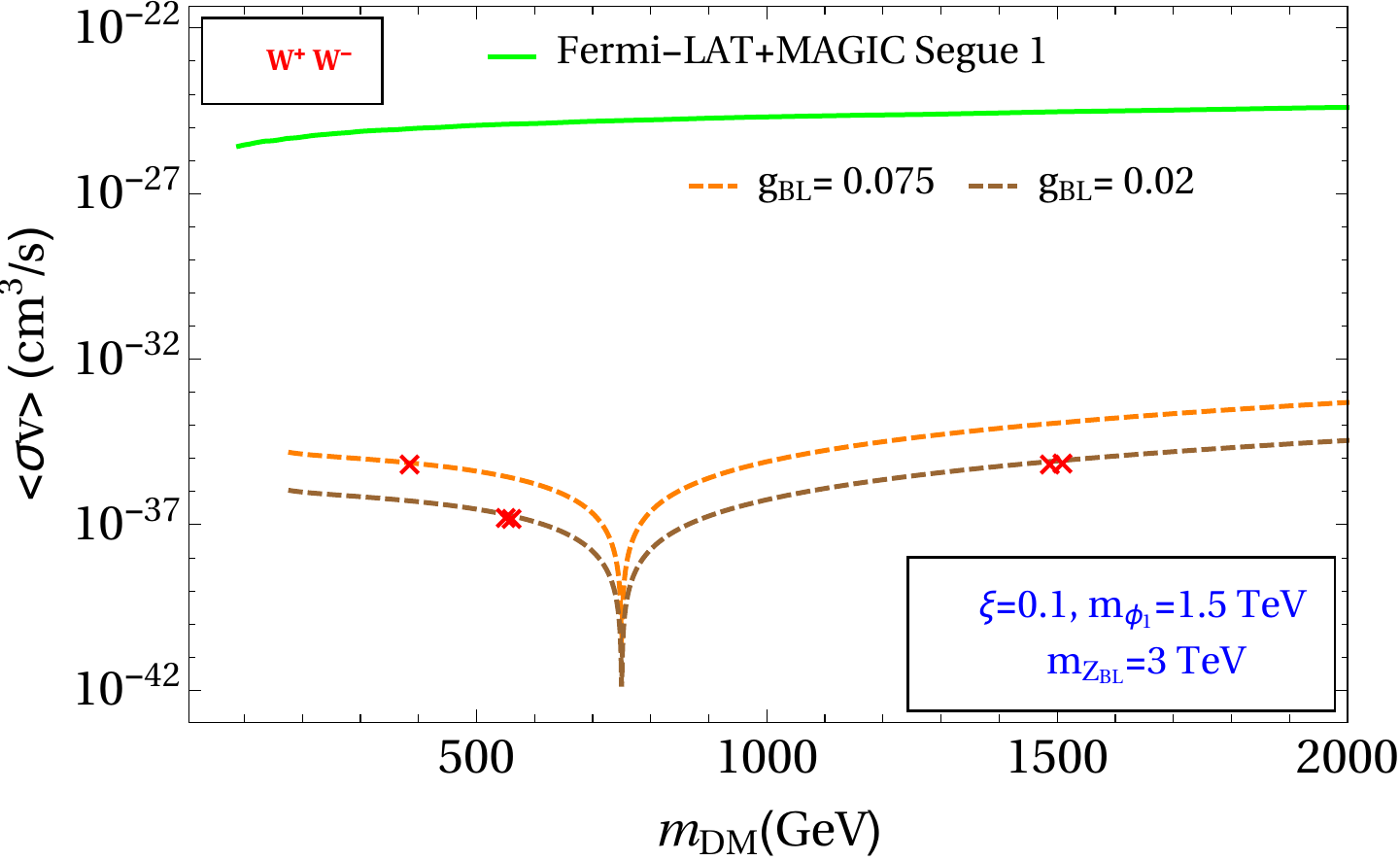,width=0.49\textwidth,clip=}
\epsfig{file=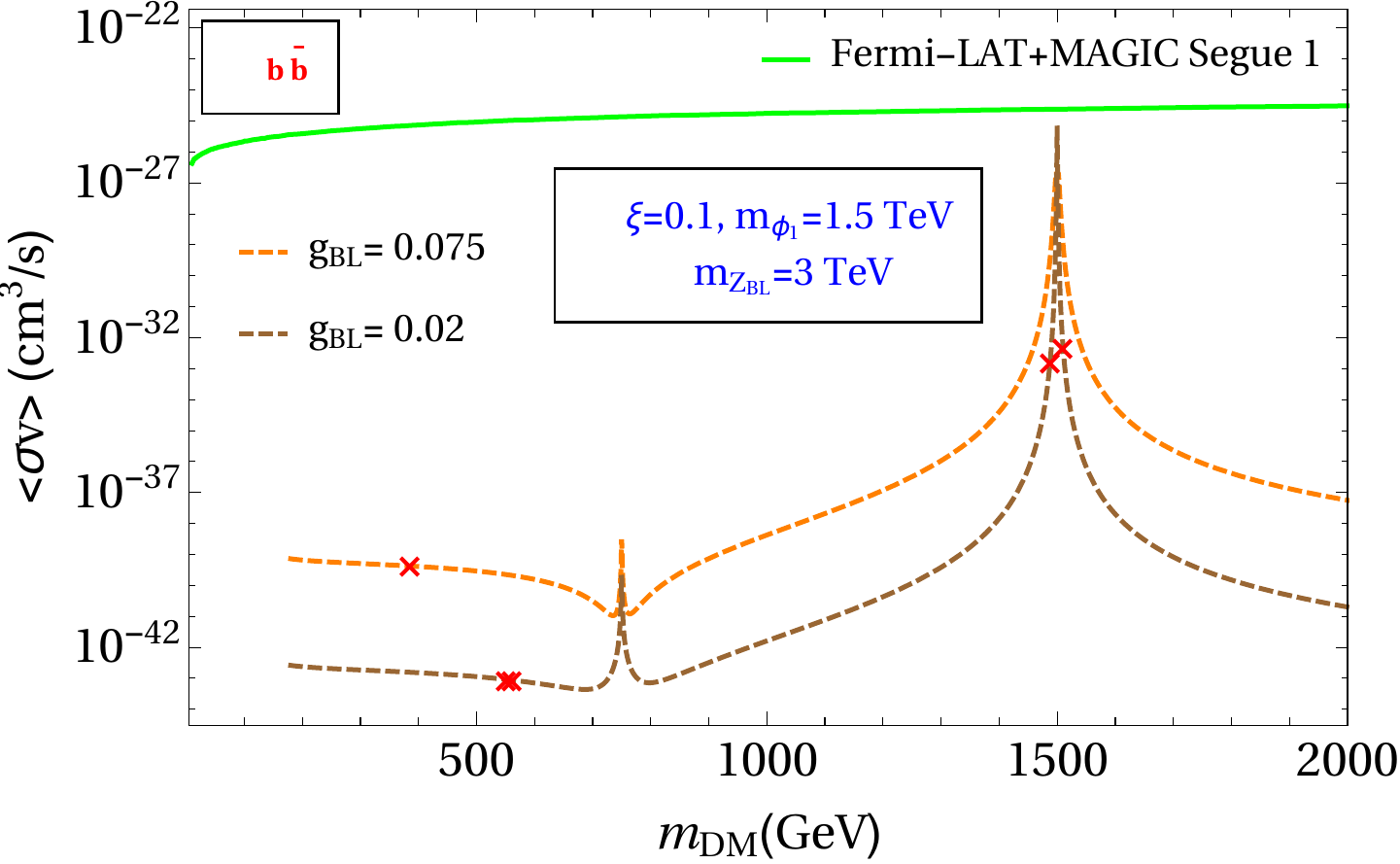,width=0.51\textwidth,clip=}
\end{tabular}
\caption{Fermion DM annihilations into $W^+ W^-, b\bar{b}$ compared against the indirect detection bounds. The points denoted by $\times$ correspond to the ones satisfying relic abundance criteria.}
\label{fig9}
\end{figure}

\begin{figure}[!h]
\centering
\begin{tabular}{cc}
\epsfig{file=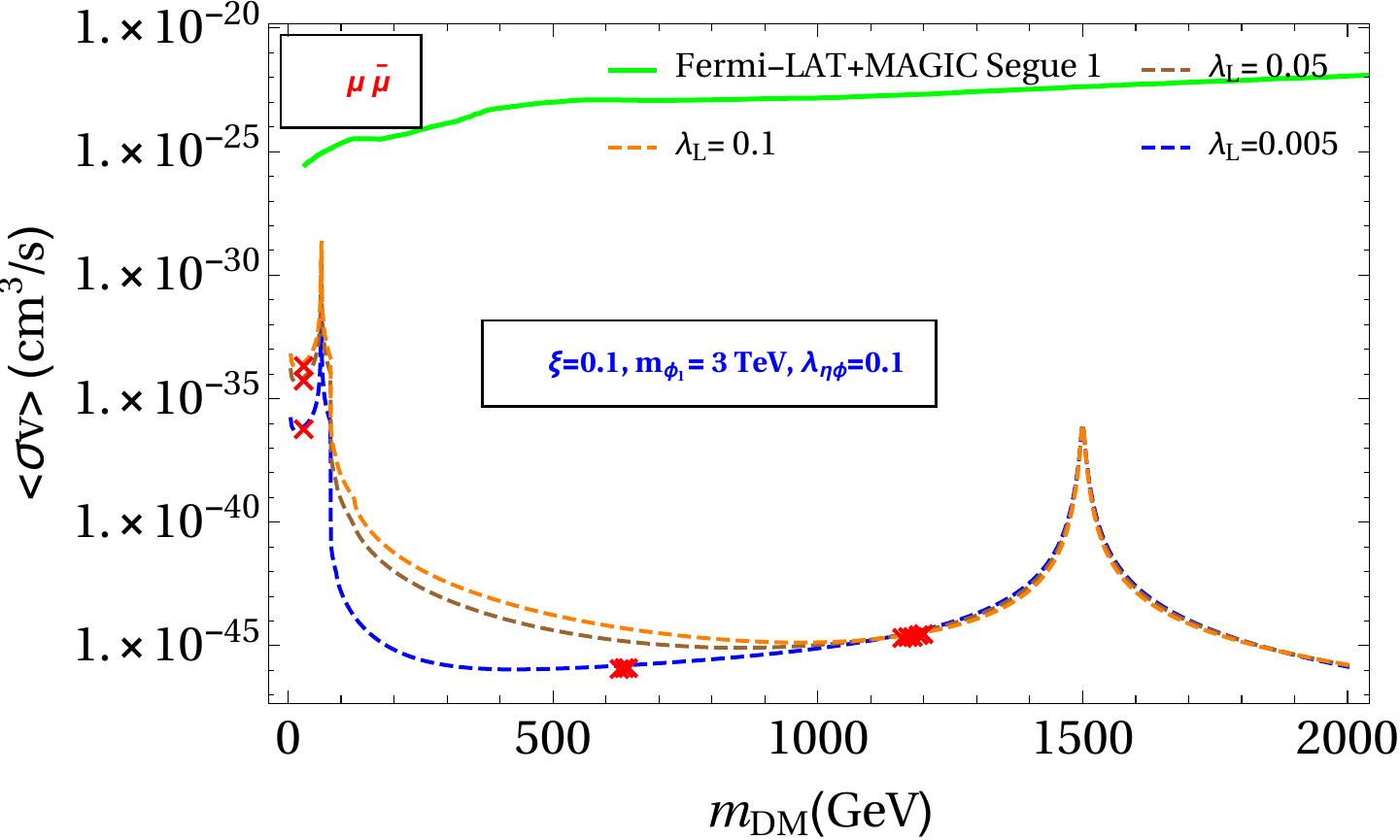,width=0.50\textwidth,clip=}
\epsfig{file=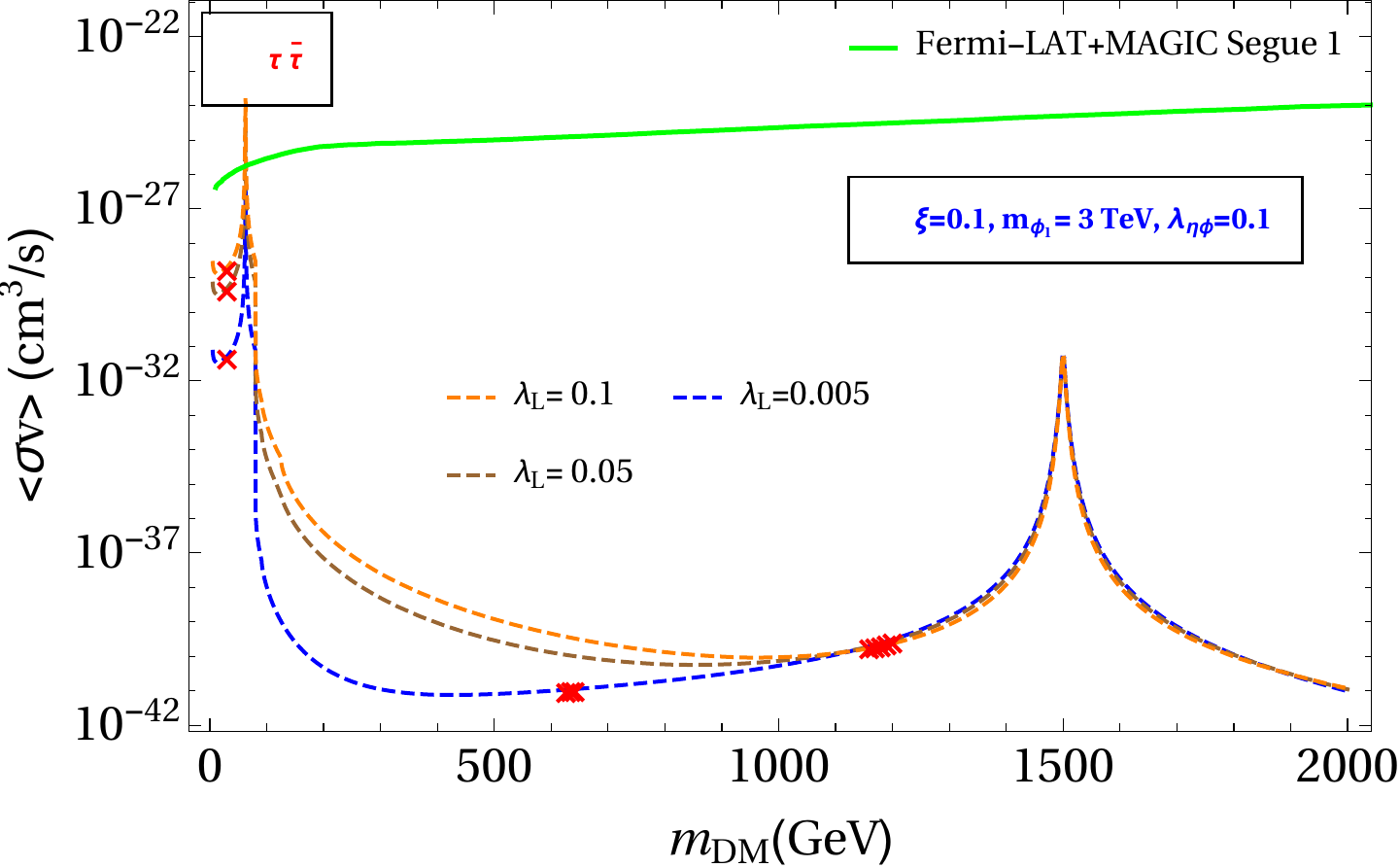,width=0.50\textwidth,clip=}
\end{tabular}
\caption{Scalar DM annihilations into $\mu^+ \mu^-, \tau^+ \tau^-$ compared against the indirect detection bounds. The mass splitting between scalar doublet components is $\Delta m_{\eta} = 5$ GeV. The points denoted by $\times$ correspond to the ones satisfying relic abundance criteria.}
\label{fig10}
\end{figure}

\begin{figure}[!h]
\centering
\begin{tabular}{cc}
\epsfig{file=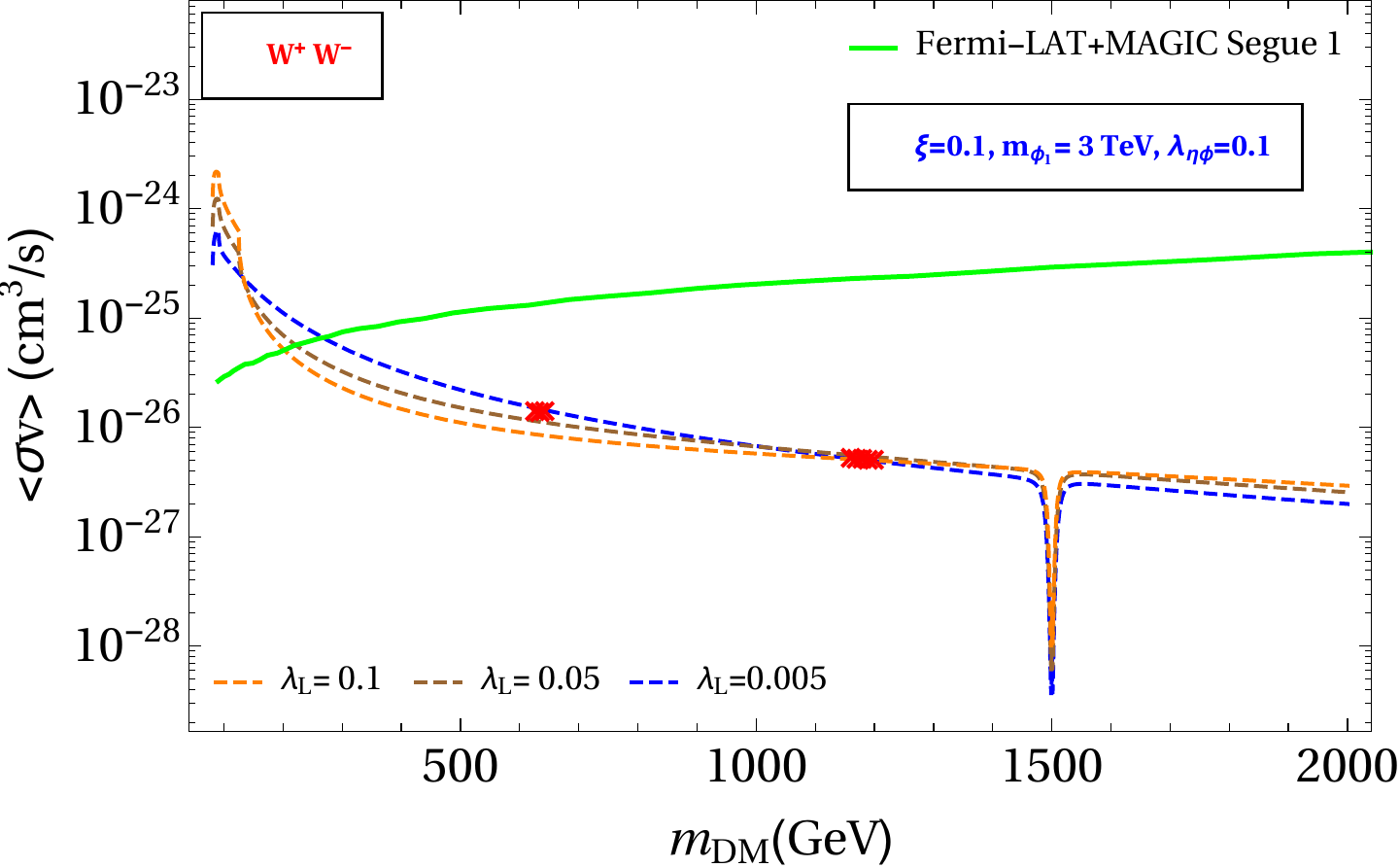,width=0.51\textwidth,clip=}
\epsfig{file=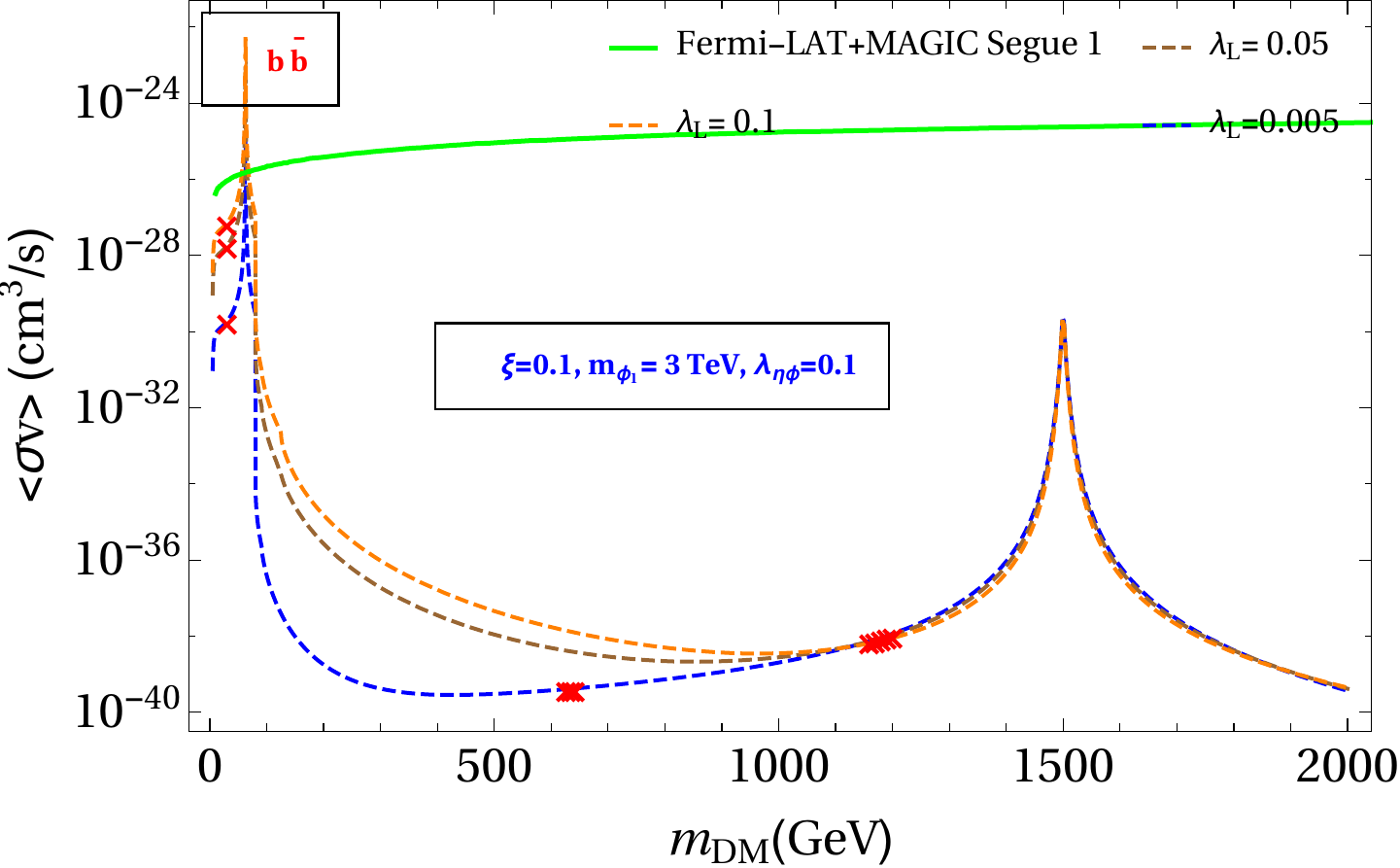,width=0.49\textwidth,clip=}
\end{tabular}
\caption{Scalar DM annihilations into $W^+ W^-, b\bar{b}$ compared against the indirect detection bounds. The mass splitting between scalar doublet components is $\Delta m_{\eta} = 5$ GeV. The points denoted by $\times$ correspond to the ones satisfying relic abundance criteria.}
\label{fig11}
\end{figure}

\begin{figure}[!h]
\centering
\epsfig{file=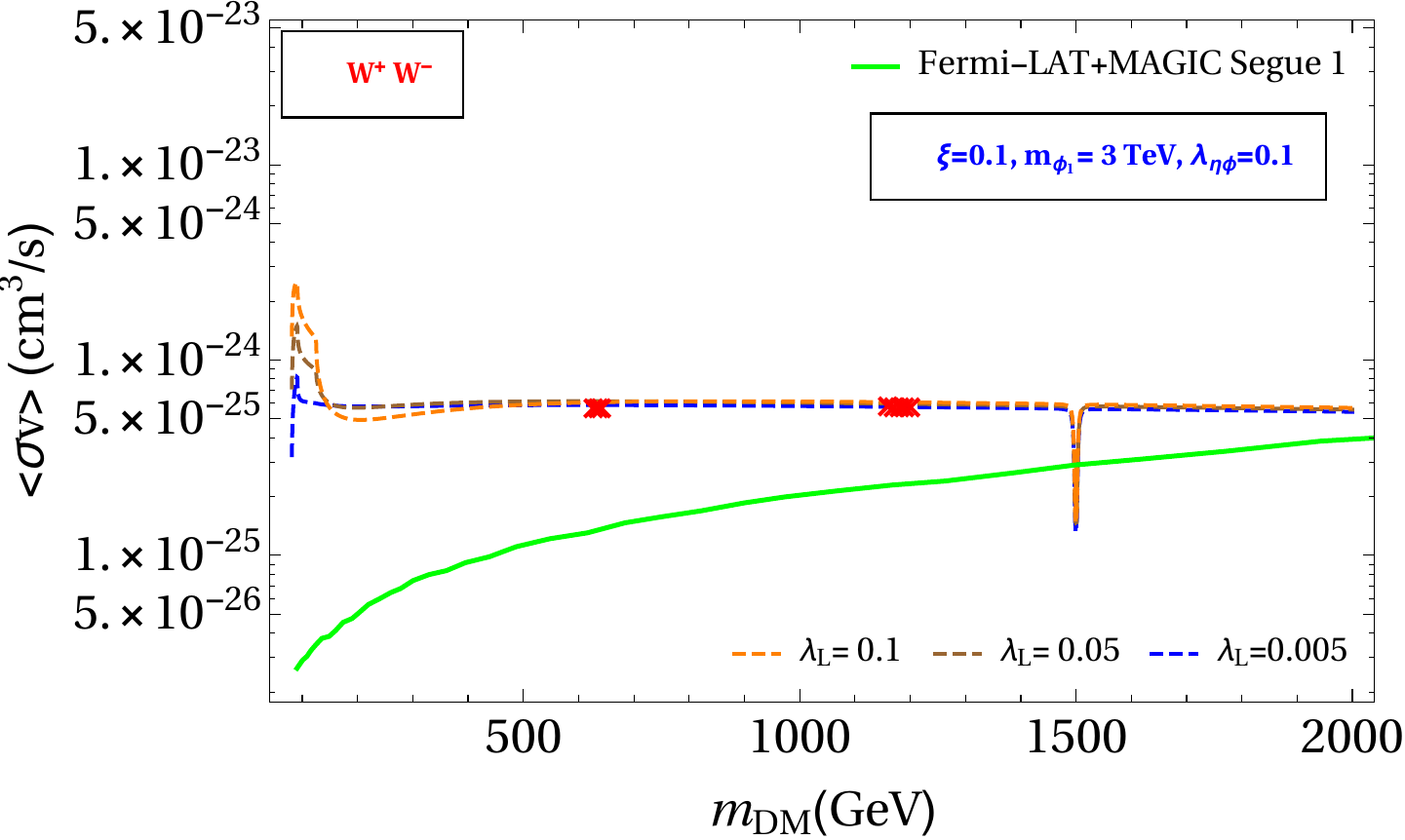,width=0.75\textwidth,clip=}
\caption{Scalar DM annihilations into $W^+ W^-$ compared against the indirect detection bounds. The mass splitting between scalar doublet components is $\Delta m_{\eta} = 50$ GeV. The points denoted by $\times$ correspond to the ones satisfying relic abundance criteria.}
\label{fig12}
\end{figure}

We first show the fermion DM annihilation to these final states in figure \ref{fig8} and \ref{fig9} for different benchmark points including the ones used to show the relic abundance as well. We also mark the points by a $\times$ which correspond to the correct relic abundance of fermion DM. It can be seen that, fermion DM having mass near the resonance region $M_{DM} = m_{\phi_1}/2$ for the scalar mediated annihilations, some of the annihilations can almost saturate the Fermi-LAT plus MAGIC bounds. On the other hand, the bounds from $W^+ W^-$ final states is weaker even for mixing between singlet scalar and SM Higgs $\xi$ as large $0.1$, as can be seen from the left panel of figure \ref{fig9}. However, as the relic abundance plots in figure \ref{fig2} shows, the DM remains under-abundant in the resonance region and hence such scenarios are still allowed from indirect detection bounds. We apply the relic bound on the indirect detection plots shown in figure \ref{fig8} and \ref{fig9} by marking those points of $m_{\text{DM}}$ which satisfy the correct relic abundance criteria, as can be seen from figure \ref{fig2}.

Similar observation is also made for the scalar DM annihilations into these final states as shown in figure \ref{fig10} and \ref{fig11}. Although the DM annihilations to $\tau^+ \tau^-, b\bar{b}$ saturates the experimental limits only near the Higgs resonance $M_{DM} = m_{h}/2$, the limit on DM annihilations into $W^+ W^-$ is more severe and it can rule out DM mass upto a few hundred GeV's depending on the mass splitting between dark scalar doublet components as well as DM-Higgs coupling $\lambda_L$. This is because of the large annihilation cross section of scalar DM into $W^+ W^-$ pairs for $m_{\text{DM}} > M_W$ that can not be suppressed due to its sole dependence on gauge couplings. However, in the high mass region $m_{\text{DM}} > M_W$, the relic remains under-abundant as seen from the figure \ref{fig6a}, specially in the range of DM mass where the constraints from $W^+W^-$ final state are very strong. Beyond a mass of 500 GeV, it is possible to satisfy relic as well as indirect detection constraints simultaneously. But for larger mass splitting say $\Delta m_{\eta}=50$ GeV, even the high mass region is ruled out by the strong constraints from indirect detection, as can be seen from figure \ref{fig12}. In the low mass region, specially near the Higgs resonance, the indirect detection bounds can be severe as can be seen clearly by comparing the relic abundance plot in figure \ref{fig5a} with the indirect detection ones in figure \ref{fig10}. We in fact denote the points which satisfy the relic abundance criteria in the indirect detection plots shown in figure \ref{fig10} and \ref{fig11} and find that all of them remain allowed currently from the Fermi-LAT plus MAGIC bounds. On the other hand, for large splitting, the points satisfying correct relic are disallowed by the indirect detection bounds as seen from figure \ref{fig12}.

It should be noted that apart from Fermi-LAT and MAGIC, there are other experiments which put equally strong bounds on DM annihilations to different charged particle final states. For example, there exists strong constraints on DM annihilations from ten years of observations with HESS experiment, as reported recently in \cite{HESS:2015cda, Abdallah:2016ygi}. As seen from \cite{Abdallah:2016ygi}, the HESS constraints on DM annihilations into $\tau^+ \tau^-$ final state is very strong and can reach the $\langle \sigma v \rangle $ values expected for DM annihilating at thermal relic cross section. However, since DM in our model remains under-abundant near the resonance regions that can saturate these indirect detection bounds, it does not rule out the allowed parameter space from relic criteria, as discussed above in the context of Fermi-LAT plus MAGIC bounds.
\section{Lepton Flavour Violation}
\label{sec7}
Charged lepton flavour violating decay is a promising process to study from BSM physics point of view. In the SM, such a process occurs at loop level and is suppressed by the smallness of neutrino masses, much beyond the current experimental sensitivity \cite{TheMEG:2016wtm}. Therefore, any future observation of such LFV decays like $\mu \rightarrow e \gamma$ will definitely be a signature of new physics beyond the SM. In the present model, such new physics contribution can come from the charged component of the additional scalar doublet $\eta$ going inside a loop along with singlet fermions. Adopting the general prescriptions given in \cite{Lavoura:2003xp}, the decay width of $\mu \rightarrow e \gamma$ can be calculated as
\begin{align}
\Gamma (\mu \rightarrow e \gamma) &= \frac{Y^4 \left(m^2_\mu - m^2_e \right)^3(m^2_\mu + m^2_e)}{4096 \pi^5 m^3_\mu m^4_{\eta^-}} \left[\frac{ \left((t-1)(t(2t+5)-1) + 6t^2\ln t\right)^2}{144 (t-1)^8}\right]
\end{align}
where $t = m^2_{N}/m^2_{\eta^-}$. The corresponding branching ratio can be found by
$$ \text{BR}(\mu \rightarrow e \gamma) \approx \frac{\Gamma (\mu \rightarrow e \gamma)}{\Gamma_{\mu}} $$
where $\Gamma_{\mu} \approx 2.996 \times 10^{-19}$ GeV denotes the total decay width of muon. The latest bound from the MEG collaboration is $\text{BR}(\mu \rightarrow e \gamma) < 4.2 \times 10^{-13}$ at $90\%$ confidence level \cite{TheMEG:2016wtm}.

\begin{figure}[!h]
\centering
\begin{tabular}{cc}
\epsfig{file=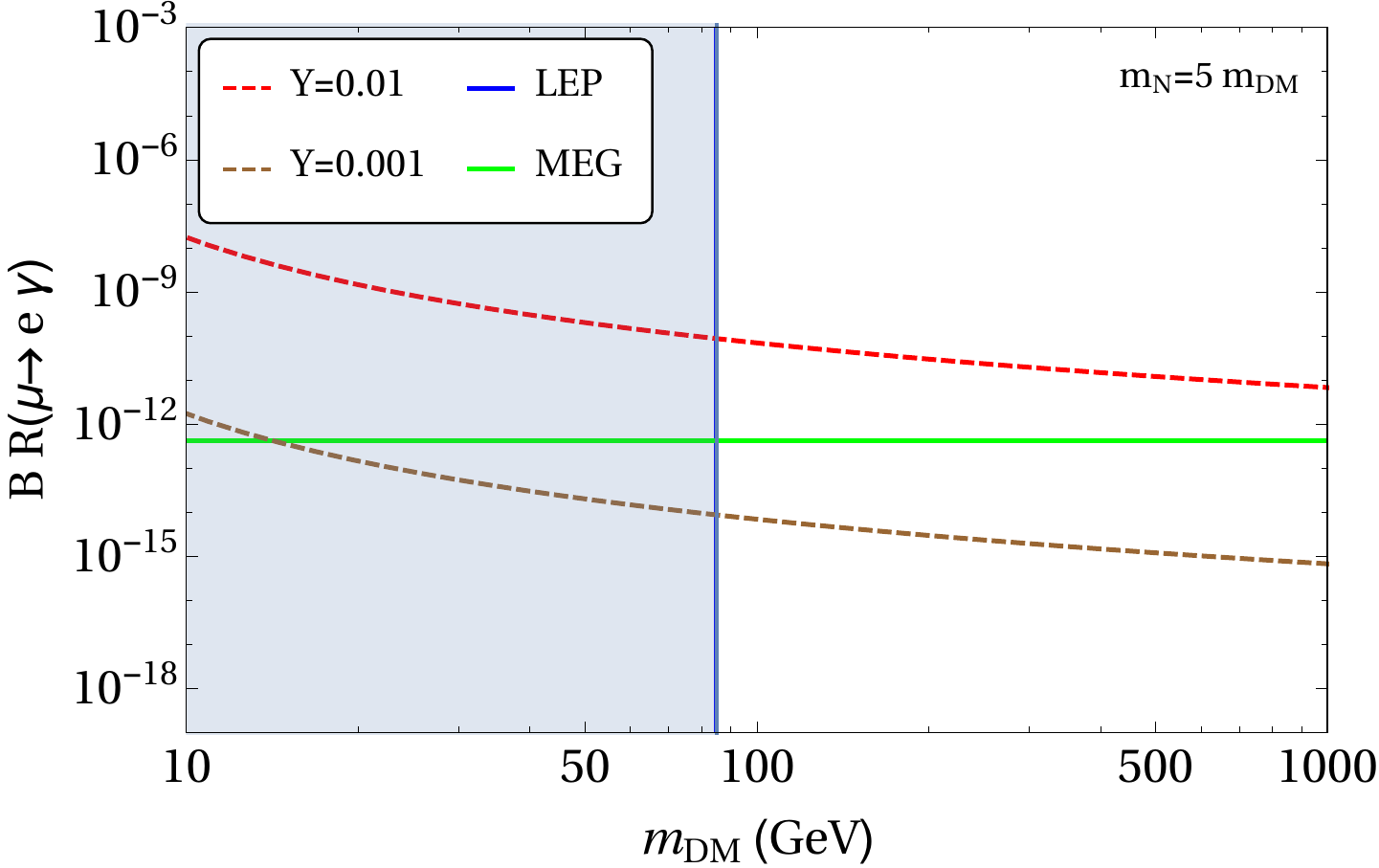,width=0.50\textwidth,clip=}
\epsfig{file=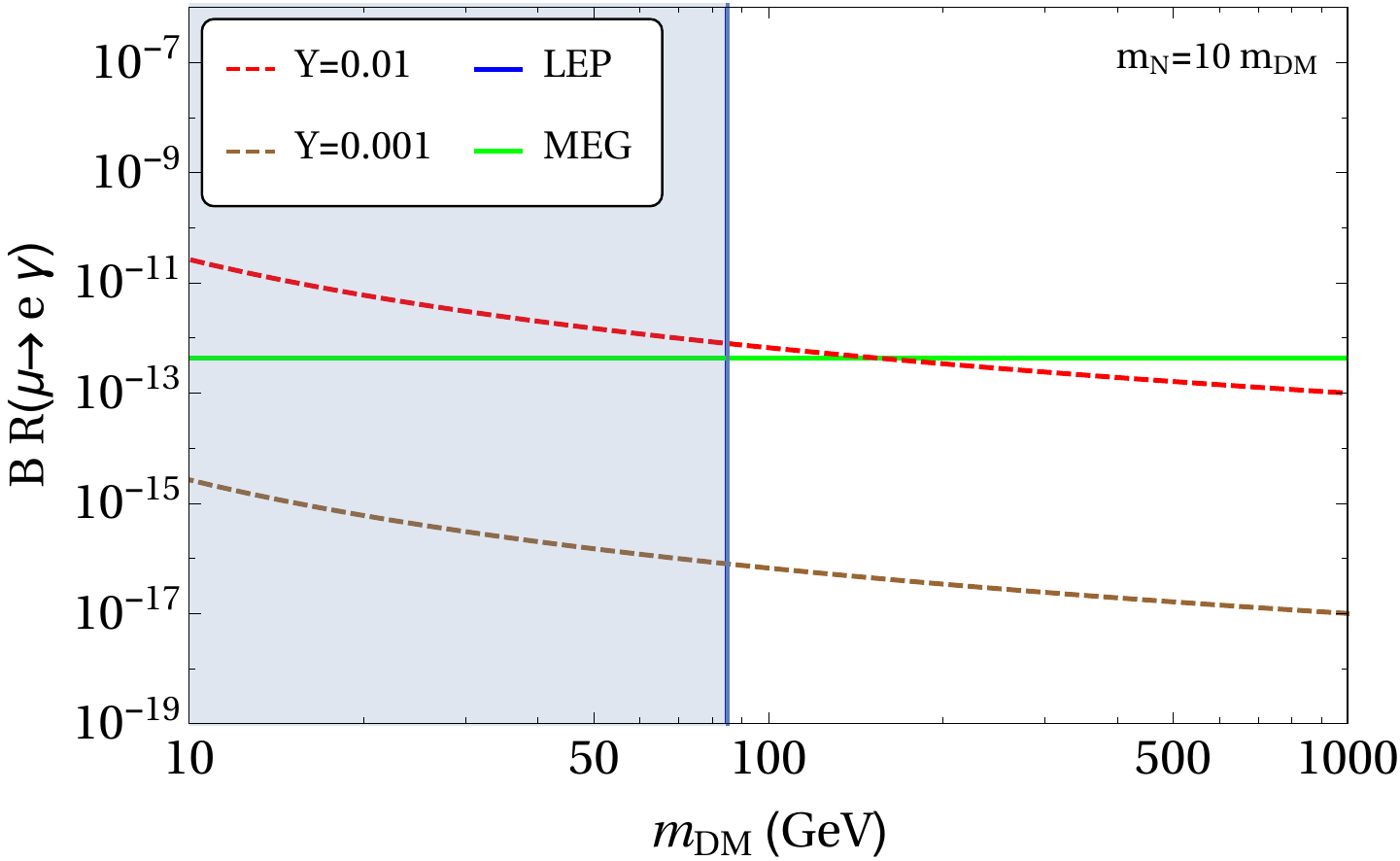,width=0.50\textwidth,clip=}
\end{tabular}
\caption{$\text{BR}(\mu \rightarrow e \gamma)$ as a function of scalar DM mass for different values of Yukawa couplings and singlet fermion mass. The mass splitting between scalar doublet components is $\Delta m_{\eta} = 5$ GeV.}
\label{fig13}
\end{figure}

We consider a scalar DM scenario so that the singlet neutrinos are heavier than $\eta$. By keeping the mass splitting within $\eta$ to be 5 GeV, we show the new physics contribution to $\mu \rightarrow e \gamma$ in figure \ref{fig13} for different benchmark values of Yukawa couplings and mass ratio between singlet fermion and scalar DM. It is seen that even for Yukawa couplings as small as $10^{-2}$, this new physics contribution can saturate the experimental upper limit from MEG. We get similar features, even if we consider singlet fermion DM instead of scalar one.

It should be noted that there are other LFV processes like $\mu^- \rightarrow e^+ e^- e^-$ or $\mu^- \rightarrow e^-$ conversion in nuclei which can get additional contribution at one loop level in this model. For details of such processes, one can refer to \cite{Toma:2013zsa}. For the chosen benchmark values of heavy singlet neutrinos and DM mass range, the contribution to $\text{BR}(\mu \rightarrow e \gamma)$ from new physics remains dominant over others, as shown by the authors of \cite{Toma:2013zsa}. Also, the constraints on $\text{BR}(\mu \rightarrow e \gamma)$ has got updated by MEG experiment recently whereas the bounds on other processes are relatively older \cite{Bellgardt:1987du, Bertl:2006up}.

Another interesting flavour observable (a lepton number violating one) is the neutrinoless double beta decay which is tightly constrained from null results at several experiments mentioned earlier \cite{KamLAND-Zen:2016pfg}. However, the present models do not have any extra contributions to this process at tree level. Therefore, the contribution to this process will be dominated by standard light neutrino mediation which can be kept within limits for hierarchical light neutrino spectrum. Since our models naturally lead to a hierarchical pattern with the lightest neutrino being massless, these bounds can be satisfied naturally.

\section{Collider Signatures}
\label{sec8}
\begin{figure}[!h]
\centering
\begin{tabular}{cc}
\epsfig{file=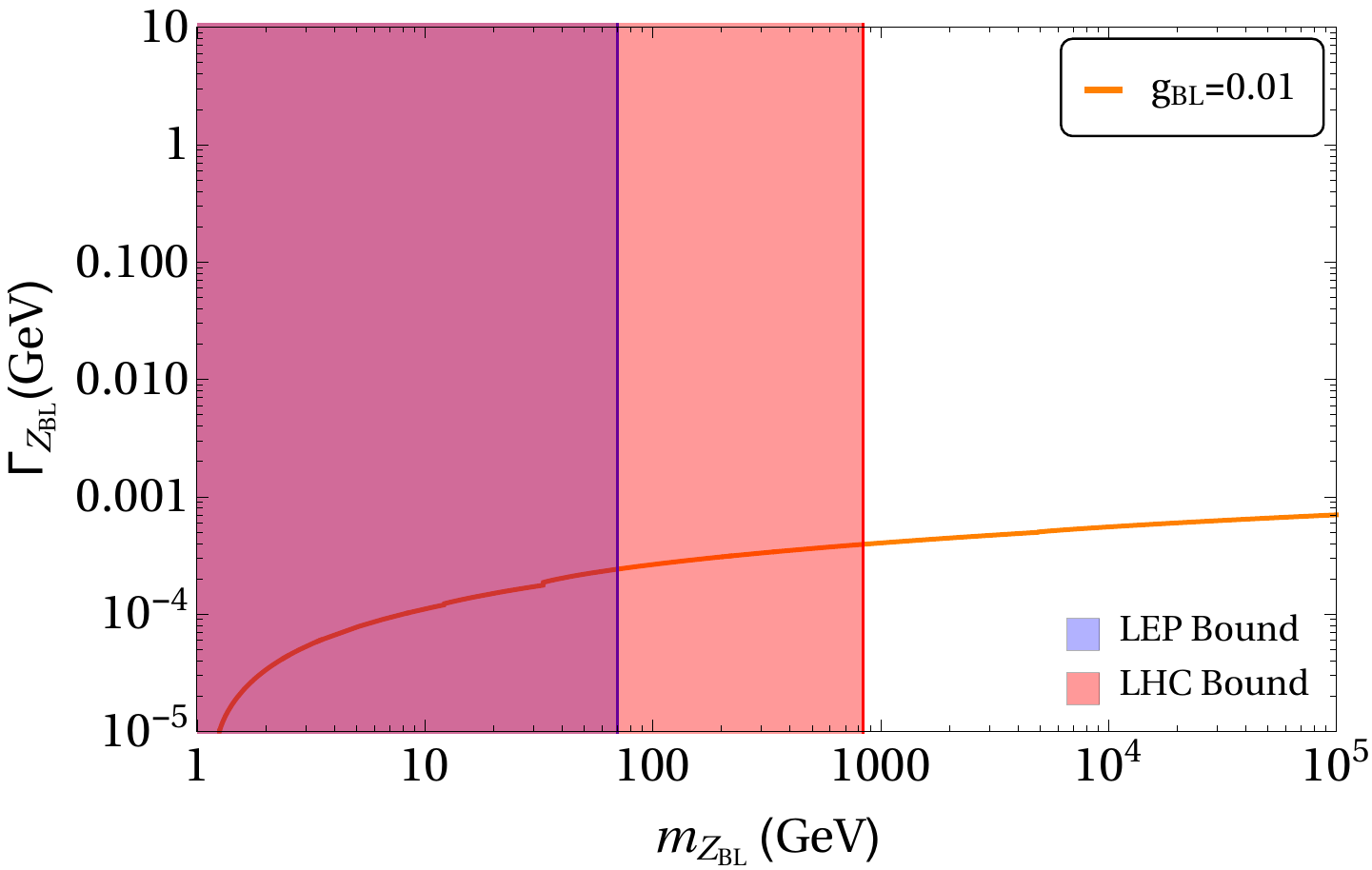,width=0.50\textwidth,clip=}
\epsfig{file=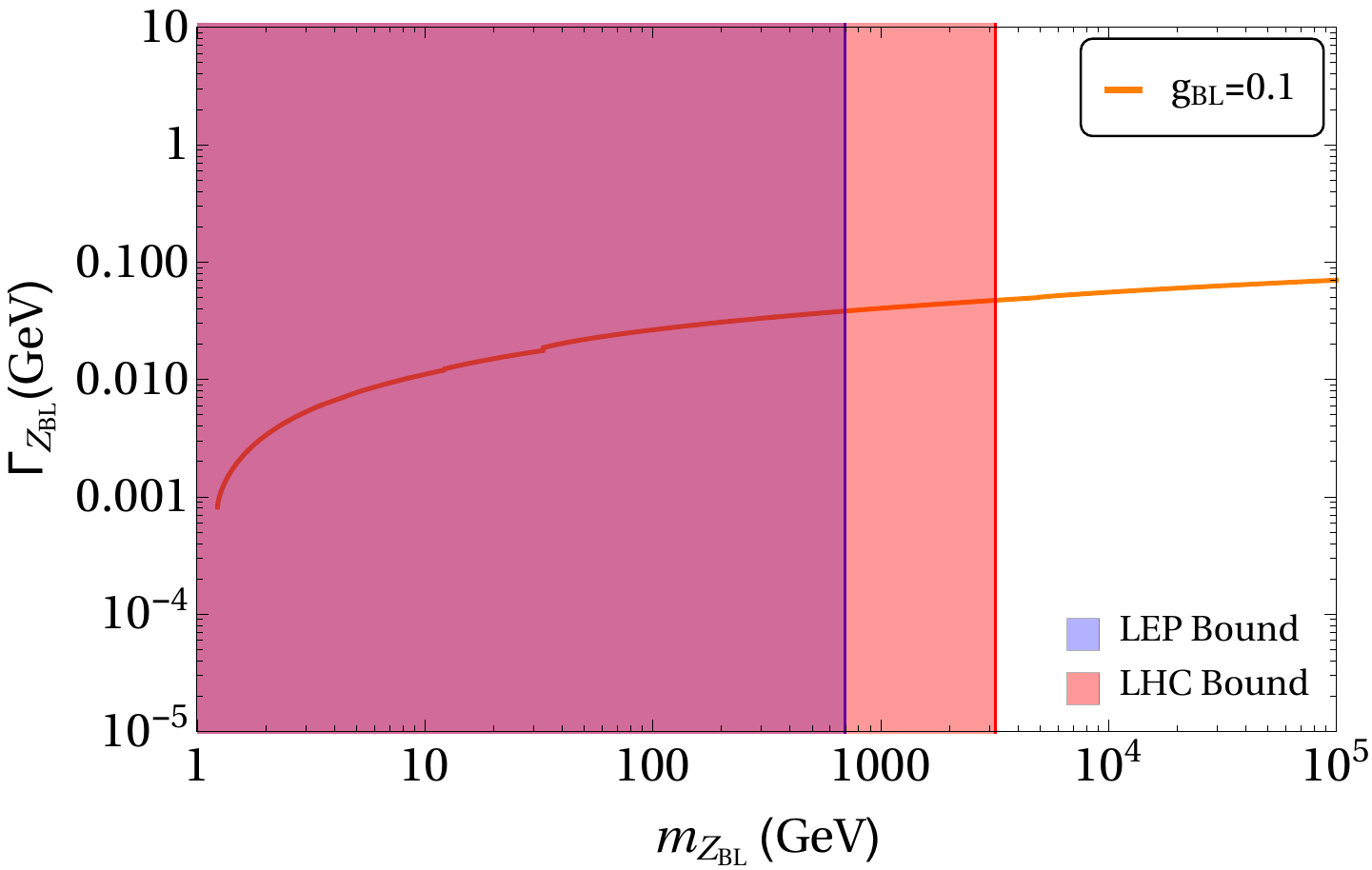,width=0.50\textwidth,clip=}\\
\epsfig{file=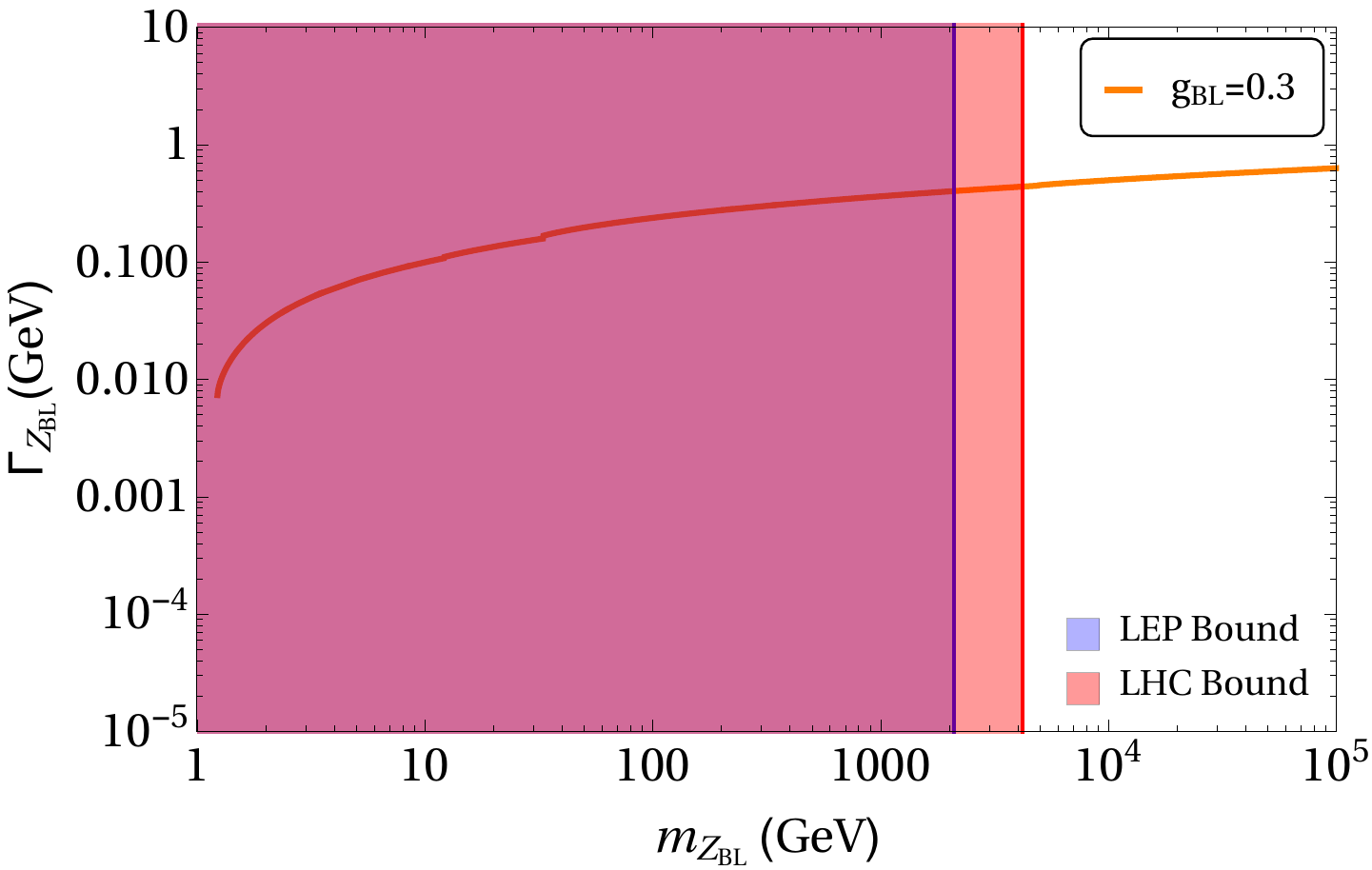,width=0.50\textwidth,clip=}
\epsfig{file=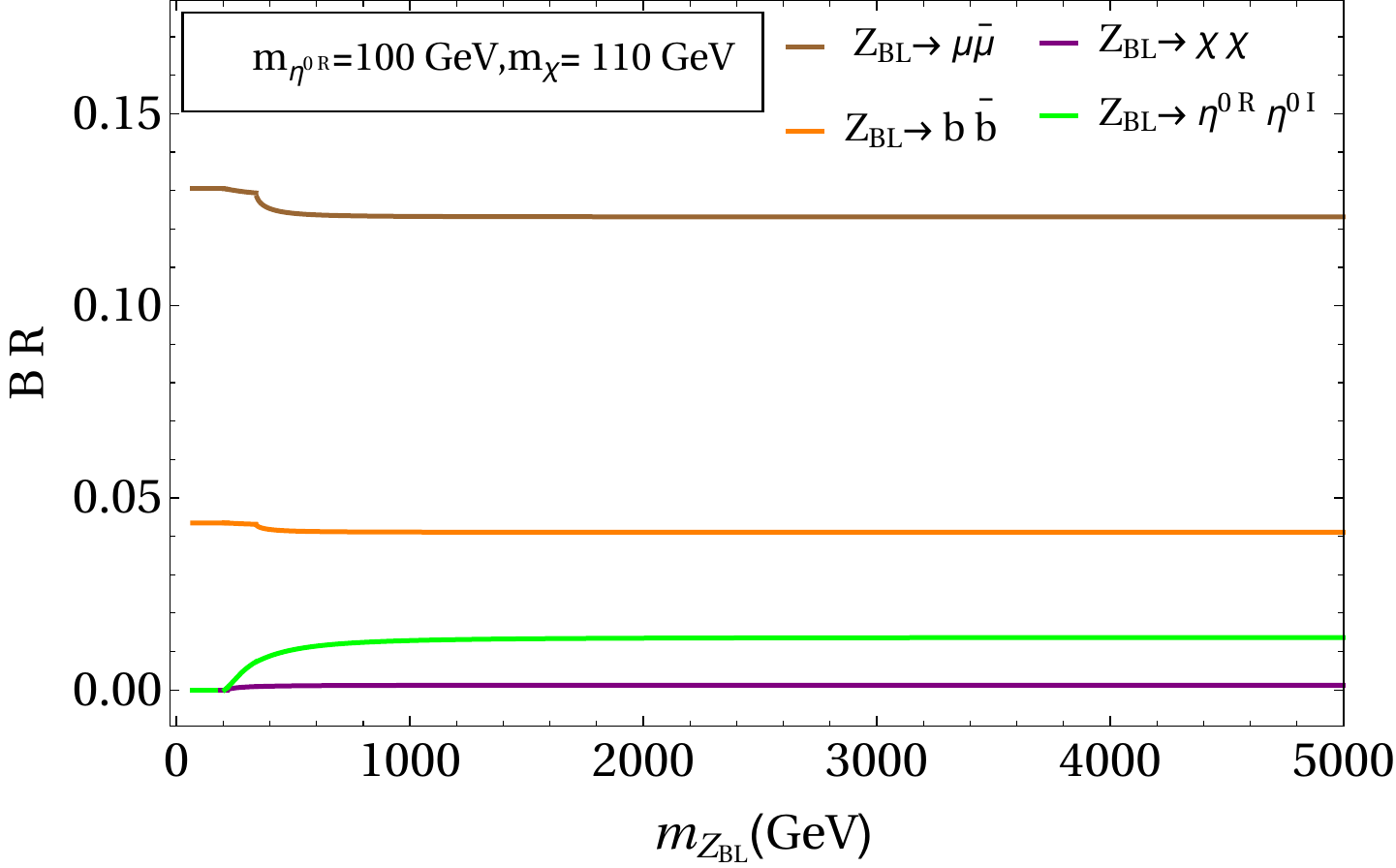,width=0.50\textwidth,clip=}\\
\end{tabular}
\caption{Decay width and branching ratio of $Z_{BL}$ into different final states. The shaded regions in the first three plots correspond to the regions ruled out by the LEP and LHC bounds.}
\label{fig14}
\end{figure}

Since all the SM fermions are charged under the $U(1)_{B-L}$ gauge interaction, there can be significant production of the corresponding $Z_{BL}$ gauge boson in proton proton collisions \cite{Okada:2016gsh, Basso:2008iv}. Such heavy gauge boson, if produced at colliders, can manifest itself as a narrow resonance through its decay into dileptons, say. Since the $U(1)_{B-L}$ charges of the leptons are three times that of quarks, the decay of $Z_{BL}$ into leptons are more in spite of the extra colour factor of quarks. This can be seen from the plot in figure \ref{fig14} showing the branching ratio of $Z_{BL}$ into different final states, along with the total decay width. 

Although the branching ratio does not depend upon the value of gauge coupling $g_{BL}$, the production of this gauge boson in proton proton collisions is sensitive to this. If this coupling is comparable to the electroweak gauge couplings, then such a gauge boson can be produced significantly in colliders. The latest measurement by the ATLAS experiment at 13 TeV LHC constrains such gauge boson mass to be heavier than $3.6-4.0$ TeV depending on whether the final state leptons are of muon or electron type \cite{Aaboud:2017buh}. The corresponding bound for tau lepton final states measured by the CMS experiment at 13 TeV LHC is slightly weaker, with the lower bound on $Z_{BL}$ mass being 2.1 TeV \cite{Khachatryan:2016qkc}. In deriving the bounds for $e^+ e^-, \mu^+ \mu^-$ final states, the corresponding gauge coupling was chosen to be $g_{BL} \approx 0.28$. Therefore, such bounds can get weaker if we consider slightly smaller values of gauge couplings. 
\begin{table}[htbp]
\caption{Cross Sections (pb) for $p p \rightarrow \mu \bar{\mu} $}
\centering
\begin{tabular}{|l|l|l|l|}

\hline
$  M_{Z_{BL}}(TeV) $ &  $g_{BL}=0.01$ & $g_{BL}=0.05$  \\
\hline
\ \ \ \ \  2 & $1.44\times 10^{-5}$ & $ 3.600\times10^{-4}$  \\
\hline
\ \ \ \ \ 3 & $ 1.29\times10^{-6}$ &$ 3.237\times 10^{-5}$  \\
\hline
\ \ \ \ \ 4 & $ 1.52\times 10^{-7}$ &$ 3.802\times 10^{-6}$  \\
\hline
\ \ \ \ \ 5 & $1.91\times 10^{-8} $& $4.776 \times 10^{-7} $   \\ 
\hline
\end{tabular}
\label{sigmamumu}
\end{table}

\begin{table}[htbp]
\caption{Cross Sections (pb) for $p p \rightarrow \chi \chi $}
\centering
\begin{tabular}{|l|l|l|l|}

\hline
$  M_{Z_{BL}} (\text{TeV}) $ &  $g_{BL}=0.01$ & $g_{BL}=0.05$  \\
\hline
\ \ \ \ \  2 &$5.680\times10^{-7}$  &$ 1.420\times10^{-5}$ \\
\hline
\ \ \ \ \ 3 & $ 5.129\times10^{-8}$ &$1.283\times10^{-6}$ \\
\hline
\ \ \ \ \ 4 &$6.040\times10^{-9}$  & $1.510\times10^{-7}$\\
\hline
\ \ \ \ \ 5 &$7.570\times10^{-10} $  &  $ 1.898\times10^{-8}$   \\ 
\hline
\end{tabular}
\label{sigmachichi}
\end{table}

\begin{table}[htbp]
\caption{Cross Sections (pb) for $p p \rightarrow \eta^{0r} \eta^{0i} $}
\centering
\begin{tabular}{|l|l|l|l|}

\hline
$  M_{Z_{BL}} (\text{TeV}) $ &  $g_{BL}=0.01$ & $g_{BL}=0.05$ \\
\hline
\ \ \ \ \  2 & $1.576\times10^{-6} $ & $3.938\times10^{-5}$ \\
\hline
\ \ \ \ \ 3 &$1.432 \times 10^{-7}$  &$4.208\times 10^{-6}$ \\
\hline
\ \ \ \ \ 4 & $ 1.684\times10^{-8}$ &$ 4.208\times 10^{-7}$   \\
\hline
\ \ \ \ \ 5 & $ 2.117\times10^{-9}$ &$ 5.294\times 10^{-8}$   \\ 
\hline
\end{tabular}
\label{sigmaeta}
\end{table}

The differential cross section with respect to the invariant final state dilepton mass $M_{ll}$ is
\begin{equation}
\frac{d\sigma}{dM_{ll}} = \sum_{a,b} \int^1_{\frac{M^2_{ll}}{E^2_{CM}}} dx \frac{2M_{ll}}{xE^2_{CM}} f_a (x, Q^2) f_b \left(\frac{M^2_{ll}}{xE^2_{CM}}, Q^2 \right) \hat{\sigma} (q \bar{q} \rightarrow Z_{BL} \rightarrow l^+ l^-),
\end{equation}
with $f_a$ being the parton distribution function for a parton denoted by 'a' and $E_{CM}=13$ TeV is the LHC centre of mass energy. The cross section $\hat{\sigma}$ is given by \cite{Okada:2016gsh}
\begin{equation}
\hat{\sigma} = \frac{1}{81} \frac{g^4_{BL}}{4\pi} \frac{M^2_{ll}}{(M^2_{ll}-M^2_{Z_{BL}})^2 + M^2_{Z_{BL}} \Gamma^2_{Z_{BL}}}
\end{equation}
Since the decay width of $Z_{BL}$ boson is narrow, the cross section is almost determined by the integral in the vicinity of the resonance. In the narrow width approximation we can use
\begin{equation}
\sigma(pp \rightarrow Z_{BL} \rightarrow l^+ l^-) = \sigma (pp \rightarrow Z_{BL}) \text{BR} (Z_{BL} \rightarrow l^+ l^-)
\end{equation}
We first calculate the production cross section $\sigma (pp \rightarrow Z_{BL}) $ for benchmark values of $g_{BL}, M_{Z_{BL}}$ at 13 TeV LHC and then multiply by the respective branching ratio to find the final state cross sections. We particularly choose some benchmark values of $g_{BL}, M_{Z_{BL}}$ which are still allowed by the latest LHC constraints \cite{Aaboud:2017buh}. The numerical values of these cross sections for different values of $g_{BL}, M_{Z_{BL}}$ are shown in table \ref{sigmamumu}, \ref{sigmachichi}, \ref{sigmaeta} for $\mu^+ \mu^-, \chi \chi$ and $\eta^{0r}, \eta^{0i}$ final states respectively. Although the last two final states just correspond to missing energies depending on fermionic or bosonic nature of DM, the leptonic final states can saturate the ATLAS bounds \cite{Aaboud:2017buh}. In fact, for $g_{BL} =0.05$, $Z_{BL}$ mass of 2 TeV can saturate the ATLAS upper bounds if we compare the results in table \ref{sigmamumu} with the results of \cite{Aaboud:2017buh}. A more detailed analysis of the model from collider point of view is beyond the scope of the present work and we leave it for future discussions.

\section{Results and Conclusion}
\label{sec9}
We have studied a $U(1)_{B-L}$ gauge extension of the SM that can explain the origin of neutrino mass and dark matter simultaneously. The new fermions that take part in making the $U(1)_{B-L}$ model anomaly free can not only provide a stable dark matter candidate, but can also give rise to light neutrino masses at one loop level. There also exists additional scalar doublets apart from the SM Higgs, that take part in the radiative generation of neutrino mass. We discuss two different versions of the model, one with fermion singlet DM and the other with scalar doublet DM. We first calculate the mass spectra for both the models after the spontaneous symmetry breaking. We find that for the minimal scalar content, both the models predict the lightest neutrino mass to be zero. 

We first calculate the relic abundance of both fermion and scalar DM and show how it depends on the model parameters, specially the ones related to the $U(1)_{B-L}$ sector. The fermion DM relic abundance crucially depends on the $U(1)_{B-L}$ gauge sector parameters $g_{BL}, M_{Z_{BL}}$, typically satisfying the relic abundance criteria near the resonance region $2M_{DM} = M_{Z_{BL}}$. There can also be Higgs portal interactions due to the existence of additional singlet scalars that not only take part in spontaneous breaking of $U(1)_{B-L}$ gauge symmetry, but can also assist in fermion DM annihilation into SM particles through its mixing with the SM Higgs boson. On the other hand, the scalar DM relic abundance is not very sensitive to the $U(1)_{B-L}$ sector physics and is mostly dictated by the Higgs portal and electroweak gauge sector interactions. The relic abundance criteria can be satisfied in two mass regions namely, the low mass regime $M_{DM} < M_W$ and high mass regime $M_{DM} > 500$ GeV.

We then study the sensitivity of direct detection experiments to both fermion and scalar DM. The spin independent DM-nucleon cross section for fermion DM is found to lie within the latest upper limit given by the Xenon1T experiment. This is observed by taking both gauge and scalar mediated DM-nucleon scatterings. A random scan of fermion DM parameter space consistent with relic abundance criteria also shows the entire DM mass range remains allowed from the direct detection constraints. In fact the Xenon1T constraints remain weaker than the LEP II bound $M_{Z_{BL}}/g_{BL} \geq 7$ TeV as well as LHC bound in fermion DM case, as can be seen from figure \ref{fig5}. The direct detection cross section of scalar DM is mostly controlled by the DM-Higgs coupling and is more sensitive to experiments like Xenon1T, primarily due to lighter mediator that is, SM Higgs. We find that for DM-Higgs coupling 0.1, scalar DM masses are ruled out even beyond 1 TeV by the Xenon1T limits. The LHC limit on the Higgs invisible decay width also rules out some region of parameter space for $M_{DM} < m_h/2$ although this constraint remains much weaker than the Xenon1T bounds. We also check the indirect detection bounds on DM annihilations into different SM final states and found that these limits can be saturated only by scalar DM for certain region of parameter space while fermion DM parameter space satisfying correct relic remains allowed. The scalar DM parameter space is more constrained by such indirect detection bounds compared to the fermion counterpart. For example, scalar DM mass can be ruled out far beyond 1 TeV for certain benchmark parameters if we incorporate the indirect detection bounds on DM annihilations into $W^+ W^-$ final states.

We then check the sensitivity of rare decay experiments like MEG looking for charged lepton flavour violating decay $\mu \rightarrow e \gamma$ to the parameter space of the model. This can arise at one loop level due to similar diagrams that give rise to light neutrino masses. Thus, the same parameters that affect DM phenomenology and light neutrino mass can also give rise to a new contribution to this rare decay process. We find that for some typical values of Yukawa couplings and DM masses, the model can saturate the MEG upper limit on the branching ratio of $\mu \rightarrow e \gamma$ and hence can be probed in near future searches. 

Finally, we briefly discuss the possibility of probing such a model at energy frontier experiments like the LHC. We calculate the decay width and branching ratio of $Z_{BL}$ into different final states like leptons, DM etc. After calculating the production cross section of $Z_{BL}$ in proton proton collisions at 13 TeV centre of mass energy of the LHC, we multiply the branching ratio into respective final states to estimate the total cross section into final states. This is possible due to the narrow decay width of the $Z_{BL}$ boson. We find that, the model can be ruled out by the LHC bounds on heavy dilepton resonance searches if $Z_{BL}$ mass is around 2 TeV and the corresponding gauge coupling is 0.05. The model can also predict other final states with missing energy that can be probed at ongoing and near future colliders. We leave a detailed collider study of such final states to future works.

In summary, we have proposed a framework for common origin of neutrino mass and dark matter within a $U(1)_{B-L}$ gauge extension of the SM so that the fermion fields responsible for radiative neutrino mass and DM phenomenology can also keep the $U(1)_{B-L}$ model anomaly free. Dark matter is stabilised by a remnant symmetry after the spontaneous breaking of the $U(1)_{B-L}$ gauge symmetry, without requiring any ad-hoc symmetry to guarantee is stability. We discuss both scalar and fermion DM in two different versions of the model and show that the model, apart from predicting a vanishing lightest neutrino mass scenario, can have signatures at cosmic, intensity as well as energy frontier experiments. Ongoing as well as near future experiments in all these three frontiers are going to probe a significant region of parameter space of the proposed model.

\begin{acknowledgments}
DN would like to thank Shibananda Sahoo, Amit Dutta Banik and Arnab Dasgupta for useful discussions. DB acknowledges the support from IIT Guwahati start-up grant (reference number: xPHYSUGIITG01152xxDB001) and Associateship Programme of IUCAA, Pune.

\end{acknowledgments}
\appendix
\section{Interaction Vertices}
\label{appen1}
Here we list the relevant vertices of different interactions, derived from the Lagrangian. They are used in calculation of different cross sections and decay widths. Some notations are defined in the table \ref{appentable1} followed by the interaction vertices in table \ref{appentable2}.
\begin{table}[h]
\caption{Relevant Parameters for Interaction Vertices}
\centering
\begin{tabular}{|l|}

\hline
$\lambda_{L}= (\lambda_{H\eta}+\lambda^{\prime}_{H\eta}) $ \\ 
\hline
$ n_{f}= U(1)_{B-L} $ charge of SM fermions \\
\hline
$ n_{\eta}=U(1)_{B-L} $ charge of $ \eta $\\
\hline
$ n_{\chi}= U(1)_{B-L} $ charge of $\chi $\\
\hline
$\lambda_{H\phi_{1}} = \frac{m^2_{ \phi_1}-m^2_{h}}{2 v u_{1}} \xi $\\
\hline
$ f = \frac{m_{\chi}}{u_{1}}  $\\
\hline
\end{tabular}
\label{appentable1}
\end{table}

\begin{table}
\caption{All new possible vertices}
\centering
\begin{tabular}{|l|l|l|l|}

\hline
Interactions &  Vertices & Interactions &  Vertices  \\
\hline
\ \ \   $\eta^{0r} \ \eta^{0r} \ h$  &  $\lambda_{L} v$ &$\eta^{0r} \ \eta^{0r} \ \phi_{1} $& $\lambda_{\eta \phi_{1}} u_{1}$\\
\hline
\ \ \  $\eta^{+} \ \eta^{-} \ h$ & $ \lambda_L v - (M^2_{\eta^{0r}} + M^2_{\eta^{0i}} -2M^2_{{\eta^{+}}})/v $  & $\eta^{+} \ \eta^{-} \ \phi_{1} $& $\lambda_{\eta \phi_{1}} u_{1}$\\
\hline
\ \ \  $\eta^{0i} \ \eta^{0i} \ h$ & $\lambda_{L}v= (\lambda_{H\eta}+\lambda^{\prime}_{H\eta}) v $ & $\eta^{0i} \ \eta^{0i} \ \phi_{1} $ & $\lambda_{\eta \phi_{1}} u_{1}$\\ 
\hline
\ \ \  $ \eta^{0i} \ \eta^{0i} \ h \ h $ & $2 \lambda _{L}$ &  $\eta^{0r}\ \eta^{0r} \ Z\ Z $ &  $(ig^{2}/2 \cos^2\theta_{w}) g_{\mu \nu}$ \\
\hline
\ \ \    $ \eta^{+} \ \eta^{-} \ h \ h $ & $ \lambda_L  - (M^2_{\eta^{0r}} + M^2_{\eta^{0i}} -2M^2_{{\eta^{+}}})/v^2 $ & $\eta^{+} \ \eta^{-}\ \gamma $ &  $-ie$ \\
\hline
\ \ \    $ \eta^{0r} \ \eta^{0r} \ h \ h $ & $2 \lambda _{L}$ & $\eta^{0I}\ \eta^{-} \ W^{+}$ &  $g/2$ \\

\hline
\ \ \  $ \phi_1 \ h\  h $ & $i (\lambda_{H\phi_{1}} \ u_{1}/2)$ & $\eta^{+}\ \eta^{-} \ W^{+}\ W^{-} $ &  $(ig^{2}/2) g_{\mu \nu}$  \\ 
\hline
\ \ \    $\eta^{0r}\ \eta^{-} \ W^{+}$ &  $g/2$  &  $\eta^{0R}\ \eta^{0I} \ Z_{BL} $ &  $ i \ n_{\eta} \  g_{BL}$ \\
\hline
\ \ \   $\eta^{+}\ \eta^{-} \ Z\ Z $ & $ (ig^{2} (1-2\sin^{2} \theta_{w})^2 / 2 \cos^2 \theta_{w}) g_{\mu \nu}$ &   $ \phi_1 \ f \ \bar{f} $ &  $(i\ \xi \ m_{f} / v) $ \\ 
\hline
\ \ \   $\eta^{+} \ \eta^{-}\ Z$ &  $(-ig (1-2\sin^2\theta_{w}) /2 \cos\theta_{w})$  &   $ \phi_1 \ W^{+} \  W^{-} $ & $i g \ m_{W}\  g_{\mu \nu}$\\ 
\hline
\ \ \   $\eta^{0I}\ \eta^{0I} \ Z\ Z $ & $(ig^{2}/2 \cos^2\theta_{w}) g_{\mu \nu}$  & $ \phi_1\  Z \ Z $ & $( i g\ m_{Z}/\cos \theta_{w})\ g_{\mu \nu}$\\ 
\hline
\ \ \   $\eta^{0r} \ \eta^{0i}\ Z$ & $(g/2 \cos\theta_{w})$ & $f \ \bar{f} \ Z_{BL}$ & $-i \ n_{f} \ g_{BL} \ \gamma ^{\mu}$ \\
\hline
\ \ \   $\eta^{0r}\ \eta^{0r} \ W^{+}\ W^{-} $ & $(ig^{2}/2) g_{\mu \nu}$ & $ \chi \ \chi \ \phi_1 $ & $-f/2$\\
\hline
\ \ \  $\eta^{0I}\ \eta^{0I} \ W^{+}\ W^{-} $ & $(ig^{2}/2) g_{\mu \nu}$ & $ \chi \ \chi\  Z_{BL}$ &  $i\ n_{\chi} \ g_{BL} \ \gamma ^{\mu}\  R $   \\
\hline
\ \ \ $\eta^{+}\ \eta^{-} \ \gamma\ \gamma $ &  $ 2i e^2 g_{\mu \nu}$ & & \\ 
\hline

\end{tabular}
\label{appentable2}
\end{table}

\newpage

\begin{thebibliography}{92}
\expandafter\ifx\csname natexlab\endcsname\relax\def\natexlab#1{#1}\fi
\expandafter\ifx\csname bibnamefont\endcsname\relax
  \def\bibnamefont#1{#1}\fi
\expandafter\ifx\csname bibfnamefont\endcsname\relax
  \def\bibfnamefont#1{#1}\fi
\expandafter\ifx\csname citenamefont\endcsname\relax
  \def\citenamefont#1{#1}\fi
\expandafter\ifx\csname url\endcsname\relax
  \def\url#1{\texttt{#1}}\fi
\expandafter\ifx\csname urlprefix\endcsname\relax\def\urlprefix{URL }\fi
\providecommand{\bibinfo}[2]{#2}
\providecommand{\eprint}[2][]{\url{#2}}

\bibitem[{\citenamefont{Fukuda et~al.}(2001)}]{Fukuda:2001nk}
\bibinfo{author}{\bibfnamefont{S.}~\bibnamefont{Fukuda}} \bibnamefont{et~al.}
  (\bibinfo{collaboration}{Super-Kamiokande}), \bibinfo{journal}{Phys. Rev.
  Lett.} \textbf{\bibinfo{volume}{86}}, \bibinfo{pages}{5656}
  (\bibinfo{year}{2001}), \eprint{hep-ex/0103033}.

\bibitem[{\citenamefont{Ahmad et~al.}(2002{\natexlab{a}})}]{Ahmad:2002jz}
\bibinfo{author}{\bibfnamefont{Q.~R.} \bibnamefont{Ahmad}} \bibnamefont{et~al.}
  (\bibinfo{collaboration}{SNO}), \bibinfo{journal}{Phys. Rev. Lett.}
  \textbf{\bibinfo{volume}{89}}, \bibinfo{pages}{011301}
  (\bibinfo{year}{2002}{\natexlab{a}}), \eprint{nucl-ex/0204008}.

\bibitem[{\citenamefont{Ahmad et~al.}(2002{\natexlab{b}})}]{Ahmad:2002ka}
\bibinfo{author}{\bibfnamefont{Q.~R.} \bibnamefont{Ahmad}} \bibnamefont{et~al.}
  (\bibinfo{collaboration}{SNO}), \bibinfo{journal}{Phys. Rev. Lett.}
  \textbf{\bibinfo{volume}{89}}, \bibinfo{pages}{011302}
  (\bibinfo{year}{2002}{\natexlab{b}}), \eprint{nucl-ex/0204009}.

\bibitem[{\citenamefont{Abe et~al.}(2008)}]{Abe:2008aa}
\bibinfo{author}{\bibfnamefont{S.}~\bibnamefont{Abe}} \bibnamefont{et~al.}
  (\bibinfo{collaboration}{KamLAND}), \bibinfo{journal}{Phys. Rev. Lett.}
  \textbf{\bibinfo{volume}{100}}, \bibinfo{pages}{221803}
  (\bibinfo{year}{2008}), \eprint{0801.4589}.

\bibitem[{\citenamefont{Abe et~al.}(2011)}]{Abe:2011sj}
\bibinfo{author}{\bibfnamefont{K.}~\bibnamefont{Abe}} \bibnamefont{et~al.}
  (\bibinfo{collaboration}{T2K}), \bibinfo{journal}{Phys. Rev. Lett.}
  \textbf{\bibinfo{volume}{107}}, \bibinfo{pages}{041801}
  (\bibinfo{year}{2011}), \eprint{1106.2822}.

\bibitem[{\citenamefont{Abe et~al.}(2012)}]{Abe:2011fz}
\bibinfo{author}{\bibfnamefont{Y.}~\bibnamefont{Abe}} \bibnamefont{et~al.}
  (\bibinfo{collaboration}{Double Chooz}), \bibinfo{journal}{Phys. Rev. Lett.}
  \textbf{\bibinfo{volume}{108}}, \bibinfo{pages}{131801}
  (\bibinfo{year}{2012}), \eprint{1112.6353}.

\bibitem[{\citenamefont{An et~al.}(2012)}]{An:2012eh}
\bibinfo{author}{\bibfnamefont{F.~P.} \bibnamefont{An}} \bibnamefont{et~al.}
  (\bibinfo{collaboration}{Daya Bay}), \bibinfo{journal}{Phys. Rev. Lett.}
  \textbf{\bibinfo{volume}{108}}, \bibinfo{pages}{171803}
  (\bibinfo{year}{2012}), \eprint{1203.1669}.

\bibitem[{\citenamefont{Ahn et~al.}(2012)}]{Ahn:2012nd}
\bibinfo{author}{\bibfnamefont{J.~K.} \bibnamefont{Ahn}} \bibnamefont{et~al.}
  (\bibinfo{collaboration}{RENO}), \bibinfo{journal}{Phys. Rev. Lett.}
  \textbf{\bibinfo{volume}{108}}, \bibinfo{pages}{191802}
  (\bibinfo{year}{2012}), \eprint{1204.0626}.

\bibitem[{\citenamefont{Adamson et~al.}(2013)}]{Adamson:2013ue}
\bibinfo{author}{\bibfnamefont{P.}~\bibnamefont{Adamson}} \bibnamefont{et~al.}
  (\bibinfo{collaboration}{MINOS}), \bibinfo{journal}{Phys. Rev. Lett.}
  \textbf{\bibinfo{volume}{110}}, \bibinfo{pages}{171801}
  (\bibinfo{year}{2013}), \eprint{1301.4581}.

\bibitem[{\citenamefont{Patrignani et~al.}(2016)}]{Olive:2016xmw}
\bibinfo{author}{\bibfnamefont{C.}~\bibnamefont{Patrignani}}
  \bibnamefont{et~al.} (\bibinfo{collaboration}{Particle Data Group}),
  \bibinfo{journal}{Chin. Phys.} \textbf{\bibinfo{volume}{C40}},
  \bibinfo{pages}{100001} (\bibinfo{year}{2016}).

\bibitem[{\citenamefont{Esteban et~al.}(2017)\citenamefont{Esteban,
  Gonzalez-Garcia, Maltoni, Martinez-Soler, and Schwetz}}]{Esteban:2016qun}
\bibinfo{author}{\bibfnamefont{I.}~\bibnamefont{Esteban}},
  \bibinfo{author}{\bibfnamefont{M.~C.} \bibnamefont{Gonzalez-Garcia}},
  \bibinfo{author}{\bibfnamefont{M.}~\bibnamefont{Maltoni}},
  \bibinfo{author}{\bibfnamefont{I.}~\bibnamefont{Martinez-Soler}},
  \bibnamefont{and} \bibinfo{author}{\bibfnamefont{T.}~\bibnamefont{Schwetz}},
  \bibinfo{journal}{JHEP} \textbf{\bibinfo{volume}{01}}, \bibinfo{pages}{087}
  (\bibinfo{year}{2017}), \eprint{1611.01514}.

\bibitem[{\citenamefont{Abe et~al.}(2015)}]{Abe:2015awa}
\bibinfo{author}{\bibfnamefont{K.}~\bibnamefont{Abe}} \bibnamefont{et~al.}
  (\bibinfo{collaboration}{T2K}), \bibinfo{journal}{Phys. Rev.}
  \textbf{\bibinfo{volume}{D91}}, \bibinfo{pages}{072010}
  (\bibinfo{year}{2015}), \eprint{1502.01550}.

\bibitem[{\citenamefont{Ade et~al.}(2016)}]{Ade:2015xua}
\bibinfo{author}{\bibfnamefont{P.~A.~R.} \bibnamefont{Ade}}
  \bibnamefont{et~al.} (\bibinfo{collaboration}{Planck}),
  \bibinfo{journal}{Astron. Astrophys.} \textbf{\bibinfo{volume}{594}},
  \bibinfo{pages}{A13} (\bibinfo{year}{2016}), \eprint{1502.01589}.

\bibitem[{\citenamefont{Gando et~al.}(2016)}]{KamLAND-Zen:2016pfg}
\bibinfo{author}{\bibfnamefont{A.}~\bibnamefont{Gando}} \bibnamefont{et~al.}
  (\bibinfo{collaboration}{KamLAND-Zen}), \bibinfo{journal}{Phys. Rev. Lett.}
  \textbf{\bibinfo{volume}{117}}, \bibinfo{pages}{082503}
  (\bibinfo{year}{2016}), \bibinfo{note}{[Addendum: Phys. Rev.
  Lett.117,no.10,109903(2016)]}, \eprint{1605.02889}.

\bibitem[{\citenamefont{Weinberg}(1979)}]{Weinberg:1979sa}
\bibinfo{author}{\bibfnamefont{S.}~\bibnamefont{Weinberg}},
  \bibinfo{journal}{Phys. Rev. Lett.} \textbf{\bibinfo{volume}{43}},
  \bibinfo{pages}{1566} (\bibinfo{year}{1979}).

\bibitem[{\citenamefont{Minkowski}(1977)}]{Minkowski:1977sc}
\bibinfo{author}{\bibfnamefont{P.}~\bibnamefont{Minkowski}},
  \bibinfo{journal}{Phys. Lett.} \textbf{\bibinfo{volume}{B67}},
  \bibinfo{pages}{421} (\bibinfo{year}{1977}).

\bibitem[{\citenamefont{Gell-Mann et~al.}(1979)\citenamefont{Gell-Mann, Ramond,
  and Slansky}}]{GellMann:1980vs}
\bibinfo{author}{\bibfnamefont{M.}~\bibnamefont{Gell-Mann}},
  \bibinfo{author}{\bibfnamefont{P.}~\bibnamefont{Ramond}}, \bibnamefont{and}
  \bibinfo{author}{\bibfnamefont{R.}~\bibnamefont{Slansky}},
  \bibinfo{journal}{Conf. Proc.} \textbf{\bibinfo{volume}{C790927}},
  \bibinfo{pages}{315} (\bibinfo{year}{1979}), \eprint{1306.4669}.

\bibitem[{\citenamefont{Mohapatra and Senjanovic}(1980)}]{Mohapatra:1979ia}
\bibinfo{author}{\bibfnamefont{R.~N.} \bibnamefont{Mohapatra}}
  \bibnamefont{and}
  \bibinfo{author}{\bibfnamefont{G.}~\bibnamefont{Senjanovic}},
  \bibinfo{journal}{Phys. Rev. Lett.} \textbf{\bibinfo{volume}{44}},
  \bibinfo{pages}{912} (\bibinfo{year}{1980}).

\bibitem[{\citenamefont{Schechter and Valle}(1980)}]{Schechter:1980gr}
\bibinfo{author}{\bibfnamefont{J.}~\bibnamefont{Schechter}} \bibnamefont{and}
  \bibinfo{author}{\bibfnamefont{J.~W.~F.} \bibnamefont{Valle}},
  \bibinfo{journal}{Phys. Rev.} \textbf{\bibinfo{volume}{D22}},
  \bibinfo{pages}{2227} (\bibinfo{year}{1980}).

\bibitem[{\citenamefont{Mohapatra and Senjanovic}(1981)}]{Mohapatra:1980yp}
\bibinfo{author}{\bibfnamefont{R.~N.} \bibnamefont{Mohapatra}}
  \bibnamefont{and}
  \bibinfo{author}{\bibfnamefont{G.}~\bibnamefont{Senjanovic}},
  \bibinfo{journal}{Phys. Rev.} \textbf{\bibinfo{volume}{D23}},
  \bibinfo{pages}{165} (\bibinfo{year}{1981}).

\bibitem[{\citenamefont{Lazarides et~al.}(1981)\citenamefont{Lazarides, Shafi,
  and Wetterich}}]{Lazarides:1980nt}
\bibinfo{author}{\bibfnamefont{G.}~\bibnamefont{Lazarides}},
  \bibinfo{author}{\bibfnamefont{Q.}~\bibnamefont{Shafi}}, \bibnamefont{and}
  \bibinfo{author}{\bibfnamefont{C.}~\bibnamefont{Wetterich}},
  \bibinfo{journal}{Nucl. Phys.} \textbf{\bibinfo{volume}{B181}},
  \bibinfo{pages}{287} (\bibinfo{year}{1981}).

\bibitem[{\citenamefont{Wetterich}(1981)}]{Wetterich:1981bx}
\bibinfo{author}{\bibfnamefont{C.}~\bibnamefont{Wetterich}},
  \bibinfo{journal}{Nucl. Phys.} \textbf{\bibinfo{volume}{B187}},
  \bibinfo{pages}{343} (\bibinfo{year}{1981}).

\bibitem[{\citenamefont{Schechter and Valle}(1982)}]{Schechter:1981cv}
\bibinfo{author}{\bibfnamefont{J.}~\bibnamefont{Schechter}} \bibnamefont{and}
  \bibinfo{author}{\bibfnamefont{J.~W.~F.} \bibnamefont{Valle}},
  \bibinfo{journal}{Phys. Rev.} \textbf{\bibinfo{volume}{D25}},
  \bibinfo{pages}{774} (\bibinfo{year}{1982}).

\bibitem[{\citenamefont{Brahmachari and Mohapatra}(1998)}]{Brahmachari:1997cq}
\bibinfo{author}{\bibfnamefont{B.}~\bibnamefont{Brahmachari}} \bibnamefont{and}
  \bibinfo{author}{\bibfnamefont{R.~N.} \bibnamefont{Mohapatra}},
  \bibinfo{journal}{Phys. Rev.} \textbf{\bibinfo{volume}{D58}},
  \bibinfo{pages}{015001} (\bibinfo{year}{1998}), \eprint{hep-ph/9710371}.

\bibitem[{\citenamefont{Foot et~al.}(1989)\citenamefont{Foot, Lew, He, and
  Joshi}}]{Foot:1988aq}
\bibinfo{author}{\bibfnamefont{R.}~\bibnamefont{Foot}},
  \bibinfo{author}{\bibfnamefont{H.}~\bibnamefont{Lew}},
  \bibinfo{author}{\bibfnamefont{X.~G.} \bibnamefont{He}}, \bibnamefont{and}
  \bibinfo{author}{\bibfnamefont{G.~C.} \bibnamefont{Joshi}},
  \bibinfo{journal}{Z. Phys.} \textbf{\bibinfo{volume}{C44}},
  \bibinfo{pages}{441} (\bibinfo{year}{1989}).

\bibitem[{\citenamefont{Zwicky}(1933)}]{Zwicky:1933gu}
\bibinfo{author}{\bibfnamefont{F.}~\bibnamefont{Zwicky}},
  \bibinfo{journal}{Helv. Phys. Acta} \textbf{\bibinfo{volume}{6}},
  \bibinfo{pages}{110} (\bibinfo{year}{1933}), \bibinfo{note}{[Gen. Rel.
  Grav.41,207(2009)]}.

\bibitem[{\citenamefont{Rubin and Ford}(1970)}]{Rubin:1970zza}
\bibinfo{author}{\bibfnamefont{V.~C.} \bibnamefont{Rubin}} \bibnamefont{and}
  \bibinfo{author}{\bibfnamefont{W.~K.} \bibnamefont{Ford},
  \bibfnamefont{Jr.}}, \bibinfo{journal}{Astrophys. J.}
  \textbf{\bibinfo{volume}{159}}, \bibinfo{pages}{379} (\bibinfo{year}{1970}).

\bibitem[{\citenamefont{Clowe et~al.}(2006)\citenamefont{Clowe, Bradac,
  Gonzalez, Markevitch, Randall, Jones, and Zaritsky}}]{Clowe:2006eq}
\bibinfo{author}{\bibfnamefont{D.}~\bibnamefont{Clowe}},
  \bibinfo{author}{\bibfnamefont{M.}~\bibnamefont{Bradac}},
  \bibinfo{author}{\bibfnamefont{A.~H.} \bibnamefont{Gonzalez}},
  \bibinfo{author}{\bibfnamefont{M.}~\bibnamefont{Markevitch}},
  \bibinfo{author}{\bibfnamefont{S.~W.} \bibnamefont{Randall}},
  \bibinfo{author}{\bibfnamefont{C.}~\bibnamefont{Jones}}, \bibnamefont{and}
  \bibinfo{author}{\bibfnamefont{D.}~\bibnamefont{Zaritsky}},
  \bibinfo{journal}{Astrophys. J.} \textbf{\bibinfo{volume}{648}},
  \bibinfo{pages}{L109} (\bibinfo{year}{2006}), \eprint{astro-ph/0608407}.

\bibitem[{\citenamefont{Taoso et~al.}(2008)\citenamefont{Taoso, Bertone, and
  Masiero}}]{Taoso:2007qk}
\bibinfo{author}{\bibfnamefont{M.}~\bibnamefont{Taoso}},
  \bibinfo{author}{\bibfnamefont{G.}~\bibnamefont{Bertone}}, \bibnamefont{and}
  \bibinfo{author}{\bibfnamefont{A.}~\bibnamefont{Masiero}},
  \bibinfo{journal}{JCAP} \textbf{\bibinfo{volume}{0803}}, \bibinfo{pages}{022}
  (\bibinfo{year}{2008}), \eprint{0711.4996}.

\bibitem[{\citenamefont{Arcadi et~al.}(2017)\citenamefont{Arcadi, Dutra, Ghosh,
  Lindner, Mambrini, Pierre, Profumo, and Queiroz}}]{Arcadi:2017kky}
\bibinfo{author}{\bibfnamefont{G.}~\bibnamefont{Arcadi}},
  \bibinfo{author}{\bibfnamefont{M.}~\bibnamefont{Dutra}},
  \bibinfo{author}{\bibfnamefont{P.}~\bibnamefont{Ghosh}},
  \bibinfo{author}{\bibfnamefont{M.}~\bibnamefont{Lindner}},
  \bibinfo{author}{\bibfnamefont{Y.}~\bibnamefont{Mambrini}},
  \bibinfo{author}{\bibfnamefont{M.}~\bibnamefont{Pierre}},
  \bibinfo{author}{\bibfnamefont{S.}~\bibnamefont{Profumo}}, \bibnamefont{and}
  \bibinfo{author}{\bibfnamefont{F.~S.} \bibnamefont{Queiroz}}
  (\bibinfo{year}{2017}), \eprint{1703.07364}.

\bibitem[{\citenamefont{Kahlhoefer}(2017)}]{Kahlhoefer:2017dnp}
\bibinfo{author}{\bibfnamefont{F.}~\bibnamefont{Kahlhoefer}},
  \bibinfo{journal}{Int. J. Mod. Phys.} \textbf{\bibinfo{volume}{A32}},
  \bibinfo{pages}{1730006} (\bibinfo{year}{2017}), \eprint{1702.02430}.

\bibitem[{\citenamefont{Akerib et~al.}(2017)}]{Akerib:2016vxi}
\bibinfo{author}{\bibfnamefont{D.~S.} \bibnamefont{Akerib}}
  \bibnamefont{et~al.} (\bibinfo{collaboration}{LUX}), \bibinfo{journal}{Phys.
  Rev. Lett.} \textbf{\bibinfo{volume}{118}}, \bibinfo{pages}{021303}
  (\bibinfo{year}{2017}), \eprint{1608.07648}.

\bibitem[{\citenamefont{Tan et~al.}(2016)}]{Tan:2016zwf}
\bibinfo{author}{\bibfnamefont{A.}~\bibnamefont{Tan}} \bibnamefont{et~al.}
  (\bibinfo{collaboration}{PandaX-II}), \bibinfo{journal}{Phys. Rev. Lett.}
  \textbf{\bibinfo{volume}{117}}, \bibinfo{pages}{121303}
  (\bibinfo{year}{2016}), \eprint{1607.07400}.

\bibitem[{\citenamefont{Cui et~al.}(2017)}]{Cui:2017nnn}
\bibinfo{author}{\bibfnamefont{X.}~\bibnamefont{Cui}} \bibnamefont{et~al.}
  (\bibinfo{collaboration}{PandaX-II}) (\bibinfo{year}{2017}),
  \eprint{1708.06917}.

\bibitem[{\citenamefont{Aprile et~al.}(2017)}]{Aprile:2017iyp}
\bibinfo{author}{\bibfnamefont{E.}~\bibnamefont{Aprile}} \bibnamefont{et~al.}
  (\bibinfo{collaboration}{XENON}) (\bibinfo{year}{2017}), \eprint{1705.06655}.

\bibitem[{\citenamefont{Mohapatra and Marshak}(1980)}]{Mohapatra:1980qe}
\bibinfo{author}{\bibfnamefont{R.~N.} \bibnamefont{Mohapatra}}
  \bibnamefont{and} \bibinfo{author}{\bibfnamefont{R.~E.}
  \bibnamefont{Marshak}}, \bibinfo{journal}{Phys. Rev. Lett.}
  \textbf{\bibinfo{volume}{44}}, \bibinfo{pages}{1316} (\bibinfo{year}{1980}),
  \bibinfo{note}{[Erratum: Phys. Rev. Lett.44,1643(1980)]}.

\bibitem[{\citenamefont{Marshak and Mohapatra}(1980)}]{Marshak:1979fm}
\bibinfo{author}{\bibfnamefont{R.~E.} \bibnamefont{Marshak}} \bibnamefont{and}
  \bibinfo{author}{\bibfnamefont{R.~N.} \bibnamefont{Mohapatra}},
  \bibinfo{journal}{Phys. Lett.} \textbf{\bibinfo{volume}{91B}},
  \bibinfo{pages}{222} (\bibinfo{year}{1980}).

\bibitem[{\citenamefont{Masiero et~al.}(1982)\citenamefont{Masiero, Nieves, and
  Yanagida}}]{Masiero:1982fi}
\bibinfo{author}{\bibfnamefont{A.}~\bibnamefont{Masiero}},
  \bibinfo{author}{\bibfnamefont{J.~F.} \bibnamefont{Nieves}},
  \bibnamefont{and} \bibinfo{author}{\bibfnamefont{T.}~\bibnamefont{Yanagida}},
  \bibinfo{journal}{Phys. Lett.} \textbf{\bibinfo{volume}{116B}},
  \bibinfo{pages}{11} (\bibinfo{year}{1982}).

\bibitem[{\citenamefont{Mohapatra and Senjanovic}(1983)}]{Mohapatra:1982xz}
\bibinfo{author}{\bibfnamefont{R.~N.} \bibnamefont{Mohapatra}}
  \bibnamefont{and}
  \bibinfo{author}{\bibfnamefont{G.}~\bibnamefont{Senjanovic}},
  \bibinfo{journal}{Phys. Rev.} \textbf{\bibinfo{volume}{D27}},
  \bibinfo{pages}{254} (\bibinfo{year}{1983}).

\bibitem[{\citenamefont{Buchmuller et~al.}(1991)\citenamefont{Buchmuller,
  Greub, and Minkowski}}]{Buchmuller:1991ce}
\bibinfo{author}{\bibfnamefont{W.}~\bibnamefont{Buchmuller}},
  \bibinfo{author}{\bibfnamefont{C.}~\bibnamefont{Greub}}, \bibnamefont{and}
  \bibinfo{author}{\bibfnamefont{P.}~\bibnamefont{Minkowski}},
  \bibinfo{journal}{Phys. Lett.} \textbf{\bibinfo{volume}{B267}},
  \bibinfo{pages}{395} (\bibinfo{year}{1991}).

\bibitem[{\citenamefont{Rodejohann and Yaguna}(2015)}]{Rodejohann:2015lca}
\bibinfo{author}{\bibfnamefont{W.}~\bibnamefont{Rodejohann}} \bibnamefont{and}
  \bibinfo{author}{\bibfnamefont{C.~E.} \bibnamefont{Yaguna}},
  \bibinfo{journal}{JCAP} \textbf{\bibinfo{volume}{1512}}, \bibinfo{pages}{032}
  (\bibinfo{year}{2015}), \eprint{1509.04036}.

\bibitem[{\citenamefont{Okada and Seto}(2010)}]{Okada:2010wd}
\bibinfo{author}{\bibfnamefont{N.}~\bibnamefont{Okada}} \bibnamefont{and}
  \bibinfo{author}{\bibfnamefont{O.}~\bibnamefont{Seto}},
  \bibinfo{journal}{Phys. Rev.} \textbf{\bibinfo{volume}{D82}},
  \bibinfo{pages}{023507} (\bibinfo{year}{2010}), \eprint{1002.2525}.

\bibitem[{\citenamefont{Dasgupta and Borah}(2014)}]{Dasgupta:2014hha}
\bibinfo{author}{\bibfnamefont{A.}~\bibnamefont{Dasgupta}} \bibnamefont{and}
  \bibinfo{author}{\bibfnamefont{D.}~\bibnamefont{Borah}},
  \bibinfo{journal}{Nucl. Phys.} \textbf{\bibinfo{volume}{B889}},
  \bibinfo{pages}{637} (\bibinfo{year}{2014}), \eprint{1404.5261}.

\bibitem[{\citenamefont{Okada and Okada}(2017)}]{Okada:2016tci}
\bibinfo{author}{\bibfnamefont{N.}~\bibnamefont{Okada}} \bibnamefont{and}
  \bibinfo{author}{\bibfnamefont{S.}~\bibnamefont{Okada}},
  \bibinfo{journal}{Phys. Rev.} \textbf{\bibinfo{volume}{D95}},
  \bibinfo{pages}{035025} (\bibinfo{year}{2017}), \eprint{1611.02672}.

\bibitem[{\citenamefont{Klasen et~al.}(2017)\citenamefont{Klasen, Lyonnet, and
  Queiroz}}]{Klasen:2016qux}
\bibinfo{author}{\bibfnamefont{M.}~\bibnamefont{Klasen}},
  \bibinfo{author}{\bibfnamefont{F.}~\bibnamefont{Lyonnet}}, \bibnamefont{and}
  \bibinfo{author}{\bibfnamefont{F.~S.} \bibnamefont{Queiroz}},
  \bibinfo{journal}{Eur. Phys. J.} \textbf{\bibinfo{volume}{C77}},
  \bibinfo{pages}{348} (\bibinfo{year}{2017}), \eprint{1607.06468}.

\bibitem[{\citenamefont{Okada and Orikasa}(2012)}]{Okada:2012sg}
\bibinfo{author}{\bibfnamefont{N.}~\bibnamefont{Okada}} \bibnamefont{and}
  \bibinfo{author}{\bibfnamefont{Y.}~\bibnamefont{Orikasa}},
  \bibinfo{journal}{Phys. Rev.} \textbf{\bibinfo{volume}{D85}},
  \bibinfo{pages}{115006} (\bibinfo{year}{2012}), \eprint{1202.1405}.

\bibitem[{\citenamefont{Guo et~al.}(2015)\citenamefont{Guo, Kang, Ko, and
  Orikasa}}]{Guo:2015lxa}
\bibinfo{author}{\bibfnamefont{J.}~\bibnamefont{Guo}},
  \bibinfo{author}{\bibfnamefont{Z.}~\bibnamefont{Kang}},
  \bibinfo{author}{\bibfnamefont{P.}~\bibnamefont{Ko}}, \bibnamefont{and}
  \bibinfo{author}{\bibfnamefont{Y.}~\bibnamefont{Orikasa}},
  \bibinfo{journal}{Phys. Rev.} \textbf{\bibinfo{volume}{D91}},
  \bibinfo{pages}{115017} (\bibinfo{year}{2015}), \eprint{1502.00508}.

\bibitem[{\citenamefont{Basak and Mondal}(2014)}]{Basak:2013cga}
\bibinfo{author}{\bibfnamefont{T.}~\bibnamefont{Basak}} \bibnamefont{and}
  \bibinfo{author}{\bibfnamefont{T.}~\bibnamefont{Mondal}},
  \bibinfo{journal}{Phys. Rev.} \textbf{\bibinfo{volume}{D89}},
  \bibinfo{pages}{063527} (\bibinfo{year}{2014}), \eprint{1308.0023}.

\bibitem[{\citenamefont{Okada and Okada}(2016)}]{Okada:2016gsh}
\bibinfo{author}{\bibfnamefont{N.}~\bibnamefont{Okada}} \bibnamefont{and}
  \bibinfo{author}{\bibfnamefont{S.}~\bibnamefont{Okada}},
  \bibinfo{journal}{Phys. Rev.} \textbf{\bibinfo{volume}{D93}},
  \bibinfo{pages}{075003} (\bibinfo{year}{2016}), \eprint{1601.07526}.

\bibitem[{\citenamefont{Carena et~al.}(2004)\citenamefont{Carena, Daleo,
  Dobrescu, and Tait}}]{Carena:2004xs}
\bibinfo{author}{\bibfnamefont{M.}~\bibnamefont{Carena}},
  \bibinfo{author}{\bibfnamefont{A.}~\bibnamefont{Daleo}},
  \bibinfo{author}{\bibfnamefont{B.~A.} \bibnamefont{Dobrescu}},
  \bibnamefont{and} \bibinfo{author}{\bibfnamefont{T.~M.~P.}
  \bibnamefont{Tait}}, \bibinfo{journal}{Phys. Rev.}
  \textbf{\bibinfo{volume}{D70}}, \bibinfo{pages}{093009}
  (\bibinfo{year}{2004}), \eprint{hep-ph/0408098}.

\bibitem[{\citenamefont{Cacciapaglia et~al.}(2006)\citenamefont{Cacciapaglia,
  Csaki, Marandella, and Strumia}}]{Cacciapaglia:2006pk}
\bibinfo{author}{\bibfnamefont{G.}~\bibnamefont{Cacciapaglia}},
  \bibinfo{author}{\bibfnamefont{C.}~\bibnamefont{Csaki}},
  \bibinfo{author}{\bibfnamefont{G.}~\bibnamefont{Marandella}},
  \bibnamefont{and} \bibinfo{author}{\bibfnamefont{A.}~\bibnamefont{Strumia}},
  \bibinfo{journal}{Phys. Rev.} \textbf{\bibinfo{volume}{D74}},
  \bibinfo{pages}{033011} (\bibinfo{year}{2006}), \eprint{hep-ph/0604111}.

\bibitem[{\citenamefont{Ma}(2006)}]{Ma:2006km}
\bibinfo{author}{\bibfnamefont{E.}~\bibnamefont{Ma}}, \bibinfo{journal}{Phys.
  Rev.} \textbf{\bibinfo{volume}{D73}}, \bibinfo{pages}{077301}
  (\bibinfo{year}{2006}), \eprint{hep-ph/0601225}.

\bibitem[{\citenamefont{Cai et~al.}(2017)\citenamefont{Cai, Herrero-GarcÌa,
  Schmidt, Vicente, and Volkas}}]{Cai:2017jrq}
\bibinfo{author}{\bibfnamefont{Y.}~\bibnamefont{Cai}},
  \bibinfo{author}{\bibfnamefont{J.}~\bibnamefont{Herrero-GarcÌa}},
  \bibinfo{author}{\bibfnamefont{M.~A.} \bibnamefont{Schmidt}},
  \bibinfo{author}{\bibfnamefont{A.}~\bibnamefont{Vicente}}, \bibnamefont{and}
  \bibinfo{author}{\bibfnamefont{R.~R.} \bibnamefont{Volkas}}
  (\bibinfo{year}{2017}), \eprint{1706.08524}.

\bibitem[{\citenamefont{Montero and Pleitez}(2009)}]{Montero:2007cd}
\bibinfo{author}{\bibfnamefont{J.~C.} \bibnamefont{Montero}} \bibnamefont{and}
  \bibinfo{author}{\bibfnamefont{V.}~\bibnamefont{Pleitez}},
  \bibinfo{journal}{Phys. Lett.} \textbf{\bibinfo{volume}{B675}},
  \bibinfo{pages}{64} (\bibinfo{year}{2009}), \eprint{0706.0473}.

\bibitem[{\citenamefont{Ma and Srivastava}(2015)}]{Ma:2014qra}
\bibinfo{author}{\bibfnamefont{E.}~\bibnamefont{Ma}} \bibnamefont{and}
  \bibinfo{author}{\bibfnamefont{R.}~\bibnamefont{Srivastava}},
  \bibinfo{journal}{Phys. Lett.} \textbf{\bibinfo{volume}{B741}},
  \bibinfo{pages}{217} (\bibinfo{year}{2015}), \eprint{1411.5042}.

\bibitem[{\citenamefont{Ma et~al.}(2015)\citenamefont{Ma, Pollard, Srivastava,
  and Zakeri}}]{Ma:2015mjd}
\bibinfo{author}{\bibfnamefont{E.}~\bibnamefont{Ma}},
  \bibinfo{author}{\bibfnamefont{N.}~\bibnamefont{Pollard}},
  \bibinfo{author}{\bibfnamefont{R.}~\bibnamefont{Srivastava}},
  \bibnamefont{and} \bibinfo{author}{\bibfnamefont{M.}~\bibnamefont{Zakeri}},
  \bibinfo{journal}{Phys. Lett.} \textbf{\bibinfo{volume}{B750}},
  \bibinfo{pages}{135} (\bibinfo{year}{2015}), \eprint{1507.03943}.

\bibitem[{\citenamefont{S·nchez-Vega et~al.}(2014)\citenamefont{S·nchez-Vega,
  Montero, and Schmitz}}]{Sanchez-Vega:2014rka}
\bibinfo{author}{\bibfnamefont{B.~L.} \bibnamefont{S·nchez-Vega}},
  \bibinfo{author}{\bibfnamefont{J.~C.} \bibnamefont{Montero}},
  \bibnamefont{and} \bibinfo{author}{\bibfnamefont{E.~R.}
  \bibnamefont{Schmitz}}, \bibinfo{journal}{Phys. Rev.}
  \textbf{\bibinfo{volume}{D90}}, \bibinfo{pages}{055022}
  (\bibinfo{year}{2014}), \eprint{1404.5973}.

\bibitem[{\citenamefont{S·nchez-Vega and Schmitz}(2015)}]{Sanchez-Vega:2015qva}
\bibinfo{author}{\bibfnamefont{B.~L.} \bibnamefont{S·nchez-Vega}}
  \bibnamefont{and} \bibinfo{author}{\bibfnamefont{E.~R.}
  \bibnamefont{Schmitz}}, \bibinfo{journal}{Phys. Rev.}
  \textbf{\bibinfo{volume}{D92}}, \bibinfo{pages}{053007}
  (\bibinfo{year}{2015}), \eprint{1505.03595}.

\bibitem[{\citenamefont{Singirala et~al.}(2017)\citenamefont{Singirala,
  Mohanta, and Patra}}]{Singirala:2017see}
\bibinfo{author}{\bibfnamefont{S.}~\bibnamefont{Singirala}},
  \bibinfo{author}{\bibfnamefont{R.}~\bibnamefont{Mohanta}}, \bibnamefont{and}
  \bibinfo{author}{\bibfnamefont{S.}~\bibnamefont{Patra}}
  (\bibinfo{year}{2017}), \eprint{1704.01107}.

\bibitem[{\citenamefont{Nomura and Okada}(2017)}]{Nomura:2017vzp}
\bibinfo{author}{\bibfnamefont{T.}~\bibnamefont{Nomura}} \bibnamefont{and}
  \bibinfo{author}{\bibfnamefont{H.}~\bibnamefont{Okada}}
  (\bibinfo{year}{2017}), \eprint{1705.08309}.

\bibitem[{\citenamefont{Wang and Han}(2015)}]{Wang:2015saa}
\bibinfo{author}{\bibfnamefont{W.}~\bibnamefont{Wang}} \bibnamefont{and}
  \bibinfo{author}{\bibfnamefont{Z.-L.} \bibnamefont{Han}},
  \bibinfo{journal}{Phys. Rev.} \textbf{\bibinfo{volume}{D92}},
  \bibinfo{pages}{095001} (\bibinfo{year}{2015}), \eprint{1508.00706}.

\bibitem[{\citenamefont{Patra et~al.}(2016)\citenamefont{Patra, Rodejohann, and
  Yaguna}}]{Patra:2016ofq}
\bibinfo{author}{\bibfnamefont{S.}~\bibnamefont{Patra}},
  \bibinfo{author}{\bibfnamefont{W.}~\bibnamefont{Rodejohann}},
  \bibnamefont{and} \bibinfo{author}{\bibfnamefont{C.~E.}
  \bibnamefont{Yaguna}}, \bibinfo{journal}{JHEP} \textbf{\bibinfo{volume}{09}},
  \bibinfo{pages}{076} (\bibinfo{year}{2016}), \eprint{1607.04029}.

\bibitem[{\citenamefont{Dupuis}(2016)}]{Dupuis:2016fda}
\bibinfo{author}{\bibfnamefont{G.}~\bibnamefont{Dupuis}},
  \bibinfo{journal}{JHEP} \textbf{\bibinfo{volume}{07}}, \bibinfo{pages}{008}
  (\bibinfo{year}{2016}), \eprint{1604.04552}.

\bibitem[{\citenamefont{Kolb and Turner}(1990)}]{Kolb:1990vq}
\bibinfo{author}{\bibfnamefont{E.~W.} \bibnamefont{Kolb}} \bibnamefont{and}
  \bibinfo{author}{\bibfnamefont{M.~S.} \bibnamefont{Turner}},
  \bibinfo{journal}{Front. Phys.} \textbf{\bibinfo{volume}{69}},
  \bibinfo{pages}{1} (\bibinfo{year}{1990}).

\bibitem[{\citenamefont{Scherrer and Turner}(1986)}]{Scherrer:1985zt}
\bibinfo{author}{\bibfnamefont{R.~J.} \bibnamefont{Scherrer}} \bibnamefont{and}
  \bibinfo{author}{\bibfnamefont{M.~S.} \bibnamefont{Turner}},
  \bibinfo{journal}{Phys. Rev.} \textbf{\bibinfo{volume}{D33}},
  \bibinfo{pages}{1585} (\bibinfo{year}{1986}), \bibinfo{note}{[Erratum: Phys.
  Rev.D34,3263(1986)]}.

\bibitem[{\citenamefont{Jungman et~al.}(1996)\citenamefont{Jungman,
  Kamionkowski, and Griest}}]{Jungman:1995df}
\bibinfo{author}{\bibfnamefont{G.}~\bibnamefont{Jungman}},
  \bibinfo{author}{\bibfnamefont{M.}~\bibnamefont{Kamionkowski}},
  \bibnamefont{and} \bibinfo{author}{\bibfnamefont{K.}~\bibnamefont{Griest}},
  \bibinfo{journal}{Phys. Rept.} \textbf{\bibinfo{volume}{267}},
  \bibinfo{pages}{195} (\bibinfo{year}{1996}), \eprint{hep-ph/9506380}.

\bibitem[{\citenamefont{Gondolo and Gelmini}(1991)}]{Gondolo:1990dk}
\bibinfo{author}{\bibfnamefont{P.}~\bibnamefont{Gondolo}} \bibnamefont{and}
  \bibinfo{author}{\bibfnamefont{G.}~\bibnamefont{Gelmini}},
  \bibinfo{journal}{Nucl. Phys.} \textbf{\bibinfo{volume}{B360}},
  \bibinfo{pages}{145} (\bibinfo{year}{1991}).

\bibitem[{\citenamefont{Griest and Seckel}(1991)}]{Griest:1990kh}
\bibinfo{author}{\bibfnamefont{K.}~\bibnamefont{Griest}} \bibnamefont{and}
  \bibinfo{author}{\bibfnamefont{D.}~\bibnamefont{Seckel}},
  \bibinfo{journal}{Phys. Rev.} \textbf{\bibinfo{volume}{D43}},
  \bibinfo{pages}{3191} (\bibinfo{year}{1991}).

\bibitem[{\citenamefont{Edsjo and Gondolo}(1997)}]{Edsjo:1997bg}
\bibinfo{author}{\bibfnamefont{J.}~\bibnamefont{Edsjo}} \bibnamefont{and}
  \bibinfo{author}{\bibfnamefont{P.}~\bibnamefont{Gondolo}},
  \bibinfo{journal}{Phys. Rev.} \textbf{\bibinfo{volume}{D56}},
  \bibinfo{pages}{1879} (\bibinfo{year}{1997}), \eprint{hep-ph/9704361}.

\bibitem[{\citenamefont{Bell et~al.}(2014)\citenamefont{Bell, Cai, and
  Medina}}]{Bell:2013wua}
\bibinfo{author}{\bibfnamefont{N.~F.} \bibnamefont{Bell}},
  \bibinfo{author}{\bibfnamefont{Y.}~\bibnamefont{Cai}}, \bibnamefont{and}
  \bibinfo{author}{\bibfnamefont{A.~D.} \bibnamefont{Medina}},
  \bibinfo{journal}{Phys. Rev.} \textbf{\bibinfo{volume}{D89}},
  \bibinfo{pages}{115001} (\bibinfo{year}{2014}), \eprint{1311.6169}.

\bibitem[{\citenamefont{Lundstrom et~al.}(2009)\citenamefont{Lundstrom,
  Gustafsson, and Edsjo}}]{Lundstrom:2008ai}
\bibinfo{author}{\bibfnamefont{E.}~\bibnamefont{Lundstrom}},
  \bibinfo{author}{\bibfnamefont{M.}~\bibnamefont{Gustafsson}},
  \bibnamefont{and} \bibinfo{author}{\bibfnamefont{J.}~\bibnamefont{Edsjo}},
  \bibinfo{journal}{Phys. Rev.} \textbf{\bibinfo{volume}{D79}},
  \bibinfo{pages}{035013} (\bibinfo{year}{2009}), \eprint{0810.3924}.

\bibitem[{\citenamefont{Pierce and Thaler}(2007)}]{Pierce:2007ut}
\bibinfo{author}{\bibfnamefont{A.}~\bibnamefont{Pierce}} \bibnamefont{and}
  \bibinfo{author}{\bibfnamefont{J.}~\bibnamefont{Thaler}},
  \bibinfo{journal}{JHEP} \textbf{\bibinfo{volume}{08}}, \bibinfo{pages}{026}
  (\bibinfo{year}{2007}), \eprint{hep-ph/0703056}.

\bibitem[{\citenamefont{Belanger et~al.}(2015)\citenamefont{Belanger, Dumont,
  Goudelis, Herrmann, Kraml, and Sengupta}}]{Belanger:2015kga}
\bibinfo{author}{\bibfnamefont{G.}~\bibnamefont{Belanger}},
  \bibinfo{author}{\bibfnamefont{B.}~\bibnamefont{Dumont}},
  \bibinfo{author}{\bibfnamefont{A.}~\bibnamefont{Goudelis}},
  \bibinfo{author}{\bibfnamefont{B.}~\bibnamefont{Herrmann}},
  \bibinfo{author}{\bibfnamefont{S.}~\bibnamefont{Kraml}}, \bibnamefont{and}
  \bibinfo{author}{\bibfnamefont{D.}~\bibnamefont{Sengupta}},
  \bibinfo{journal}{Phys. Rev.} \textbf{\bibinfo{volume}{D91}},
  \bibinfo{pages}{115011} (\bibinfo{year}{2015}), \eprint{1503.07367}.

\bibitem[{\citenamefont{Barbieri et~al.}(2006)\citenamefont{Barbieri, Hall, and
  Rychkov}}]{Barbieri:2006dq}
\bibinfo{author}{\bibfnamefont{R.}~\bibnamefont{Barbieri}},
  \bibinfo{author}{\bibfnamefont{L.~J.} \bibnamefont{Hall}}, \bibnamefont{and}
  \bibinfo{author}{\bibfnamefont{V.~S.} \bibnamefont{Rychkov}},
  \bibinfo{journal}{Phys. Rev.} \textbf{\bibinfo{volume}{D74}},
  \bibinfo{pages}{015007} (\bibinfo{year}{2006}), \eprint{hep-ph/0603188}.

\bibitem[{\citenamefont{Lopez~Honorez and Yaguna}(2011)}]{LopezHonorez:2010tb}
\bibinfo{author}{\bibfnamefont{L.}~\bibnamefont{Lopez~Honorez}}
  \bibnamefont{and} \bibinfo{author}{\bibfnamefont{C.~E.}
  \bibnamefont{Yaguna}}, \bibinfo{journal}{JCAP}
  \textbf{\bibinfo{volume}{1101}}, \bibinfo{pages}{002} (\bibinfo{year}{2011}),
  \eprint{1011.1411}.

\bibitem[{\citenamefont{Peskin and Takeuchi}(1992)}]{Peskin:1991sw}
\bibinfo{author}{\bibfnamefont{M.~E.} \bibnamefont{Peskin}} \bibnamefont{and}
  \bibinfo{author}{\bibfnamefont{T.}~\bibnamefont{Takeuchi}},
  \bibinfo{journal}{Phys. Rev.} \textbf{\bibinfo{volume}{D46}},
  \bibinfo{pages}{381} (\bibinfo{year}{1992}).

\bibitem[{\citenamefont{Aad et~al.}(2015)}]{Aad:2015pla}
\bibinfo{author}{\bibfnamefont{G.}~\bibnamefont{Aad}} \bibnamefont{et~al.}
  (\bibinfo{collaboration}{ATLAS}), \bibinfo{journal}{JHEP}
  \textbf{\bibinfo{volume}{11}}, \bibinfo{pages}{206} (\bibinfo{year}{2015}),
  \eprint{1509.00672}.

\bibitem[{\citenamefont{Berlin et~al.}(2014)\citenamefont{Berlin, Hooper, and
  McDermott}}]{Berlin:2014tja}
\bibinfo{author}{\bibfnamefont{A.}~\bibnamefont{Berlin}},
  \bibinfo{author}{\bibfnamefont{D.}~\bibnamefont{Hooper}}, \bibnamefont{and}
  \bibinfo{author}{\bibfnamefont{S.~D.} \bibnamefont{McDermott}},
  \bibinfo{journal}{Phys. Rev.} \textbf{\bibinfo{volume}{D89}},
  \bibinfo{pages}{115022} (\bibinfo{year}{2014}), \eprint{1404.0022}.

\bibitem[{\citenamefont{Junnarkar and Walker-Loud}(2013)}]{Junnarkar:2013ac}
\bibinfo{author}{\bibfnamefont{P.}~\bibnamefont{Junnarkar}} \bibnamefont{and}
  \bibinfo{author}{\bibfnamefont{A.}~\bibnamefont{Walker-Loud}},
  \bibinfo{journal}{Phys. Rev.} \textbf{\bibinfo{volume}{D87}},
  \bibinfo{pages}{114510} (\bibinfo{year}{2013}), \eprint{1301.1114}.

\bibitem[{\citenamefont{Aaboud et~al.}(2017)}]{Aaboud:2017buh}
\bibinfo{author}{\bibfnamefont{M.}~\bibnamefont{Aaboud}} \bibnamefont{et~al.}
  (\bibinfo{collaboration}{ATLAS}) (\bibinfo{year}{2017}), \eprint{1707.02424}.

\bibitem[{\citenamefont{Giedt et~al.}(2009)\citenamefont{Giedt, Thomas, and
  Young}}]{Giedt:2009mr}
\bibinfo{author}{\bibfnamefont{J.}~\bibnamefont{Giedt}},
  \bibinfo{author}{\bibfnamefont{A.~W.} \bibnamefont{Thomas}},
  \bibnamefont{and} \bibinfo{author}{\bibfnamefont{R.~D.} \bibnamefont{Young}},
  \bibinfo{journal}{Phys. Rev. Lett.} \textbf{\bibinfo{volume}{103}},
  \bibinfo{pages}{201802} (\bibinfo{year}{2009}), \eprint{0907.4177}.

\bibitem[{\citenamefont{Mambrini}(2011)}]{Mambrini:2011ik}
\bibinfo{author}{\bibfnamefont{Y.}~\bibnamefont{Mambrini}},
  \bibinfo{journal}{Phys. Rev.} \textbf{\bibinfo{volume}{D84}},
  \bibinfo{pages}{115017} (\bibinfo{year}{2011}), \eprint{1108.0671}.

\bibitem[{\citenamefont{Ahnen et~al.}(2016)}]{Ahnen:2016qkx}
\bibinfo{author}{\bibfnamefont{M.~L.} \bibnamefont{Ahnen}} \bibnamefont{et~al.}
  (\bibinfo{collaboration}{Fermi-LAT, MAGIC}), \bibinfo{journal}{JCAP}
  \textbf{\bibinfo{volume}{1602}}, \bibinfo{pages}{039} (\bibinfo{year}{2016}),
  \eprint{1601.06590}.

\bibitem[{\citenamefont{Abramowski et~al.}(2015)}]{HESS:2015cda}
\bibinfo{author}{\bibfnamefont{A.}~\bibnamefont{Abramowski}}
  \bibnamefont{et~al.} (\bibinfo{collaboration}{H.E.S.S.}),
  \bibinfo{journal}{Phys. Rev. Lett.} \textbf{\bibinfo{volume}{114}},
  \bibinfo{pages}{081301} (\bibinfo{year}{2015}), \eprint{1502.03244}.

\bibitem[{\citenamefont{Abdallah et~al.}(2016)}]{Abdallah:2016ygi}
\bibinfo{author}{\bibfnamefont{H.}~\bibnamefont{Abdallah}} \bibnamefont{et~al.}
  (\bibinfo{collaboration}{H.E.S.S.}), \bibinfo{journal}{Phys. Rev. Lett.}
  \textbf{\bibinfo{volume}{117}}, \bibinfo{pages}{111301}
  (\bibinfo{year}{2016}), \eprint{1607.08142}.

\bibitem[{\citenamefont{Baldini et~al.}(2016)}]{TheMEG:2016wtm}
\bibinfo{author}{\bibfnamefont{A.~M.} \bibnamefont{Baldini}}
  \bibnamefont{et~al.} (\bibinfo{collaboration}{MEG}), \bibinfo{journal}{Eur.
  Phys. J.} \textbf{\bibinfo{volume}{C76}}, \bibinfo{pages}{434}
  (\bibinfo{year}{2016}), \eprint{1605.05081}.

\bibitem[{\citenamefont{Lavoura}(2003)}]{Lavoura:2003xp}
\bibinfo{author}{\bibfnamefont{L.}~\bibnamefont{Lavoura}},
  \bibinfo{journal}{Eur. Phys. J.} \textbf{\bibinfo{volume}{C29}},
  \bibinfo{pages}{191} (\bibinfo{year}{2003}), \eprint{hep-ph/0302221}.

\bibitem[{\citenamefont{Toma and Vicente}(2014)}]{Toma:2013zsa}
\bibinfo{author}{\bibfnamefont{T.}~\bibnamefont{Toma}} \bibnamefont{and}
  \bibinfo{author}{\bibfnamefont{A.}~\bibnamefont{Vicente}},
  \bibinfo{journal}{JHEP} \textbf{\bibinfo{volume}{01}}, \bibinfo{pages}{160}
  (\bibinfo{year}{2014}), \eprint{1312.2840}.

\bibitem[{\citenamefont{Bellgardt et~al.}(1988)}]{Bellgardt:1987du}
\bibinfo{author}{\bibfnamefont{U.}~\bibnamefont{Bellgardt}}
  \bibnamefont{et~al.} (\bibinfo{collaboration}{SINDRUM}),
  \bibinfo{journal}{Nucl. Phys.} \textbf{\bibinfo{volume}{B299}},
  \bibinfo{pages}{1} (\bibinfo{year}{1988}).

\bibitem[{\citenamefont{Bertl et~al.}(2006)}]{Bertl:2006up}
\bibinfo{author}{\bibfnamefont{W.~H.} \bibnamefont{Bertl}} \bibnamefont{et~al.}
  (\bibinfo{collaboration}{SINDRUM II}), \bibinfo{journal}{Eur. Phys. J.}
  \textbf{\bibinfo{volume}{C47}}, \bibinfo{pages}{337} (\bibinfo{year}{2006}).

\bibitem[{\citenamefont{Basso et~al.}(2009)\citenamefont{Basso, Belyaev,
  Moretti, and Shepherd-Themistocleous}}]{Basso:2008iv}
\bibinfo{author}{\bibfnamefont{L.}~\bibnamefont{Basso}},
  \bibinfo{author}{\bibfnamefont{A.}~\bibnamefont{Belyaev}},
  \bibinfo{author}{\bibfnamefont{S.}~\bibnamefont{Moretti}}, \bibnamefont{and}
  \bibinfo{author}{\bibfnamefont{C.~H.} \bibnamefont{Shepherd-Themistocleous}},
  \bibinfo{journal}{Phys. Rev.} \textbf{\bibinfo{volume}{D80}},
  \bibinfo{pages}{055030} (\bibinfo{year}{2009}), \eprint{0812.4313}.

\bibitem[{\citenamefont{Khachatryan et~al.}(2017)}]{Khachatryan:2016qkc}
\bibinfo{author}{\bibfnamefont{V.}~\bibnamefont{Khachatryan}}
  \bibnamefont{et~al.} (\bibinfo{collaboration}{CMS}), \bibinfo{journal}{JHEP}
  \textbf{\bibinfo{volume}{02}}, \bibinfo{pages}{048} (\bibinfo{year}{2017}),
  \eprint{1611.06594}.

\end{thebibliography}

\end{document}